%% file: EURETILE-D6-2-HW-IPdev-ebook.tex
\documentclass{euretile-deliverables-ebook}
\workpackagenumber{WP6}
\workpackagename{Innovations on HW Intellectual Properties}

\deliverablenumber{D6.2}
\deliverablename{Second Report on Innovations on HW Intellectual Properties}

\leadcontractor{INFN Roma}
\duedatesubmission{2013-02-28}

\authorinstitutionA{Istituto Nazionale di Fisica Nucleare - sezione di
  Roma (INFN)}
\authornameA{Roberto Ammendola, Andrea Biagioni, Ottorino Frezza,
  Michela Giovagnoli, Francesca Lo Cicero, Alessandro Lonardo, Pier
  Stanislao Paolucci, Davide Rossetti, Francesco Simula, Laura
  Tosoratto, Piero Vicini}

\authorinstitutionB{RWTH Aachen University - ISS \& SSS (RWTH)}
\authornameB{Jovana Jovic, Rainer Leupers, Luis Murillo, Christoph
  Schumacher, Jan Henrik Weinstock}

\authorinstitutionC{ETH Zurich - The Swiss Federal Insitute of
  Technology Zurich (ETHZ)}
\authornameC{Iuliana Bacivarov, Devendra Rai, Lars Schor, Lothar
  Thiele, Hoeseok Yank}

\authorinstitutionD{Université Joseph Fourier - Grenoble Institute of
  Technology - TIMA Laboratories (UJF/INP TIMA)}
\authornameD{Ashraf El Antably, Ikbel Belaid, Clément Deschamps,
  Nicolas Fournel, Mohamad Jaber, Julian Michaud, Frédéric Petrot,
  Etienne Ripert, Frédéric Rousseau}

\authorinstitutionE{Target Compiler Technologies (TARGET)}
\authornameE{Erik Brockmeyer, Koen Danckaert, Jos Decoster, Gert
  Goossens, Werner Geurts}

\usepackage{multirow, array}
\usepackage{eurosym}
\usepackage{xspace}
\usepackage{longtable}
\usepackage{placeins}
\usepackage{rotating}
\usepackage[font=footnotesize]{caption}
\newcommand{\hw}[1]{\texttt{#1}}
\newcommand{\sw}[1]{\texttt{#1}}
\newcommand{\us}{\,$\mu$s\xspace}
\newcommand{\quong}{QUonG\xspace}
\newcommand{\apenet}{APEnet\xspace}
\newcommand{\apenetp}{APEnet+\xspace}

\newcommand{\etc}{\textit{etc.}\xspace}
\newcommand{\cpp}{C$+$$+$\xspace}

\newcommand{\muC}{$\mu$C\xspace}
\newcommand{\euretile}{EURETILE\xspace}
\newcommand{\lofamo}{LO{\textbar}FA{\textbar}MO\xspace}
\newcommand{\lifama}{\texttt{LiFaMa}\xspace}
\newcommand{\PtoP}{\textit{peer-to-peer}\xspace}
\newcommand{\nios}{\texttt{Nios~II}\xspace}
\newcommand{\ptoptx}{\texttt{GPU\_P2P\_TX}\xspace}
\newcommand{\PCIe}{PCIe\xspace}
\newcommand{\nvidia}{NVIDIA\xspace}

\newcommand{\dwr}{\texttt{DNP Watchdog Register}\xspace}
\newcommand{\hwr}{\texttt{Host Watchdog Register}\xspace}
\newcommand{\txlifama}{\textit{TX LiFaMa}\xspace}
\newcommand{\rxlifama}{\textit{RX LiFaMa}\xspace}
\newcommand{\ldm}{\texttt{LiFaMa Diagnostic Message}\xspace}
\newcommand{\hfm}{\textit{Host Fault Manager}\xspace}
\newcommand{\dfm}{\texttt{DNP Fault Manager}\xspace}

\newcommand{\header}{\texttt{Header}\xspace}
\newcommand{\footer}{\texttt{Footer}\xspace}
\newcommand{\payload}{\texttt{Payload}\xspace}
\newcommand{\credit}{\texttt{Credit}\xspace}
\newcommand{\magic}{\texttt{Magic}\xspace}
\newcommand{\start}{\texttt{Start}\xspace}

\newcommand{\CQ}{\textit{Completion Event Queues}\xspace}
\newcommand{\TXaevent}{\textit{Channel 0 TX Event}\xspace}
\newcommand{\TXbevent}{\textit{Channel 1 TX Event}\xspace}
\newcommand{\RXevent}{\textit{RX Event}\xspace}
\newcommand{\GPUTXevent}{\textit{GPU TX Event}\xspace}
\newcommand{\niosevent}{\textit{NIOS Event}\xspace}
\newcommand{\SIprotocol}{\textit{Software Interface Protocol}\xspace}

\newcommand{\ringbuffer}{\texttt{Ring Buffer}\xspace}
\newcommand{\writepointer}{\texttt{Write Pointer}\xspace}
\newcommand{\readpointer}{\texttt{Read Pointer}\xspace}
\newcommand{\wrptr}{\texttt{wr\_ptr}\xspace}
\newcommand{\rdptr}{\texttt{rd\_ptr}\xspace}

\newcommand{\malloc}{\texttt{malloc()}\xspace}

\newcommand{\checkalign}{\textit{Check Align}\xspace}
\newcommand{\restartalign}{\textit{Restart Align}\xspace}

\begin{document}

% Project Title
\maketitle

% Deliverable Title
\makesubtitle

{\Large
  \textbf{Abstract}
}

This is the second of a planned collection of four yearly volumes
describing the deployment of a heterogeneous many-core platform for
experiments on scalable custom interconnects and management of fault
and critical events, applied to many-process applications.
This volume covers several topics, among which: 1- a system for
awareness of faults and critical events (named LO|FA|MO) on
experimental heterogeneous many-core hardware platforms; 2- the
integration and test of the experimental hardware heterogeneous
many-core platform QUoNG, based on the APEnet+ custom interconnect; 3-
the design of a Software-Programmable Distributed Network Processor
architecture (DNP) using ASIP technology; 4- the initial stages of
design of a new DNP generation onto a 28nm FPGA.
These developments were performed in the framework of the EURETILE
European Project under the Grant Agreement no. 247846.
\clearpage

% Table of content
\setcounter{tocdepth}{2}
\tableofcontents
\cleardoublepage

\input{chapter1}

\clearpage
\input{chapter2}
\clearpage
%
%%\input{sections/chapter3}
%%\clearpage
%
\input{chapter4}

\clearpage
\input{chapter5}
\clearpage
\input{glossary}
\clearpage
%

%\input{sections/summary}
%\clearpage
%
%\input{sections/appendix}
%\clearpage
\bibliographystyle{unsrt}
\bibliography{bibliography}
\cleardoublepage

\lhead{}

%%\vspace*{\fill}

%%\begin{left}
\small{\textbf{A heterogeneous \mbox{many-core} platform for
    experiments on scalable custom interconnects and management of
    fault and critical events, applied to \mbox{many-process}
    applications: Vol. II, 2012 technical report}}
%%\end{left}

\vspace{1.0cm}

\small{Roberto Ammendola, Andrea Biagioni, Ottorino Frezza, Werner
  Geurts, Gert Goossens, Francesca Lo Cicero, Alessandro Lonardo, Pier
  Stanislao Paolucci, Davide Rossetti, Francesco Simula, Laura
  Tosoratto, Piero Vicini}

\vspace{8.0cm}

\small{The EURETILE project, started in 2010, is funded by the
  European Commission through the Grant Agreement no.~247846, Call:
  FP7-ICT-2009-4 Objective FET-ICT-2009.8.1 Concurrent
  \mbox{Tera-device} Computing (see \url{http://www.euretile.eu}).}

%%\vspace*{\fill}
\cfoot{}
%%ci sono?

\end{document}

%% file: chapter1.tex
\section{Introduction}
\subsection{Project abstract}
\euretile investigates and implements \mbox{brain-inspired} and
\mbox{fault-tolerant} foundational innovations to the system
architecture of massively parallel tiled computer architectures and
the corresponding programming paradigm.

The execution targets are either a \mbox{many-tile} HW platform and a
\mbox{many-tile} simulator.
A set of SW process - HW tile mapping candidates is generated by the
holistic SW toolchain using a combination of analytic and
\mbox{bio-inspired} methods.
The Hardware dependent Software is then generated, providing
OS services with maximum efficiency/minimal overhead.
The \mbox{many-tile} simulator collects profiling data, closing the
loop of the SW tool chain.

Fine-grain parallelism inside processes is exploited by optimized
\mbox{intra-tile} compilation techniques, but the project focus is
above the level of the elementary tile.
The elementary HW tile is a \mbox{multi-processor}, which includes a
fault tolerant Distributed Network Processor (for \mbox{inter-tile}
communication) and ASIP accelerators.

Furthermore, \euretile investigates and implements the innovations for
equipping the elementary HW tile with high-bandwidth,
\mbox{low-latency} \mbox{brain-like} \mbox{inter-tile} communication
emulating 3 levels of connection hierarchy --- namely neural columns,
cortical areas and cortex --- and develops a dedicated cortical
simulation benchmark: DPSNN-STDP (Distributed Polychronous Spiking
Neural Net with synaptic Spiking Time Dependent Plasticity).
\euretile leverages on the \mbox{multi-tile} HW paradigm and SW
toolchain developed by the FET-ACA SHAPES Integrated Project
(2006-2009).

For an overview of the full project scope and its activities during
the first three years, see the \euretile 2010-2012
summary:\cite{EURETILE_2010_2012_summary}.

\subsection{Roles of the project partners}
The APE Parallel Computing Lab of INFN Roma is in charge of the
\euretile HW Design (\quong system~\cite{ammendola2011quong}/\apenetp
board~\cite{ammendola2011apenet+,ammendola2012apenet+}/DNP
(Distributed Network Processor~\cite{DNP2012}) and Scientific
Application Benchmarks.

The Computer Engineering and Networks Laboratory (TIK) of ETH Zurich
(Swiss Federal Institute of Technology) designs the \mbox{high-level}
explicit parallel programming and automatic mapping tool (DOL/DAL) and
a set of ''Embedded Systems'' benchmarks.

The Software for Systems on Silicon (SSS) of the ISS institute of RWTH
Aachen, investigates and provides the parallel simulation technology
and scalable \mbox{simulation-based} profiling/debugging support.

The TIMA Laboratory of the University Joseph Fourier in Grenoble
explores and deploys the HdS (Hardware dependent Software) including
the distributed OS architecture.

TARGET Compiler Technologies, the Belgian leading provider of
retargetable software tools and compilers for the design, programming,
and verification of \mbox{application-specific} processors (ASIPs), is
in charge of the HW/SW Co-design tools for custom components of the
\euretile architecture.

\subsection{Structure of this document}
This publication is a derivative of the project deliverable document
D6.2 and it is the second of a planned collection of four yearly
volumes reporting the disclosable activities of Workpackage 6 (WP6)
which is dedicated to the deployment of a heterogeneous
\mbox{many-core} platform for experiments on scalable custom
interconnects and management of fault and critical events, applied to
\mbox{many-process} applications. All project deliverable documents
have been \mbox{peer-reviewed} by a committee of four experts of the
field.

The 2012 activities of \textbf{WP6: Second Report on Innovations on HW
Intellectual Properties} planned to achieve multiple goals in the areas
of fault tolerance, experimental HPC Platform integration and design of
programmable DNP.

Fault tolerance issues focused on the implementation of the \lofamo
approach to fault awareness on the HPC experimental platform.
The activity, reported in chapter \ref{sec:lofamo} (see also
\cite{LOFAMO2013}), delivered a consolidated specification of the
\lofamo architecture, the preliminary integration of the VHDL code of
DNP Fault Manager, the introduction of a controller of the diagnostic
messages flow over the 3D Torus Network (the Link Fault Manager,
LiFaMa), and a first implementation of the Host Fault Manager as a
Linux daemon.

The HPC platform integration and test activity (detailed in chapter
\ref{sec:quong}) provided the mass production of the \apenetp boards 
and their integration in a 16 nodes,
16 Kcores, 32 TFlops \quong system, selected as the \euretile HPC experimental 
hardware platform. Extensive test sessions, based on synthetic benchmarks
and scientific application kernels, validated the architecture and provided
measurements of scalability and global performance analysis.
Furthermore, software development activities were
devoted to the support of the HW innovations and of the complete 
\euretile toolchain on the IP (\apenetp) aiming to achieve the best 
performances and usability for the hardware platform. Lastly, several
exploitation initiatives were launched in the area of high-level trigger
systems of the current and future High Energy Physics colliders.

In chapter \ref{sec:target2} we report on one activity related to the
design of ASIPs in the \euretile platform. This activity, jointly
carried out by INFN and TARGET, was the design of a
software-programmable DNP architecture using ASIP technology trying to
overcome the current latency limitations of the communication
structure in the HPC platform.

Finally, chapter \ref{sec:28nm} details the preliminary exploration of
the achievable performance of the last generation 28nm high-end FPGA.
Understanding the real capabilities of such components in our system
is the key point to plan future developments and designs.  Preliminary
porting of the DNP IP on the ALTERA Stratix V development kit allowed
to measure the target torus link speed, \mbox{micro-controller} top
frequency and hardware resources occupation.

\pagebreak
\section{Awareness of faults and critical events on the Experimental HW Platform} % - Laura
\label{sec:lofamo}
Fault-tolerance issues have been addressed in EURETILE since 2011 with
the \lofamo HW/SW approach~\cite{LOFAMO2013}, whose principles were
presented in D6.1~\cite{euretile:D6_1} have been applied in the DNP
SystemC model and lately demonstrated over the VEP simulator at the
2012 review meeting (see D5.2~\cite{euretile:D5_2}).\\
This section describes the implementation of \lofamo on \quong, the
EURETILE HW Platform, introduced by a complete and consolidated
specification of the \lofamo approach.
\label{sec:LOFAMOmain}
\subsection{Consolidated Specification}  %- Laura
\subsubsection{Introduction}
Features like resilience, power consumption, and system availability
strongly depend both on the complexity of individual components
(e.g. the gate count of each chip) and the number of components in the
system.  HPC systems in the peta/hexa-scale especially require
techniques that aim to maintain a low Failure In Time (FIT) ratio to
guarantee reasonable functionality.  A first step is the adoption of
hardware design techniques which improve the individual components to
reduce their FIT.  This design trend is already clear in the
transition between past tera-scale systems, adopting commodity
processors with 0.1\textdiv~0.5 fails per year per socket, to
peta-scale systems, where the failure rate could be reduced to 0.001
fails per year per socket \cite{Kogge:2008:exa} adopting hardware
design techniques like memory and bus encoding, memory scrubbing,
provision of spare processors and memories. Even considering those
reduced FIT rates and a very limited number of components, mission
critical/life support systems mandate for architectures adopting
double or triple redundancy. In practice, petaFLOPS designs based on
resilient sockets adopting such countermeasures are characterized by a
rate of system-stopping features in the range of a few days, while a
system failure rate in the range of few hours is displayed by systems
mounting less resilient sockets. \cite{Elnozahy:2009:res}.  Without
additional measures, the FIT rate of exa-scale systems becomes
unacceptable due to the scaling in the number of
components. Analogously, for what regards power and thermal issues,
each socket and component is nowadays designed keeping the energetic
concerns as key drivers but systemic countermeasures are required due
to the numerosity of components.  In our vision, a necessary feature
on larger scale architectures is the detection and collation of
relevant information about faults and critical events and, due to the
distributed nature of the system, the reliable propagation of this
awareness up through the system hierarchy. In other words, the system
must be rendered fault-aware to be able to choose and enact the actual
system fault response.

Based on these considerations, the \euretile project starts bottom-up
proposing a mechanism that creates a systemic awareness of fault and
critical events, the \lofamo design: a distributed, mutual watchdog
paradigm which enables the management of faults and critical events in
large scale systems.  The \lofamo design can complement a pro-active
mitigation approach, \ie the enforcing of preventive actions (\eg
preemptive process migration, dynamic clock frequency reduction for
individual components, dynamic routing strategies, \etc) before the
failure occurs, so as to avoid faults that can reasonably be expected
or minimize the impact of those that can not.

In this document, we will mainly focus on the requirements imposed by
exa-scale systems \cite{Savage:2010:qcd}, \ie an assembly of tens or
hundreds of thousands of processors, hundreds of I/O nodes and
thousands of disks, and future many-tile sockets; however, similar
techniques could be applied to large networks of independent,
autonomous devices.  Our approach should mitigate the performance
penalty and productivity reduction due to work loss and failure
recovery, obtainable using exclusively the conventional approach to
fault-tolerance (checkpointing/failure/rollback), which is foreseen to
be problematic \cite{Kogge:2008:exa}.  With the assistance of some
hardware components located in the DNP (Distributing Network
Processor) - the core of our \apenetp - the \lofamo design paradigm
employs some 'watchdog' techniques for reciprocal fault diagnosis
between the DNP itself and a companion 'host processing element'
within a node or on nodes which are neighbouring in the \apenetp mesh
topology; moreover, it employs a number of best-effort heuristics for
delivering the diagnostic data even in case of faulty or broken links
along an auxiliary support network. This network is a possibly
low-speed, but highly reliable, diagnostic-dedicated, independent one,
which leans on its high-speed 3D toroidal companion mesh in the
extreme case of its own failing.  Once complemented with diagnostic
facilities that monitor whatever metrics are deemed relevant to the
prediction of faults (\eg temperature and voltage/current probes, BER
counters, \etc), \lofamo is a keystone upon which a fault management
system can draw inferences that drive its strategies and actions to
keep the system up and running.

\subsubsection{Terminology and global picture}
First of all, from now on we will use \emph{fault} as abbreviation for
\emph{fault} and \emph{critical event}.

Then, we split the fault-tolerance problem in two major key areas:
\textit{fault awareness} and \textit{fault reactivity}.
\begin{itemize}
\item \textbf{Fault awareness} is the capability of the system to
assess its own health status, in order to acknowledge faults that have
already appeared or to make guesses about those likely to occur. Going
bottom-up, this 'introspection' can be reduced to two aspects:
\begin{itemize}
\item \textbf{Local fault detection}, the capability of a device to
perform a number of HW and SW tests to detect a condition of fault in
itself or other contiguous devices.
\item \textbf{Systemic fault awareness}, the collation of diagnostics
propagated throughout the whole network by the local detecting
sub-systems to compose a global picture of the system's health.
\end{itemize}  
\item \textbf{Fault reactivity} is the range of initiatives that the
system enacts, under the presumptions it can make when its own global
health is known to it, to prevent a fault situation which is about to
occur or to gracefully degrade its performance instead of bringing the
whole system to a stop when the fault has occurred. Going top-down,
this 'self-adjustment' can be reduced to two aspects:
\begin{itemize}
\item \textbf{Systemic response}, the set of strategies that the
system can choose to apply, following inferences that it can make from
its own diagnostic self-image, to prevent and counter the faults.
\item \textbf{Local readjustment}, the set of readjustments that can
be locally enacted to prevent and counter the faults, \eg reduction in
clock frequency, changes to the routing tables to bypass a faulty
link, remapping the assignment of tasks to nodes, \etc.
\end{itemize}
\end{itemize}
It is clear that a complete design of a fault-tolerant architecture
must give detailed specifications in each of the above-mentioned
areas. On the other hand, the challenging part for the most
interesting fault-tolerant features is the actual implementation,
which cannot be detached from a low-level specification of the host
architecture. For example, task migration capabilities are derived
from process management features of the host operating system;
application checkpointing is strictly bound to storage options
available to the host node; protection from memory errors by ECC is a
low-level addition to the host memory architecture, \etc.

\begin{figure}[!hbt]
  \centering
  \includegraphics[width=0.8\textwidth]{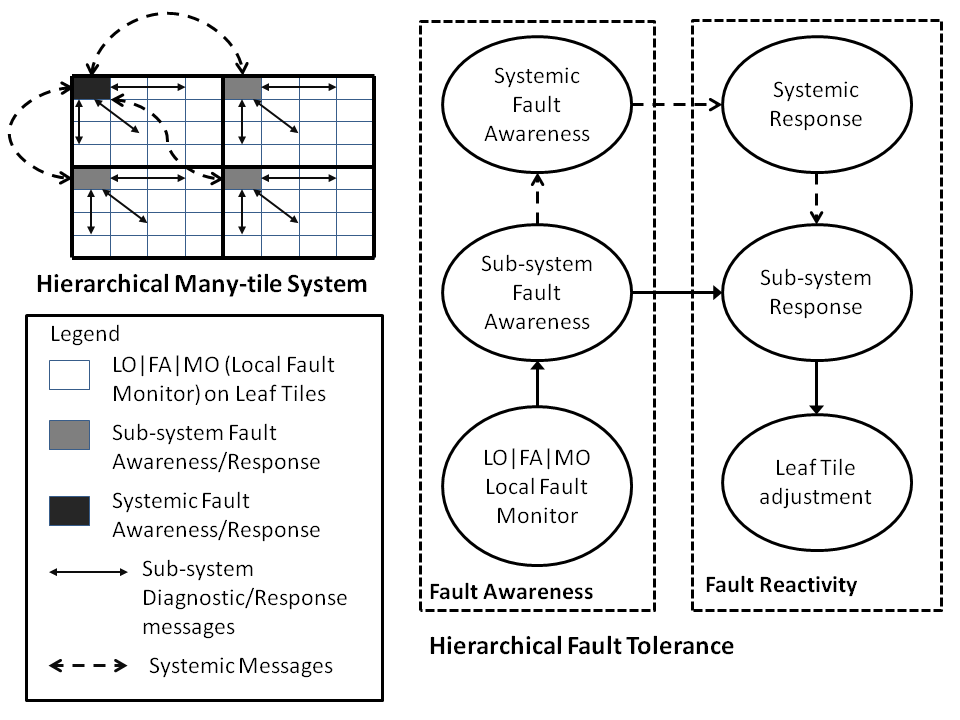}
  \caption{The Network Processor of each leaf in the many-tile HW
system is equipped with its own \lofamo components. The local Fault
Awareness is propagated towards the upper hierarchy levels, creating
systemic Fault Awareness. Fault reactions could be autonomously
initiated by sub-system controllers or could require a systemic
response.}
  \label{fig:LOFAMO:hier}
\end{figure}

Our idea with \lofamo is that of a fault-tolerant framework which is
as host-agnostic as possible. By encapsulating as many features as can
be accommodated independently from such host specifications in
\lofamo, we strive to achieve a clear separation of problems - with
the hope this leads to easier solution - and a degree of design reuse.
By saying this, we make clear from the start that \lofamo, by its very
nature, has to be restricted to the side of fault-awareness.  In a
hierarchical system the \lofamo approach is then integrated as shown
in figure \ref{fig:LOFAMO:hier}.  The local Fault Awareness is
propagated towards the upper hierarchy levels, creating a Systemic
Fault Awareness.  The reactivity starts from a Systemic Response and
follows the inverse path from the upper layers, that have a global
picture of the system status, towards the lower ones, where actions are
enacted to adjust the system components behaviour to the fault
condition.  The many-tiled structure also requires the blocks able to
reach Fault Awareness to be distributed and the presence of paths to
propagate the Awareness towards some centralized intelligence able to
make decisions.

We give here a sketch of an architecture where \lofamo is employed. We
assume a computing mesh where every node is a combination of the
hardware supporting \lofamo, \ie a DNP, mated to a host processing
element; beyond the communication facilities provided by the DNP, the
host exposes another communication interface towards an auxiliary
'service network' (more details in section \ref{sec:snet}).

Failures can generally be of \emph{commission} and \emph{omission}
type: the former encompasses the case of failing elements performing
their tasks in an incorrect or inconsistent way (\eg corruption in
node memory, corruption in transmitted messages, packet misrouting,
\etc); the latter deals with the case of failing elements skipping
their tasks altogether (\eg node stops responding due to crash
failure, power outage or burn-out, message passing does not progress
due to link disconnection, \etc).  The most general kind of faults are
those where the behaviour of a faulty component is assumed to possibly
be completely random as to its correctness; in literature,
fault-tolerance to this kind of faults is defined \emph{Byzantine
fault-tolerance} \cite{Castro:2002:byz}, Byzantine failures can be
seen as undetectable commission failures or, where possible, as
malicious activity by some agent which is trying to sabotage the
network. This kind of failures is explicitly not covered here.

With this restriction, detectable commission failures signal either a
component that is about to break or keeps on working wrong, while
omission failures, when permanent, mostly stand for an already broken
or disconnected component.  In this picture, the \lofamo design is
charged of polling the supplied sources of diagnostic data; any
inconsistent value, being it any value beyond a certain threshold or a
timed-out update of a watchdog counter, is a failure to
report. \lofamo attempts then to push this report along the service
network - which means, in emergency cases, leaning against the
neighbouring DNP's - towards an upper layer Fault Supervisor.  As per
previous definitions, we remark that the only behaviours \lofamo
foresees for a failing component are two:
\begin{itemize}
\item \textbf{sick} - the component has a rate of detected commission
failures beyond the compatibility threshold of normal operativity
$\rightarrow$ this may need action;
\item \textbf{failed} - the component has a permanent commission
failure (it keeps on working wrong) or simply stops participating in
the network, \ie it has a permanent omission failure (it has broken)
$\rightarrow$ this needs action.
\end{itemize}

\subsubsection{LOFAMO specification}
\label{sec:lofamospec}
The Local Fault Monitor (\lofamo) is the mechanism chosen to obtain
the fault-awareness; it implements health self-tests for a number of
hardware devices and takes care of propagating the deriving
information. Moreover, the devices are able to monitor other
contiguous devices and communicate their faulty status.
Synthetically, each device is able to:
\begin{itemize}
\item check/elaborate/store/transmit its own status; 
\item monitor other
contiguous devices.  
\end{itemize}
In the \euretile platform, the actors of the described mechanism are
the DNP/\apenetp and the host sub-system (Intel for the \quong
platform, a RISC-like model for the Virtual \euretile Platform). The
DNP is able to run self-tests on its own links and logic, as well as
to retrieve information from its own temperature and electrical
sensors. All information pertaining to the sub-systems status is
gathered by the \lofamo-appointed component inside the DNP itself and
stored in the \dwr (see section \ref{sec:lofamowd}). A second register
inside the DNP is dedicated to the surveillance of the health status
of the host, with \lofamo performing periodic checks of the \hwr. In
the event the host on one or more nearest neighbouring nodes were
faulty, a third register, the \hw{Remote Fault Descriptor Register}
, would end up containing information about the nature of the
remote fault. The self-test capabilities of the DNP links and logic
allow mutual monitoring between nearest neighbour DNP's, all of them
acting as watchdog for one another. The key points for this \lofamo
implementation are:
\begin{itemize}
\item the presence of a \sw{HOST FAULT MANAGER} (HFM), a software
process running on the host that is aware of the host local status, is
able to read/write the DNP internal registers and the \dwr and \hwr
and can send messages through the Service Network.
\item the presence of a \hw{DNP FAULT MANAGER} (DFM), a component
residing on the DNP that is able to collect the information about the
DNP health status, to read/write the \dwr and \hwr and to send
messages through the 3D Network.
\item a 3D toroidal network connecting all the nodes, implemented by
  the DNPs
\item a service network, connecting all the nodes, as a primary and
  redundant path to communicate diagnostic messages (offloading the 3D
  network from services duties when it's not strictly required).
\item a Fault Supervisor, an high level software layer that can gather
  diagnostic information from all the system components, composing the
  global picture of the system status, and that can make decision
  about how to react to faults.
\end{itemize}

Thanks to these key components, the \lofamo mechanism allows the
system to cope with faults and critical events, also under the
following peculiar circumstances:
\begin{itemize}
\item The DNP breaks down, consequently the DNP does not respond to
  queries from the host - the host acknowledges the omission fault and
  signals it via the service network to the Fault
  Supervisor (this does not differ from the ordinary condition).
\item The host breaks down, consequently the host does not respond to
queries from the DNP - although from that node the service network is
inaccessible, the DNP has a last chance of relaying its reports along
to its neighbours in the high-speed 3D mesh and from there, all
receiving DNP's can relay the data to their own host and then on to
the Fault Supervisor.
\item In the showstopping event of both host and DNP breaking
down in a node, the system has a way to become aware of the situation:
no more activity from the node means that all the neighbouring nodes
in the 3D mesh become eventually aware of a permanent omission fault
in one of their channels; as soon as reports of this fact reach the
Fault Supervisor, this latter can infer the node has died and take
relevant action.
\end{itemize}

\begin{figure}[!hbt]
  \centering
  \includegraphics[trim=0mm 20mm 0mm 20mm,clip,width=0.9\textwidth]{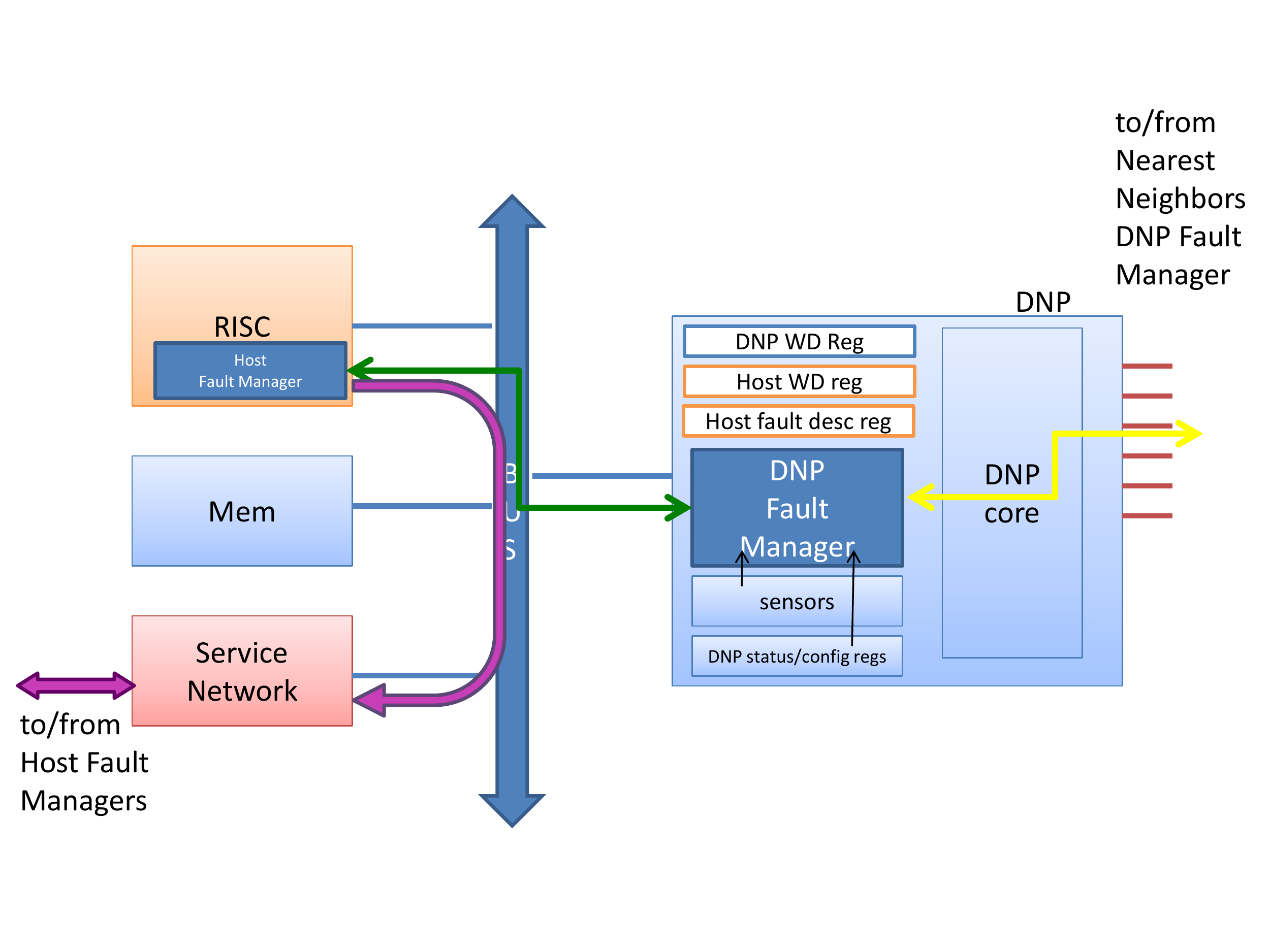}
  \caption{The basic EURETILE platform configuration showing the
position and the communication paths of the \sw{HOST FAULT MANAGER}
and the \hw{DNP FAULT MANAGER} in a \lofamo implementation.  The \hwr
stores info about Host Faults. The \dwr stores info about local or
global faults of the networking system. The \hw{Host Remote Fault
Descriptor Register} contains information about the nature of remote
host Faults.}
  \label{fig:LOFAMO:general}
\end{figure}

Figure \ref{fig:LOFAMO:general} illustrates the basic platform
configuration detailing the position and the communication paths of
the \sw{HOST FAULT MANAGER} and the \hw{DNP FAULT MANAGER}. Keeping to
the definitions given above, the task of \lofamo thus encloses the
whole of Local fault detection and the interface to the Fault
Awareness system.

\subsubsubsection{Fault Supervisor}
\label{sec:faultsupv}
The Fault Supervisor is the generic term that encompasses the set of
processes receiving the output of the \lofamo machinery; its duty is
to create systemic Fault Awareness and to issue appropriate systemic
Fault Responses.  For a small number of nodes, the Fault Supervisor
could be implemented as a single software process running on an
appointed 'master node' of the system; for larger systems, a process
cloud residing on a subset of nodes participating in a hierarchy would
certainly be more scalable.  The Fault Supervisor is kept timely fed
by the set of \hw{DNP FAULT MANAGER} and \sw{HOST FAULT MANAGER}, with
periodic updates about their health. This supervisor is the 'systemic
intelligence' that embodies the fault awareness for the system and
drives its response; all information (or lack thereof, in case of
omission failures from faulty nodes, which is information as well) is
brought by the \lofamo network to the Fault Supervisor system, so that
it can choose any fault prediction, prevention and reaction strategy
it deems feasible.
 
We don't give here a detailed specification for the Fault Supervisor;
we acknowledge its presence - after all, it is the target of all
communications from the \lofamo components - and we briefly describe
(see section \ref{sec:hfm}) its implementation for the \quong Platform
for what regards the obtaining of the systemic Fault Awareness. A more
detailed specification would require a discussion on the Fault
Reactivity strategies and on how HW and SW interact to apply them, but
at the moment this is out of the scope of this document.

\subsubsubsection{Service Network}
\label{sec:snet}
Besides \apenetp's high speed 3D mesh, \lofamo expects the system
nodes to partake in a secondary, diagnostic-dedicated network fabric
to which only the host has access. In ordinary conditions, the DNP
relays the gathered diagnostic data to its host companion which,
through this network, in turn relays them to the Fault Supervisor. In
this way, the high speed network is unencumbered from dealing with the
added traffic of the health status reports.  We expect this service
network to be a relatively inexpensive local interconnect. On the HPC
market, Ethernet is a mature technology, mostly ubiquitous presence
for any architecture we think to match the \apenetp board with - \eg,
our \quong platform cluster node prototype is a Supermicro R board
equipped with dual GbE. For this reason, the presence of such service
network is regarded as a rather unconstraining addition on the HPC
flavour of the \euretile architecture. On many-tile embedded systems,
represented in our case by the VEP platform, we maintain at this stage
open the definition of the service network.  We are positing that the
bulk of diagnostic data does need neither high bandwidth nor extremely
low latency. This means that performance concerns are not overtly
constraining in the building of this service network and all effort
can be instead put in pushing its reliability, by means of ruggedness
of components (for the switches, routers, NICs, cabling, \etc) or some
kind of redundancy; reliable Ethernet is a wide ranging subject with
many possible approaches \cite{Golash:2006:ethrel}.  However, this
diagnostic network is a system element itself subject to failure. So,
the problem must be raised of how to deliver fault awareness data in
presence of criticality of the service network itself or the
DNP's. First, we analyze the case of host and DNP not affected by
simultaneous fail, then the case of simultaneous fail of the DNP and
host on a tile. The hypothesis we put forward is that the probability
for a node of host and DNP simultaneously failing is smaller than
their individual failure rate. This means that having the host and the
DNP mutually cross-checking each other, \lofamo has meaningful escape
routes in both the scenarios: when the DNP breaks down or when the
Host breaks down. If both the components are completely faulty a
Systemic Awareness can be reached thanks to the fact that the
neighbouring nodes in the 3D mesh can sense the non-operative status
of the channels towards the faulty node and spread the information.

\subsubsubsection{Watchdog implementation}
\label{sec:lofamowd}
One of the foundation of the \lofamo design is the mutual watchdog
mechanism, that for the \euretile platform is implemented in the
following way: the DNP acts as watchdog for the host, \ie it
periodically monitors the host status as reported in the \hwr updated
by the host itself; the host acts as watchdog for the DNP, \ie it
periodically monitors the DNP status as reported in the \dwr updated
by the DNP itself. Although both the mentioned registers are located
inside the DNP, they are written (updated) and validated by their
'owner' and read and invalidated by the other device.
Validation/invalidation consists of setting the Valid Bit to 1 or 0,
respectively. The update period is such that $T_{write} < T_{read}$,
in this way is guaranteed that the reader always founds a valid status
and viceversa, unless a destructive omission fault occurs that makes
the writer unable to update its status register (see section
\ref{sec:lofamoreg}).

\begin{figure}[!hbt]
  \centering
  \includegraphics[trim=0mm 30mm 0mm 15mm,clip,width=0.9\textwidth]{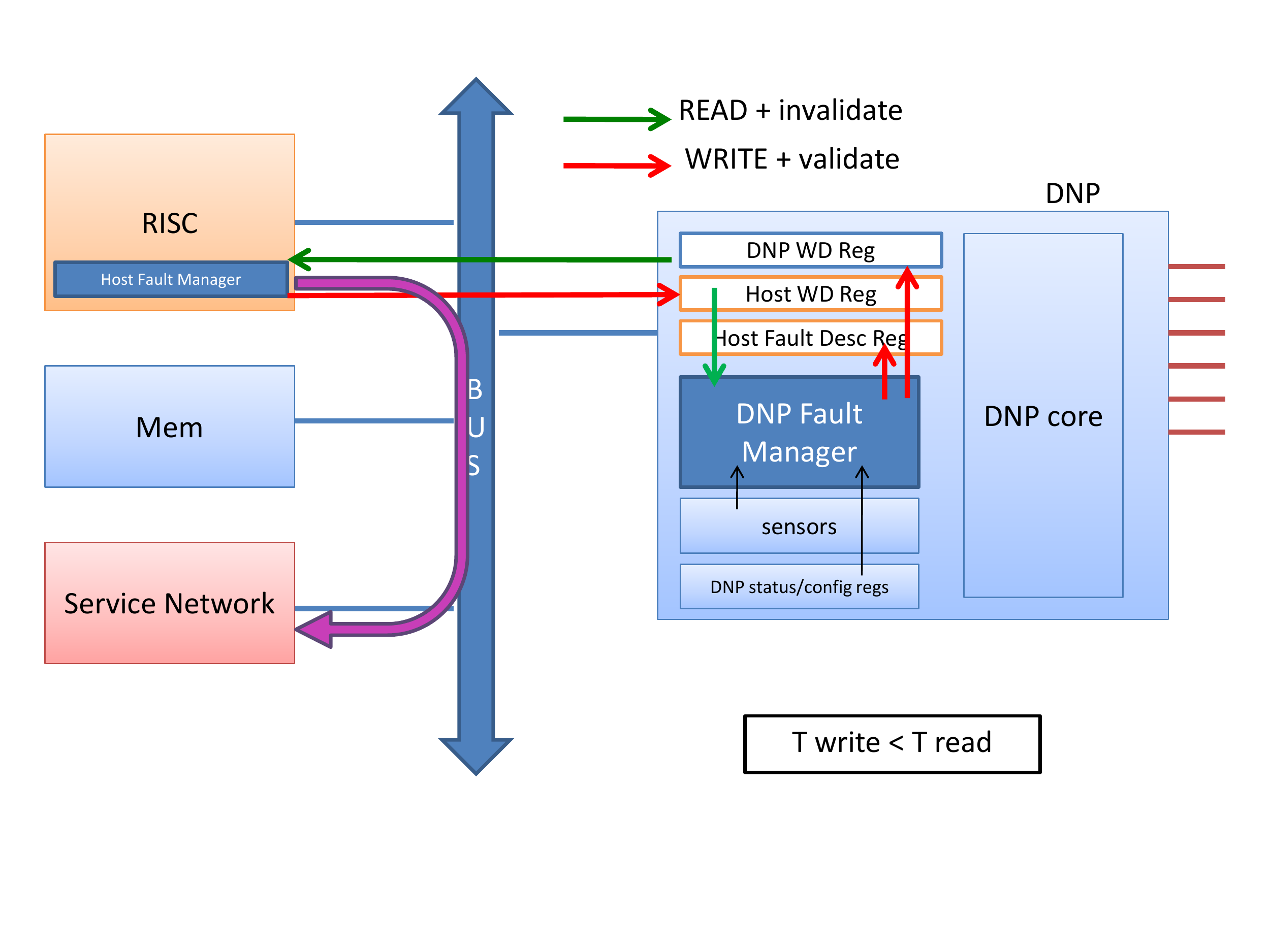}
  \caption{The host and the DNP act as reciprocal watchdogs. Even if
both the host and DNP WD Registers are located inside the DNP, they
are periodically updated by their \emph{owner} and invalidated by the
other device.}
  \label{fig:LOFAMO:wd_impl}
\end{figure}

\subsubsubsection{Faults, Fault detection and paths to local/global 
awareness}
Table \ref{tab:FaultDetect} shows for each kind of fault that we chose
here to consider which component is responsible to detect the fault
and how the \lofamo mechanism allows to reach a local/global
awareness.  A description of each fault listed in table is reported
here:
\begin{itemize}
\item \textbf{Link 'sick'}
\begin{itemize} 
\item \textit{Fault Description}: the DNP link (physical link) 
malfunctions and introduces errors during the packet transmission.
\item \textit{Fault Detection}: Errors are detected by the links logic
using CRC: the sender DNP calculates the CRC word of the packet and
puts such word into the packet footer; the receiving DNP calculates
the CRC word by itself and checks the matching with the CRC word in
the packet footer, in case it doesn't match exception bits are raised
in the proper register.
\end{itemize}
\item \textbf{Link broken}
\begin{itemize} 
\item \textit{Fault Description}: the DNP link is broken, the packet 
can't be transmitted.
\item \textit{Fault Detection}: the RX side of the linked DNP sends
continuously, back along the cable and towards the TX side,
information (called credits) about the space available in FIFO to
receive the data transfer. In this way the TX logic can sense whether
the cable is ok or not (credits not received and timeout reached).
\end{itemize}
\item \textbf{Temperature, power, voltage over/under threshold}
\begin{itemize} 
\item \textit{Fault Description}: the DNP temperature (or power, or
voltage) is over (or under) the threshold values, contained in the
registers and set on the basis of the values for a correct hardware
operativity.
\item \textit{Fault Detection}: the hardware provides sensors to
monitor temperature, power consumption and voltage. The control logic
that reads the sensors and checks if the values are within the limits
is part of the \hw{DNP FAULT MANAGER}.
\end{itemize}
\item \textbf{Host side faults}
\begin{itemize} 
\item \textit{Fault Description}: the Host is broken or a peripheral
in tile is faulty.
\item \textit{Fault Detection}: the \hw{DNP FAULT MANAGER} is able to
detect these situations on the basis of the watchdog mechanism.
\end{itemize}
\end{itemize}

\begin{table}[!ht]
\centering
\footnotesize
\setlength\extrarowheight{13pt}
\begin{tabular}{|m{0.7cm}|m{1.2cm}|p{1.8cm}|m{4cm}|m{3.5cm}|m{3.2cm}|}
%\begin{tabular}{|c|c|c|c|c|c|}
\hline
\hline
\textbf{Fault class} & \textbf{Faulty Elem} & \textbf{Faults} & \textbf{Detector} & \textbf{Diagnostic info storage} & \textbf{Diagnostic info path}\\
\hline
\hline
\multirow{7}{0.7cm}{DNP side} & \multirow{2}{1.2cm}{Link} & Link sick
(CRC errs) & receiving Link self-test $\rightarrow$ receiving DFM & \multirow{2}{3.5cm}{bits in
 \dwr} & \multirow{5}{3.2cm}{DFM $\rightarrow$ Host $\rightarrow$
Service Net}\\
\cline{3-4}
& & Link broken &  DFM on both the receiving and transmitting side (if link logic still working), otherwise on a single side & & \\
\cline{2-5}
& \multirow{3}{1.2cm}{DNP $^{\circ}$C/W/V alert} & Temperature over/under threshold & \multirow{3}{1.8cm}{Sensors
$\rightarrow$  DFM} & \multirow{3}{3.2cm}{bits in \dwr}& \\
\cline{3-3}
&&Power over/under threshold &&&\\
\cline{3-3}
&&Voltage over/under threshold &&&\\
\cline{2-6}
& \multirow{2}{1.2cm}{DNP logic} & DNP core sick & DFM & bits in \dwr
&\multirow{3}{3.2cm}{Host $\rightarrow$ Service Net}\\
\cline{3-5}
& & DNP core meltdown & Host (DFM misses to update \dwr) & bit in \dwr
&\\
\cline{1-5}
\multirow{3}{0.7cm}{Host side}&\multirow{3}{1.2cm}{Host
  fault}&\multirow{2}{1.8cm}{Memory, Peripherals (\ldots, Service Net)}&
\multirow{2}{4cm}{Machine dependent but managed by the HFM}&\multirow{2}{3.5cm}{Spare room in \hwr}&\\
\cline{6-6}
&&&&&\multirow{2}{3.2cm}{DNP $\rightarrow$ 3D net $\rightarrow$ Neighbour
  DFM $\rightarrow$ Neighbour Host $\rightarrow$ Service Net}\\
\cline{3-5}
&&Bus or total Host breakdown & DFM (Host misses to update \hwr)& bit in RISC watchdog register &\\
\hline
\hline
\end{tabular}
\caption{A summary of how faults are detected (which component is
responsible to detect them) and how the information is conveyed
bottom-up to obtain the Global Fault
Awareness. (Abbreviations: DFM = DNP Fault Manager; HFM= Host
fault manager; Net = Network)}
\label{tab:FaultDetect}
\end{table}

\subsubsubsection{\lofamo registers}
\label{sec:lofamoreg}
There are three key registers used by the \lofamo components: 
\begin{itemize}
\item \dwr. It contains information about: 
  \begin{itemize}
  \item the status of the local DNP
  \item the status of the hosts on first neighbouring tiles;
  \end{itemize}  
\item \hwr. It contains the local host status; 
\item \hw{Remote Fault descriptor Register}. In case of one or
more host(s) on first neighbour tiles are faulty, it contains
information about the nature of the fault.
\end{itemize}

More registers can be used in the different implementations to control
the behaviour of the various \lofamo components.

\begin{figure}[htbp]
\centering
%\begin{minipage}[t]{\textwidth}
%\hspace{-12pt}
\centering
\includegraphics[trim=10mm 20mm 10mm 5mm,clip,width=0.7\textwidth]{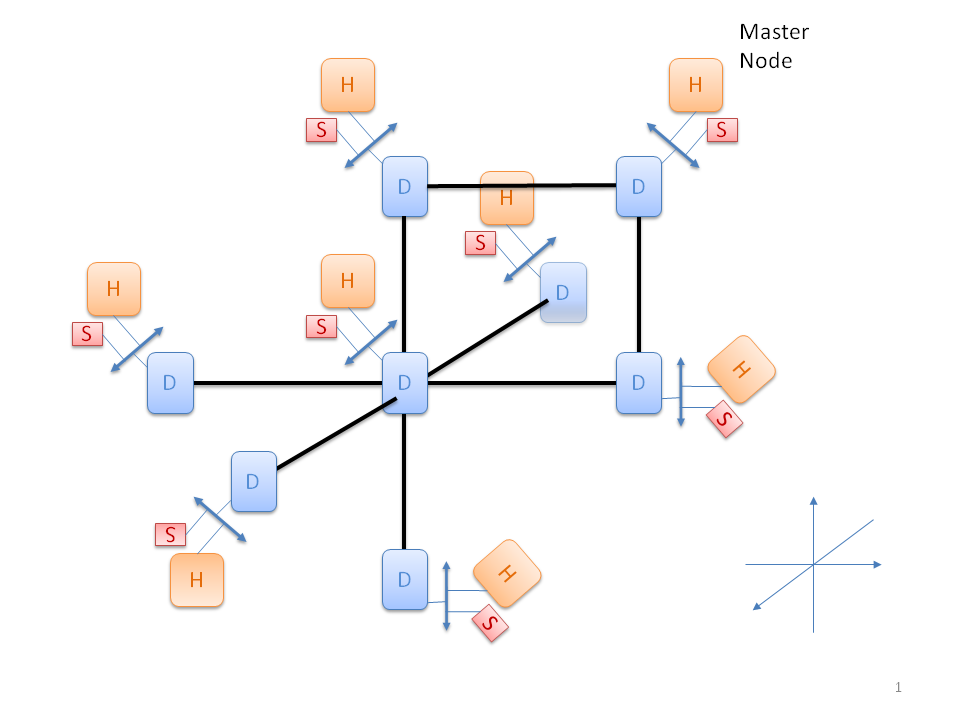}
\caption{\textbf{A} - Implementing the \lofamo approach on 8 Nodes
  connected through the DNPs (D) equipped with a \dfm. Each Host (H)
  runs an \hfm which has access to the Service Network (S) to send
  diagnostic messages to the \hfm running on the Master Node.}
\label{fig:LOFAMO3dA}
\end{figure}

\begin{figure}[htbp]
%\vspace{1cm}
%\hspace{-12pt}
\centering
\includegraphics[trim=10mm 20mm 10mm 5mm,clip,width=0.7\textwidth]{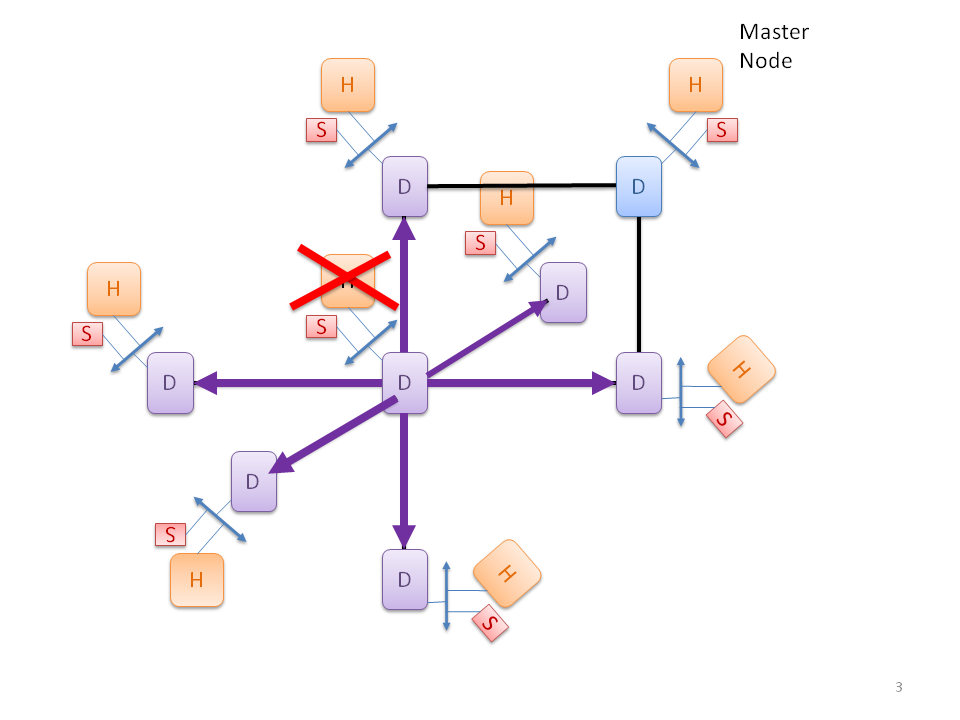}
\caption{\textbf{B} - If one of the hosts comes down, the DNP in the
  same tile is able to detect the fault thanks to the watchdog
  mechanism and it sends a diagnostic message to its first
  neighbouring DNPs via the 3D network.}
\label{fig:LOFAMO3dC}
\end{figure}

\begin{figure}[tbp]
%\vspace{1cm}
%\hspace{-12pt}
\centering
\includegraphics[trim=10mm 5mm 10mm 0mm,clip,width=0.7\textwidth]{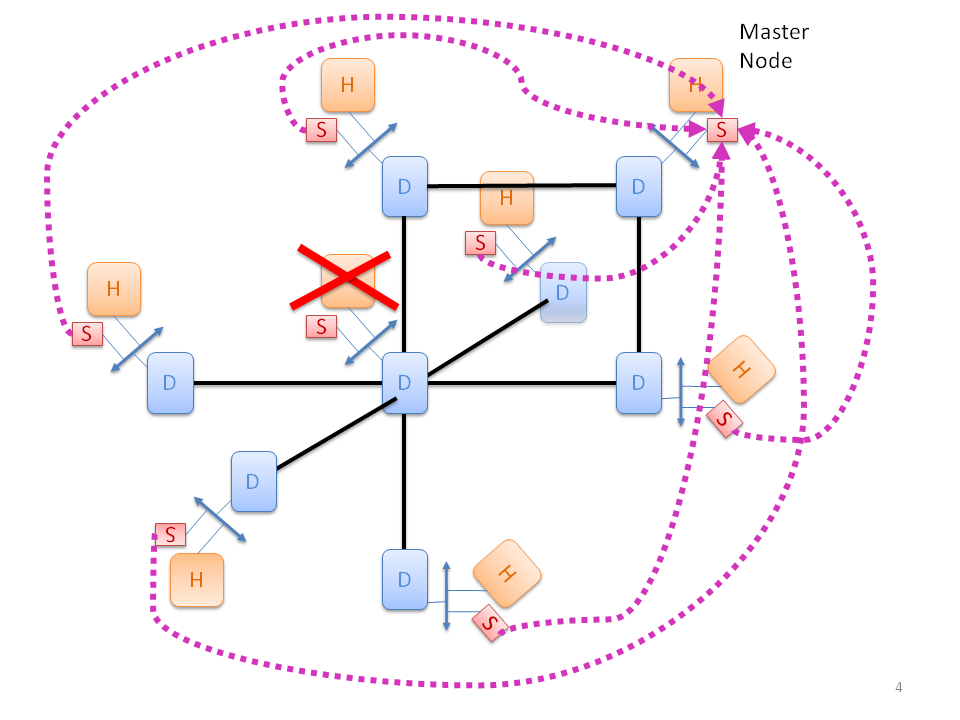}
\caption{\textbf{C} - As each host exchanges status information with
  his companion DNP in the watchdog mechanism, the Hosts first
  neighbours of the faulty Host become aware of the fault and can
  signal it to the Master node via Service Network.}
\label{fig:LOFAMO3dD}
\end{figure}

\subsubsubsection{Example}
An example of how the \lofamo approach is applied in case of Host
Breakdown fault is shown by figures \ref{fig:LOFAMO3dA},
\ref{fig:LOFAMO3dC}, \ref{fig:LOFAMO3dD}.
\FloatBarrier

%\end{minipage}
%
%\hfill
%% \begin{figure}[htbp]
%% %\hspace{-12pt}
%% \centering
%% \includegraphics[width=0.7\textwidth]{LOFAMO3DexampleB.png}
%% \caption{}
%% \label{fig:LOFAMO3dB}
%% \end{figure}
%
%
%\hfill

\FloatBarrier
\subsubsection{\lofamo on the \quong platform}
In the following we give details on how to specialize the \lofamo
mechanism over the QUonG platform peculiarities, \ie each host is a
x86\_64 Linux based machine mounting an \apenetp~card.
\begin{itemize}
\item \textbf{Registers}: 
\begin{itemize}
\item \underline{\dwr}. It has got 3 macro fields: the local part,
  containing informations about the status of the Local DNP; the remote
  part, which contains the status of the first neighbours
  hosts (thus it can be propagated by the local and well functioning
  node) in terms of faulty/non-faulty; the valid field, that states
  the aliveness of the local DNP and certifies the reported DNP status
  as up to date.
\item \underline{\hwr}. It has got 2 macro fields: the (local) host
  status and the valid field. The host status is specified for each
  meaningful host device or periphery like memory or (service)
  network.  The valid field is used to state that the host is alive
  and the status contained in the other field up to date.
\item \underline{\hw{Remote Fault Descriptor Registers}}. One or
  more registers used to specify the nature of the fault occurred on 
  remote host signaled as faulty in the global (remote) part of the \dwr..
\item \underline{Temperature/Voltage/Power consumption thresholds
  registers}. Registers used to set the boundaries for normal,
  warning and alarm ranges for temperature, voltage and DNP/\apenetp power
  consumption. The thresholds must be set
  coherently with the ranges of correct operativity of the board.
\item \underline{Link Related Thresholds}. Other registers can be used
  to set the thresholds to define the correct behaviour ranges of the
  3D channels, \eg maximum number of hops, the threshold for the
  $\text{CRC errors} / \text{transferred bytes}$ ratio, \etc.
\item \underline{\lofamo timer register}. Used to set the watchdog
  write and read period on the DNP side. The \hw{DNP FAULT MANAGER}
  uses this information to update the \dwr and read the \hwr at the
  right pace.
\item \underline{\lofamo masks register}. Allows to control the
  behaviour of the \hw{DNP FAULT MANAGER}, \eg to mask or unmask the
  signaling of fault type.
\end{itemize}
\item \textbf{\hw{DNP FAULT MANAGER} (DFM)}: it's an hardware
  component within the DNP core of \apenetp; it can access the board
  sensors and link control logic, gathering information about the DNP
  status; it can as well access periodically the DNP registers to
  implement the mutual watchdog together with the host as defined in
  section \ref{sec:lofamowd}. More details on VHDL implementation of
  the \hw{DNP Fault Manager} for \quong is reported in section
  \ref{sec:lofaman}.
\item \textbf{\sw{HOST FAULT MANAGER} (HFM)}: it's a software running
  on each host able to access the \apenetp device through calls to its
  driver handles. It's also able to test the host itself to check its
  health status and retrieve and collect information about faults that
  may affect the host and his peripherals.\\ One or more hosts in the
  network are initially configured as Master Nodes and the HFM running
  on them is also responsible to collect the fault awareness
  information and store them. The initial configuration of the HFMs
  must match the topology programmed in the DNP device.\\ The HFM is
  also able to configure the behaviour of the DNP Local Fault Manager,
  by properly setting the \lofamo mask register on the DNP. In case of
  fault (either DNP or host class), once the information reach the HFM
  or the master HFM, a mechanism of acknowledge is necessary to
  shutdown the alarms and avoid network (service network or 3D)
  congestion due to the diagnostic messages. The implementation of the
  HFM for \quong is detailed in section \ref{sec:hfm}.
\end{itemize}

\begin{table}[!hbt]
\centering
\footnotesize
\setlength\extrarowheight{3pt}
\begin{tabular}{|p{7cm}|p{1.5cm}|p{1.3cm}|p{5.5cm}|}
%%\begin{longtable}{|l|c|c|c|}
\hline
\textbf{Register Name} & \textbf{address (BAR5) }&\textbf{\#reg (BAR5)}&\textbf{Description}\\
\hline
\lofamo DNP local/global watchdog & 0x474& 29  & \dwr\\
\hline
\lofamo Host watchdog & 0x478 & 30 & \hwr\\
\hline
\lofamo Remote Fault Descriptor +X &  0x44C  & 19  & Remote fault descriptor\\
\hline
\lofamo Remote Fault Descriptor -X & 0x450  & 20  & Remote fault descriptor\\
\hline
\lofamo Remote Fault Descriptor +Y & 0x454 & 21 & Remote fault descriptor\\
\hline
\lofamo Remote Fault Descriptor -Y & 0x458 & 22 & Remote fault descriptor\\
\hline
\lofamo Remote Fault Descriptor +Z & 0x45C & 23 & Remote fault descriptor\\
\hline
\lofamo Remote Fault Descriptor -Z & 0x460 & 24 & Remote fault descriptor\\
\hline
\lofamo thresholds & 0x46C   &  27 & Boundaries for the \emph{normal},
\emph{warning},
\emph{alarm} temperature ranges \\
\hline
\lofamo timer & 0x464  &  25 & Set the watchdog R/W periods\\
\hline
\lofamo mask & 0x468 & 26 & Mask or unmask fault signaling\\
\hline
\end{tabular}
\caption{\apenetp registers meaningful for fault detection and awareness.}
\label{tab:FaultDetRegs}
\end{table}

\subsection{VHDL coding of DNP FAULT MANAGER}
\label{sec:lofaman}

The \hw{DNP FAULT MANAGER} is an hardware component within the DNP
core. It mainly implements, together with the \hw{HOST FAULT MANAGER},
the Mutual Watchdog mechanism. It communicates to the host the Neighbours Status,
on-board sensors values, DNP Core self test results and Link Status,
updating the \dwr (DWR). 
Moreover it collects information
about the local host status, Host Memory and Service Network, reading
the \hwr (HWR) updated by the \hw{HOST FAULT MANAGER}.

Furthermore in case of Service Network faults the \hw{DNP FAULT
MANAGER} can initialize Link Fault Manager (\lifama) transactions to
communicate the fault status to the six neighbour nodes. This
additional fault communication approach can be requested also when the
Service Network is operating without reporting faults to exploit a
redundancy of the critical status communication.

\begin{figure}[!h]
  \centering
  \includegraphics[trim=15mm 30mm 15mm 30mm,clip,width=0.7\textwidth]{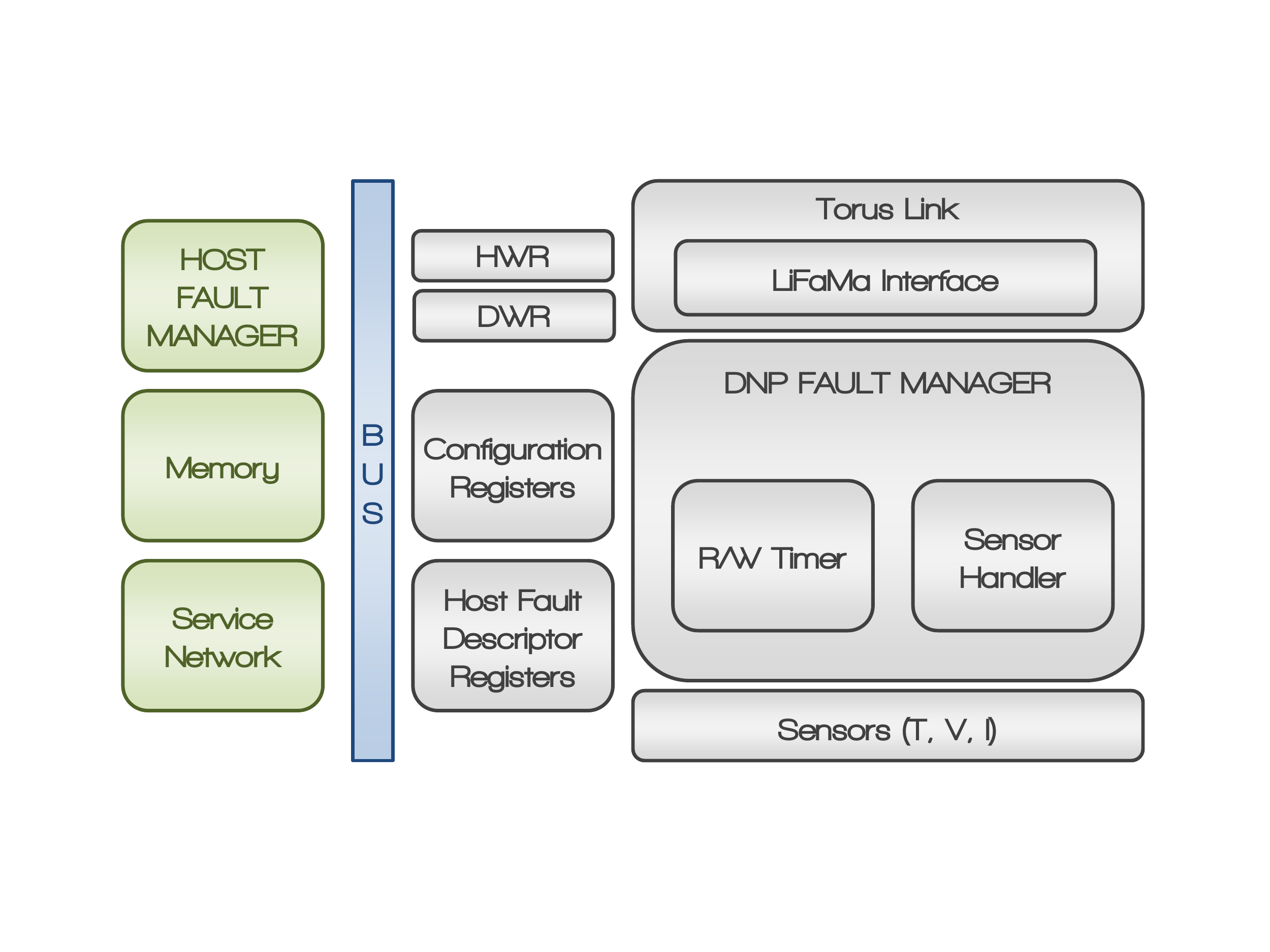}
  \caption{\lofamo Overview}
  \label{fig:lofamo}
\end{figure}

Referring to figure~\ref{fig:lofamo} the \hw{DNP FAULT MANAGER} 
implements two additional elementary blocks:

\begin{itemize}
\item the \hw{R/W TIMER} is a programmable timer to set the frequency
of \hwr reading process and \dwr writing process. Current
implementation allows to vary the interval between two consecutive
operations from 1ms up to 65s depending on the accuracy required.
\item the \hw{SENSOR HANDLER} compares the measured Temperature,
Voltage and Current values with programmable threshold stored in the
\hw{Configuration Register} (see table~\ref{tab:statconf}) and
evaluates the health status of the board.
\end{itemize}

In order to transfer information about the local board and remote node
status to the local host, the \hw{DNP FAULT MANAGER} performs the
following tasks:

\begin{itemize}
\item It collects \rxlifama messages containing the remote nodes
status and it updates the \hw {Remote Fault Descriptor Register} 
(see table~\ref{tab:statconf}).
\item It collects information about the board local status (DNP core
status, Sensor measured values, Local Link Status).
\item It updates the \dwr at the frequency set in the \hw{R/W TIMER}     
with the information stressed in table~\ref{tab:DWR}.     
\end{itemize}
\begin{table}[!htb]
\centering
\setlength\extrarowheight{2pt}
\begin{tabular}{|c|c|c|l|}
\hline
\hline
\multicolumn{4}{|c|}{\textbf{DNP Watchdog Register (DWR)}}\\
\hline 
\hline
\textbf{\# Bits} & \textbf{Bit Range} & \textbf{Field Name}         & \textbf{Protocol}                           \\
\hline
1                &  0                 & Valid                       & 1=DWR content valid; 0=DWR content not valid\\
\hline 
1                &  1                 & Z- Neighbours Status          & 1=fails; 0=healthy;                         \\ 
1                &  2                 & Z+ Neighbours Status          &                                             \\
1                &  3                 & Y- Neighbours Status          &                                             \\
1                &  4                 & Y+ Neighbours Status          &                                             \\
1                &  5                 & X- Neighbours Status          &                                             \\
1                &  6                 & X+ Neighbours Status          &                                             \\
\hline
2                &  8 -  7            & DNP core Status             & 00=normal; 01=sick; 10=broken               \\
\hline
2                & 10 -  9            & Current Status              & 00=normal; 01=warning; 10=alarm             \\ 
\hline
2                & 12 - 11            & Voltage Status              & 00=normal; 01=warning; 10=alarm             \\ 
\hline
2                & 14 - 13            & Temperature Status          & 00=normal; 01=warning; 10=alarm             \\ 
\hline
2                & 16 - 15            & Z- Link Status              & 00=normal; 01=sick; 10= broken              \\
2                & 18 - 17            & Z+ Link Status              & 00=normal; 01=sick; 10= broken              \\
2                & 20 - 19            & Y- Link Status              & 00=normal; 01=sick; 10= broken              \\
2                & 22 - 21            & Y+ Link Status              & 00=normal; 01=sick; 10= broken              \\
2                & 24 - 23            & X- Link Status              & 00=normal; 01=sick; 10= broken              \\
2                & 26 - 25            & X+ Link Status              & 00=normal; 01=sick; 10= broken              \\
\hline
4                & 30 - 27            & Spare                       &                                             \\
\hline
1                & 31                 & LiFaMa Busy                 & 1=LiFaMa Busy; 0=LDM Sent; see table \ref{tab:ldm}                 \\
\hline
\hline
\end{tabular}
\caption{DNP  watchdog register layout}
\label{tab:DWR}
\end{table}

In order to communicate the local host and board status to the six
neighbour nodes, the \hw{DNP FAULT MANAGER} performs the following
tasks:

\begin{itemize}
\item It evaluates the health status of the host verifying the regular
updating of the \hwr at the frequency set by the \hw{R/W TIMER}
(\hw{DNP FAULT MANAGER} checks the valid bit status, see
table~\ref{tab:HWR}).
\item It collects the status of the Service Network, Memory and
Peripheral reading \hwr.
\item It collects information about the board local status (DNP core
status, Sensor measured values, Local Link Status).
\item It initializes \txlifama transfers to communicate the node local
status to the six neighbour nodes.
\end{itemize}
\begin{table}[!htb]
\centering
\setlength\extrarowheight{3pt}
\begin{tabular}{|c|c|c|l|}
\hline
\hline
\multicolumn{4}{|c|}{\textbf{Host Watchdog Register (HWR)}}\\
\hline 
\hline
\textbf{\# Bits} & \textbf{Bit Range} & \textbf{Field Name}     & \textbf{Protocol}                \\
\hline                                                          
1                &  0                  & Valid                  & 1=HWR content valid; 0=HWR content not valid \\
\hline
2                &  2 - 1              & Service Net Status     & 00=normal; 01=sick; 10=broken;   \\ 
\hline
2                &  4 - 3              & Memory Status          & 00=normal; 01=sick; 10=broken;   \\
\hline
2                &  6 - 5              & Peripheral Status      & 00=normal; 01=sick; 10=broken;   \\
\hline
24               &  30 - 7             & Spare                  &                                  \\
\hline
1                &  31                 & Send LDM               & see table \ref{tab:ldm}          \\
\hline
\hline
\end{tabular}
\caption{Host Watchdog Register layout}
\label{tab:HWR}
\end{table}
\begin{table}[!htb]
\centering
\setlength\extrarowheight{3pt}
\begin{tabular}{|c|c||l|}
\hline
\hline
\multicolumn{3}{|c|}{\textbf{Remote Fault Descriptor Registers}}\\
\hline 
\textbf{\# Bits} & \textbf{Field Name}        & \textbf{Description}             \\
\hline             
32               & X+ Neighbour Node Status   &  see table~\ref{tab:ldm} for details \\ 
32               & X- Neighbour Node Status   &  see table~\ref{tab:ldm} for details \\
32               & Y+ Neighbour Node Status   &  see table~\ref{tab:ldm} for details \\
32               & Y- Neighbour Node Status   &  see table~\ref{tab:ldm} for details \\
32               & Z+ Neighbour Node Status   &  see table~\ref{tab:ldm} for details \\
32               & Z- Neighbour Node Status   &  see table~\ref{tab:ldm} for details \\
\hline             
\multicolumn{3}{|c|}{\textbf{Configuration Registers}}\\
\hline 
32               & Mask Enable                &  activate the \hw{DNP FAULT MANAGER}       \\ 
32               & Sensor Threshold           &  set threshold for the \hw{SENSOR HANDLER} \\
32               & Timer Configuration        &  set frequency of read/write operation     \\
32               & Emulation                  &  test the \lofamo approach                 \\
\hline             
\hline             
\end{tabular}
\caption{Remote Fault Descriptor and Configuration Registers of the 
\hw{DNP FAULT MANAGER}.}
\label{tab:statconf}
\end{table}

\subsection{Link Fault Manager (LiFaMa)} %- Andrea

As explained above the \lofamo approach leverages on the redundancy of
the fault communication mechanism. The first method expects that the
host communicates the occurrence of faults through the Service
Network.
An additional fault communication mechanism is to exploit the 3D Torus
Network to transfer fault informations. The latter approach is
necessary in case of fault of the host and/or of the Service Network and it
guarantees a fast broadcast of critical status to the neighbour nodes.

\begin{figure}[!hbt]
  \centering
  \includegraphics[trim=50mm 50mm 50mm 45mm,clip,width=0.7\textwidth]{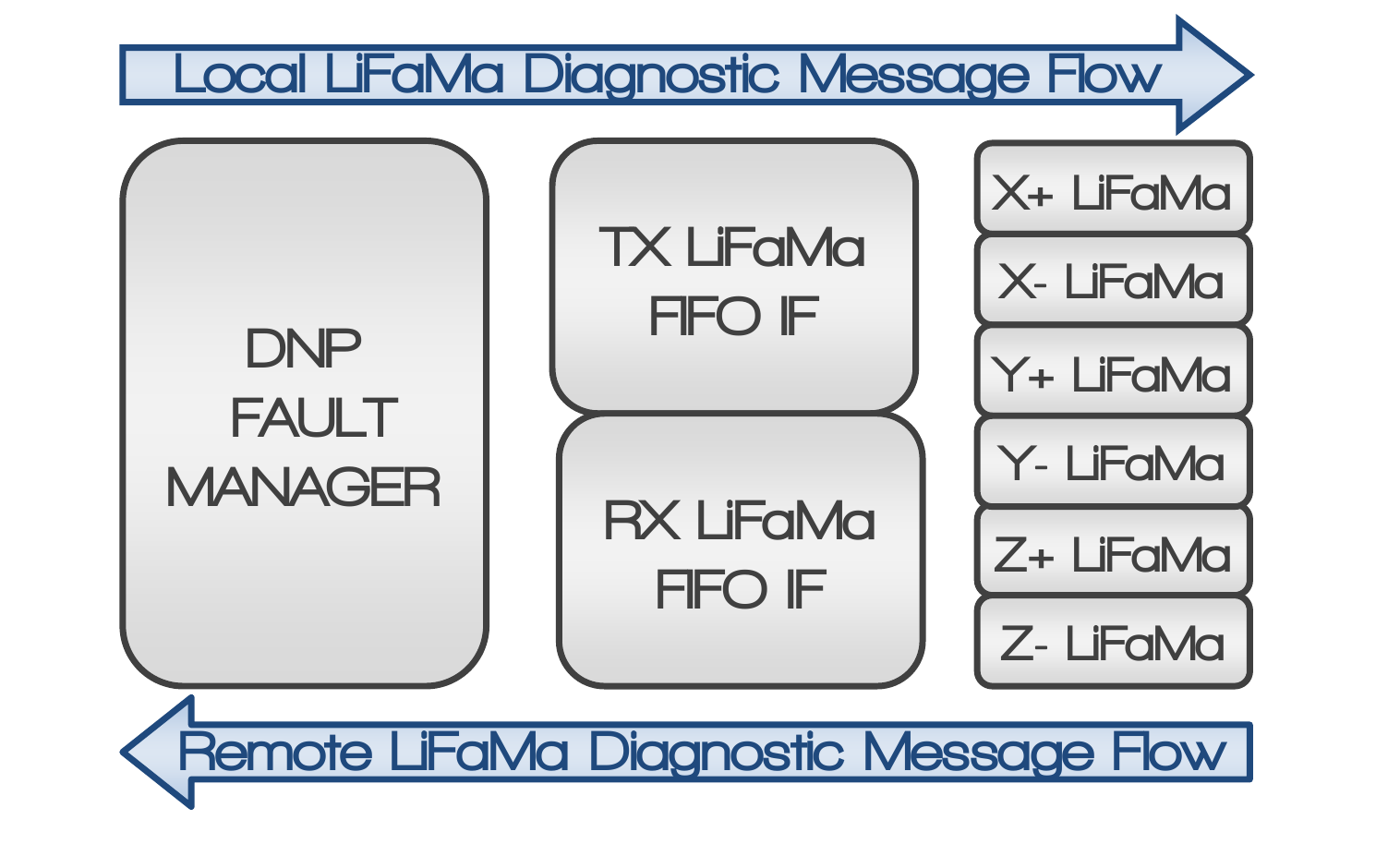}
  \caption{Link Fault Manager (\lifama).}
  \label{fig:lifama}
\end{figure}
The Link Fault Manager (\lifama) is an hardware component in charge of
handle the Diagnostic Message Flow as shown in
figure~\ref{fig:lifama}.
As explained in section~\ref{sec:link_bw_lat} the adopted transmission
protocol provides the communication of the \hw{RX LINK FIFO} status to
avoid the FIFOs overflow:
the \hw{LINK CTRL} sends to the neighbour node a 128-bit word called
\credit containing information about the status of its receiving
FIFOs.
In the same section we also motivate the requirement of stopping the
data transmission to update the \hw{RX LINK FIFO} status every $C$
cycles.
In order to avoid performance degradation, \ldm (see
table~\ref{tab:ldm}) are embedded in the spare bits of the \credit
without adding protocol overhead.

In the current implementation \lifama performs two main tasks:

\begin{itemize}
\item On the transmit side it handles the reading process of the
\hw{TX LiFaMa FIFO}. It updates the \ldm only when the \hw{LINK CTRL}
completes sending a \credit containing the previous one. This protocol
guarantees the integrity of fault communication.
\item On the receive side it simply collects \ldm into the \hw{RX
LiFaMa FIFO} avoiding the overflow.
\end{itemize}
\begin{table}[!htb]
\centering
\setlength\extrarowheight{3pt}
\begin{tabular}{|c|c|c|c|}
\hline
\hline
\multicolumn{4}{|c|}{\textbf{LiFaMa Diagnostic Message (LDM)}}\\
\hline 
\hline
\textbf{\# Bits} & \textbf{Bit Range} & \textbf{Field Name}        & \textbf{Protocol}             \\
\hline
2                &  1 -  0            & Service Network Status     & 00=normal; 01=sick; 10=broken \\ 
2                &  3 -  2            & Memory Status              & 00=normal; 01=sick; 10=broken \\
2                &  5 -  4            & Peripheral Status          & 00=normal; 01=sick; 10=broken \\
\hline
2                &  7 -  6            & DNP Core Status            & 00=normal; 01=sick; 10=broken \\
\hline
2                &  9 -  8            & Current Status             & 00=normal; 01=sick; 10=broken \\
2                & 11 - 10            & Voltage Status             & 00=normal; 01=sick; 10=broken \\
2                & 13 - 12            & Temperature Status         & 00=normal; 01=sick; 10=broken \\
\hline
2                & 15 - 14            & Z- Link Status             & 00=normal; 01=sick; 10=broken \\
2                & 17 - 16            & Z+ Link Status             & 00=normal; 01=sick; 10=broken \\
2                & 19 - 18            & Y- Link Status             & 00=normal; 01=sick; 10=broken \\
2                & 21 - 20            & Y+ Link Status             & 00=normal; 01=sick; 10=broken \\
2                & 23 - 22            & X- Link Status             & 00=normal; 01=sick; 10=broken \\
2                & 25 - 24            & X+ Link Status             & 00=normal; 01=sick; 10=broken \\
\hline
5                & 30 - 26            & Spare                      &                               \\
\hline
1                & 31                 & Valid                      &                               \\
\hline
\hline
\end{tabular}
\caption{\ldm Layout}
\label{tab:ldm}
\end{table}

\subsection{Host Fault Manager implementation}
%laura
\label{sec:hfm}
As described in section~\ref{sec:lofamospec} the \sw{HOST FAULT
MANAGER} is the software component in the \lofamo approach that is
responsible to gather information about the health status of the host;
that implements, together with the \hw{DNP FAULT MANAGER}, the
Watchdog mechanism described in section~\ref{sec:lofamowd} and that
creates a global awareness in case of fault thanks to diagnostic
messages sent via a Service Network towards the Fault Supervisor.

\begin{figure}[!hbt]
  \centering
  \includegraphics[width=\textwidth]{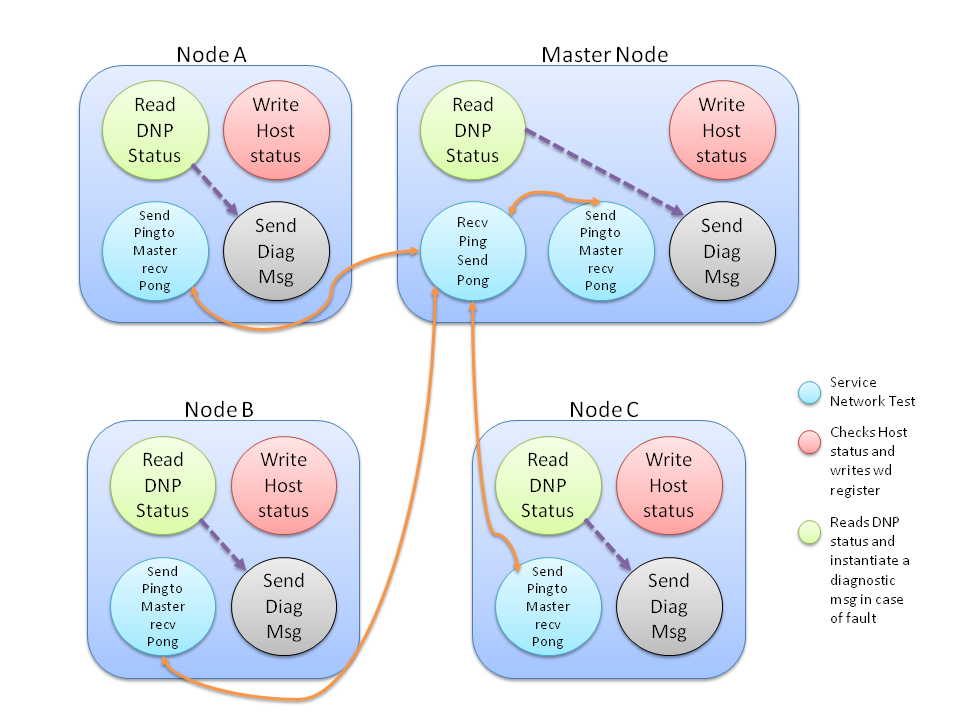}
  \caption{Structure of the HFM daemon running on different nodes,
  where one is the Master node. Each bubble is a Pthread.}
  \label{fig:LOFAMO:HFMthreads}
\end{figure}

On the \quong platform the HFM implementation is characterized by the
following features:

\begin{itemize}
\item it is a Linux daemon, \ie a program running in background and
using system calls to interact with the OS;
\item it spawns Pthreads to take care in parallel of the different
duties as shown in figure~\ref{fig:LOFAMO:HFMthreads};
\item it uses sockets to communicate via Ethernet network (Service
Network) to the HFM running on the Master node;
\item it accesses the \apenetp \lofamo registers using the \apenetp
API.
\end{itemize}

For what regards the checking of the host and the Service Network
status, the HFM can parse \sw{/proc/*} files to check - for example -
for memory errors (\sw{/proc/meminfo}), Ethernet interface errors
(\sw{/proc/net/dev}), \etc. A number of tools can be used to check for
more host status parameters, \eg temperature. All these techniques,
together with the watchdog mechanism, allow to obtain awareness of the
host Sick condition.\\ Moreover, each HFM instance is able to
determine if the Service Network is completely cut off on the node it
runs on: a HFM thread sends a UDP message (\sw{snet\_ping}) to the
Master node every \sw{SNET\_MON\_PING\_TMOUT} seconds (3 for example,
but it is configurable), if the Master node does not reply
(\sw{snet\_pong}) it waits more \sw{SNET\_MON\_PING\_TMOUT} seconds
and retries. If a \sw{snet\_pong} is not received by then, the Service
Network is set as \emph{broken} on that node, situation that is
managed by the watchdog mechanism.\\ In case of complete breakout of
the host the watchdog mechanism is alerted by the missing update of
the \hwr, and managed as described before.

To summarize, table~\ref{tab:HFMthr} describes the work done by each
thread in the HFM:
\begin{table}[!htb]
\centering
\footnotesize
\begin{tabular}{|l|p{12cm}|}
\hline
\hline
\textbf{host\_wd\_thread} & Gathers the Host status and writes the \hwr\\
\hline
\textbf{DNP\_wd\_thread} & Reads the \dwr and enqueues a diagnostic message in case of faulty DNP\\
\hline
\textbf{snet\_monitor\_thread} & Continuously pings the master to check for the snet status.\\
\hline
\textbf{snet\_master\_thread} & Receives the pings from the nodes and replies with a pong; receives and manages diagnostic messages.\\
\hline
\textbf{snet\_fault\_notifier\_thread} & Sends diagnostic messages to the Master node.\\
\hline
\hline
\end{tabular}
\caption{HFM Pthreads and their function.}
\label{tab:HFMthr}
\end{table}

Note that the HFM does not make decisions about how to react to faults
and it does not have the big picture of the system status, but as part
of the \lofamo approach, it is a means to spread information about
faults and critical events and make the software high level layers
(like DAL) to obtain \emph{systemic fault awareness}.

Every time the HFM is launched --- typically at the platform start up
--- the \sw{main()} function initializes the \lofamo mechanism on the
basis of a configuration file. The initialization procedure requires
the following steps:
\begin{itemize}
\item acquire the system topology and wait for the links to be
aligned;
\item set DFM masks and thresholds;
\item set watchdog timers;
\item enable DFM;
\end{itemize}

The \lofamo registers mapping for \apenetp/\quong is reported in table
\ref{tab:FaultDetRegs}.

%%Laura: probably there won't be a lofamo demo this year
%%\subsection{Design of Systemic Awareness Demo on Experimental HW Platform (at least 2 apenet+ - Laura}

%% file: chapter2.tex
\section{Experimental HW Platform Integration and Test}
\label{sec:quong}
	% It is a first phase of the activity, other work next year.

During the 2012 the HPC platform integration and test activity
provided the massive production of the \apenetp boards, their
integration to the full \quong~\cite{ammendola2011quong} tower and the
validation of the system through synthetic benchmarks.

In this year we implemented architectural improvement of \apenetp with
the aim of achieving the best performances and usability for the
hardware platform (sections \ref{sub:TxAcceleration}, \ref{sec:eventq} and
\ref{sec:link_bw_lat}).  Furthermore, we developed the VHDL coding of
the DNP Fault Manager required to interface to the \apenetp's sensors
and to transfer the information to the DNP status registers
(section \ref{sec:sensor}).

The software development consisted in the implementation of the
necessary support to the hardware innovations as well as a continuous
tuning to reduce the latency due to the driver itself
(section \ref{sec:swstack}).  Besides the fundamental middleware needed to
implement a working API for communication with the board, a large
number of small test codes (mostly in C supplemented by scripts in
Bash/TCSH) were written (section \ref{sec:testsuite}).

\subsection{\apenetp}

\begin{figure}[!hbt]
  \centering
  \includegraphics[trim=60mm 25mm 60mm 30mm,clip,width=0.75\textwidth]{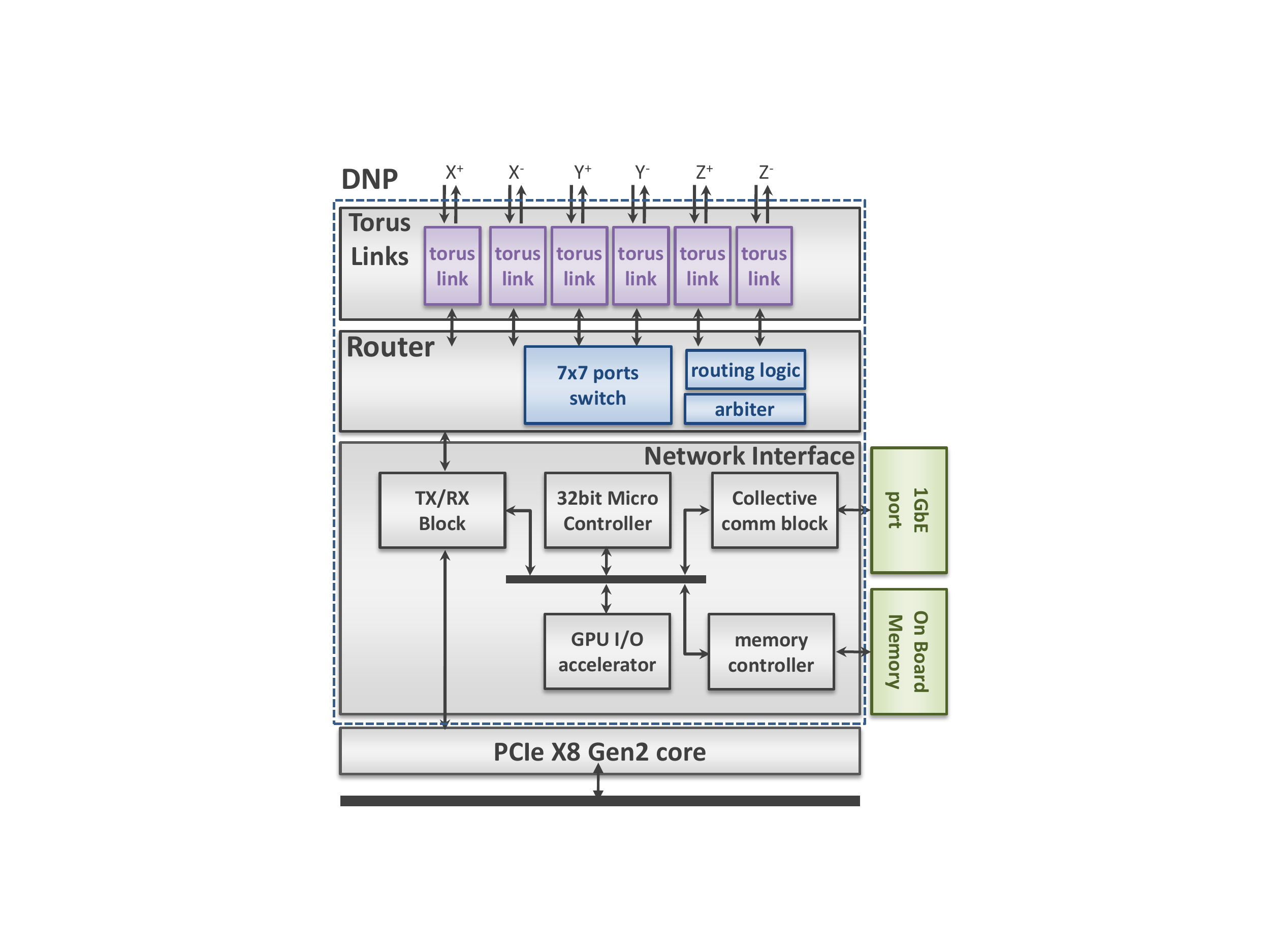}
  \caption{Overview of APEnet+. The DNP is the core of the
architecture - composed by the Torus Links, Router and Network
Interface macro blocks - implemented on the FPGA. The system interfaces
to the host CPU through the PCIe bus.}
  \label{fig:architecture}
\end{figure}

%\apenetp is a standard full length, full height, two I/O slots wide PCI Express card;
%its goal is the development of a low-latency and high-speed network with a 
%3D-torus structure and point-to-point links. 
The \apenetp network
architecture~\cite{rossetti2011apenet+,ammendola2011apenet+,ammendola2012apenet+}
has, at its core, the Distributed Network Processor
(DNP)~\cite{DNP2012}.
This acts as an off-loading network engine for the computing node,
performing inter-node data transfers.  The DNP hardware blocks
structure, depicted in figure \ref{fig:architecture}, is split into a
Network interface (the packet injection/processing logic comprising
host interface, TX/RX logic and Microcontroller), a Router component
and six Torus Links.

The network interface block has basically two main tasks: on the
transmit data path, it gathers data coming in from the PCIe port -
either from host or GPU memory - and forwards them to the relevant
destination ports; on the receiving side, it provides hardware support
for the Remote Direct Memory Access (RDMA) protocol, allowing remote
data transfer over the network without involvement of the CPU of the
remote node.

The router establishes dynamic links among the ports of the DNP,
managing conflicts on shared resources.

The torus link block manages the data flow and allows point-to-point,
full-duplex connection of each node with its 6 neighbours with
coordinates both increasing and decreasing along the X, Y and Z axes,
forming a 3D torus.

\subsubsection{Architectural Improvements}
 \label{subsec:arch_impr}

%In this section we describe the \apenetp's architectural improvements

\subsubsubsection{Link Bandwidth and Latency}
   \label{sec:link_bw_lat}
In this section we show the performance achieved with the current
implementation of the \hw{TORUS LINK} and we will argue the results of
the tests, analyzing the limits being reached and providing a set of
possible changes to overcome them.

Referring to the Open Systems Interconnection (OSI) model, the
\hw{TORUS LINK} plays the dual role of the Physical Layer and Data
Link Layer.

The Physical Layer is constituted by Altera transceivers of the
Stratix IV, able to reach 8.5 Gbps per line (each link is formed by 4
lines for a total of 34 Gbps).
Dynamic reconfiguration of Physical Medium Attachment's analog
settings are guaranteed by \hw{Reconfig Block}.
We have performed various trials to increase the operating frequency
of transceivers, minimizing the noise level of the channel. At present
transceivers operate at a frequency of 350 MHz (28 Gbsp) to maximize
the reliability of data transmissions.
The Altera transceiver ensures DC-balancing on the serial data
transmitted exploiting the 8B10B encoder. In addition it implements a
\hw{Word Aligner Block} to restore the word boundary lost during the
operation of serialization and deserialization and a \hw{Byte Ordering
Block}.
Finally the bonding of four independent lanes on a single a channel is
gained with four \hw{DESKEW FIFO}s.

The Data Link Layer is constituted by \hw{LINK CTRL}. It manages the
data flow by encapsulating packets into a light, low-level,
word-stuffing protocol able to detect transmission errors via CRC.

\begin{figure}[!hbt]
  \centering
  \includegraphics[trim=0mm 40mm 0mm 40mm,clip,width=\textwidth]{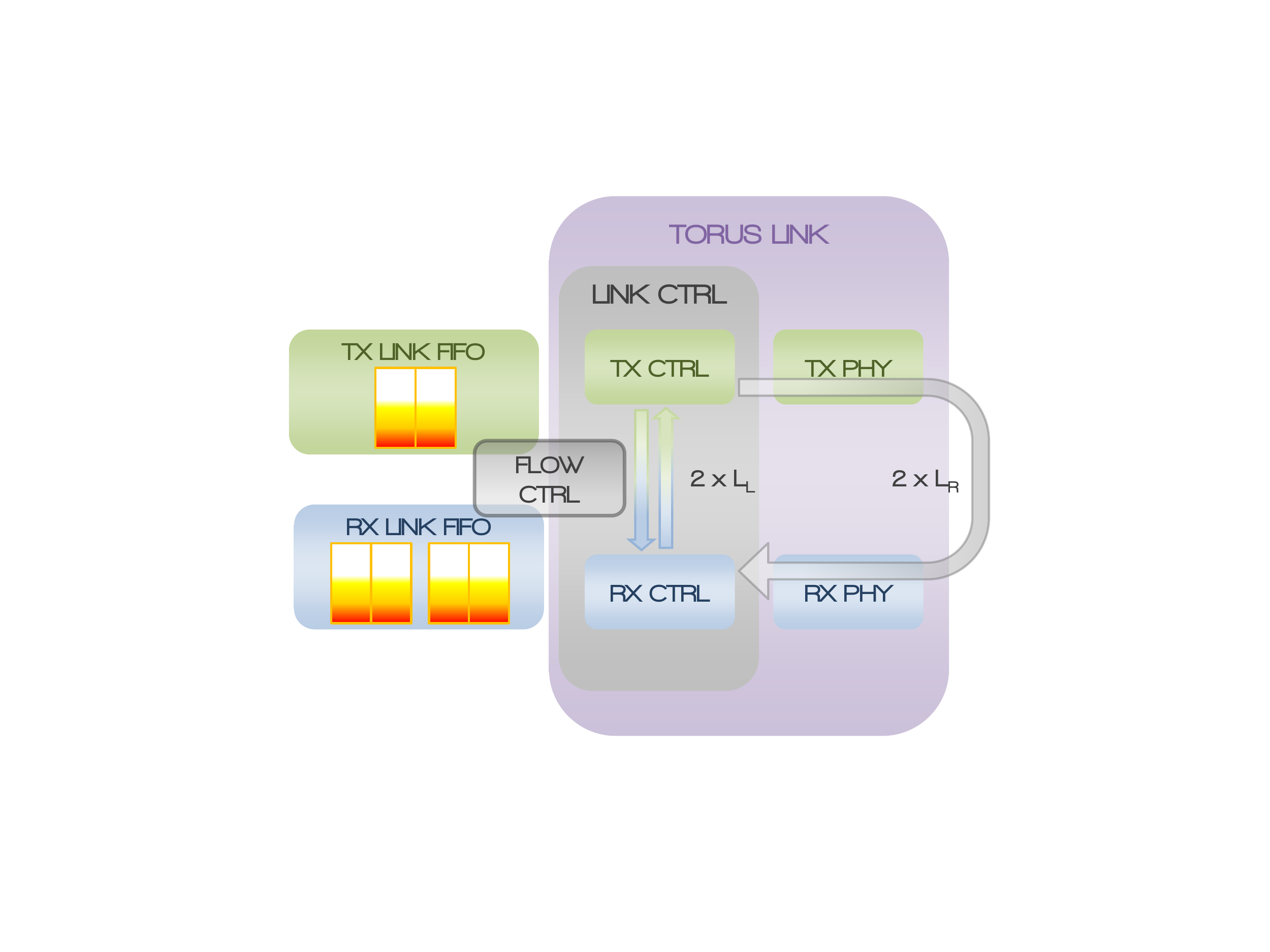}
  \caption{\hw{TORUS LINK} Overview. $L_{R}$ and $L_{L}$ are the
  latency observed during a data-transaction between two neighbour
  nodes.}
  \label{fig:link}
\end{figure}

\begin{figure}[!hbt]
  \centering
  \includegraphics[trim=0mm 20mm 0mm 20mm,clip,width=\textwidth]{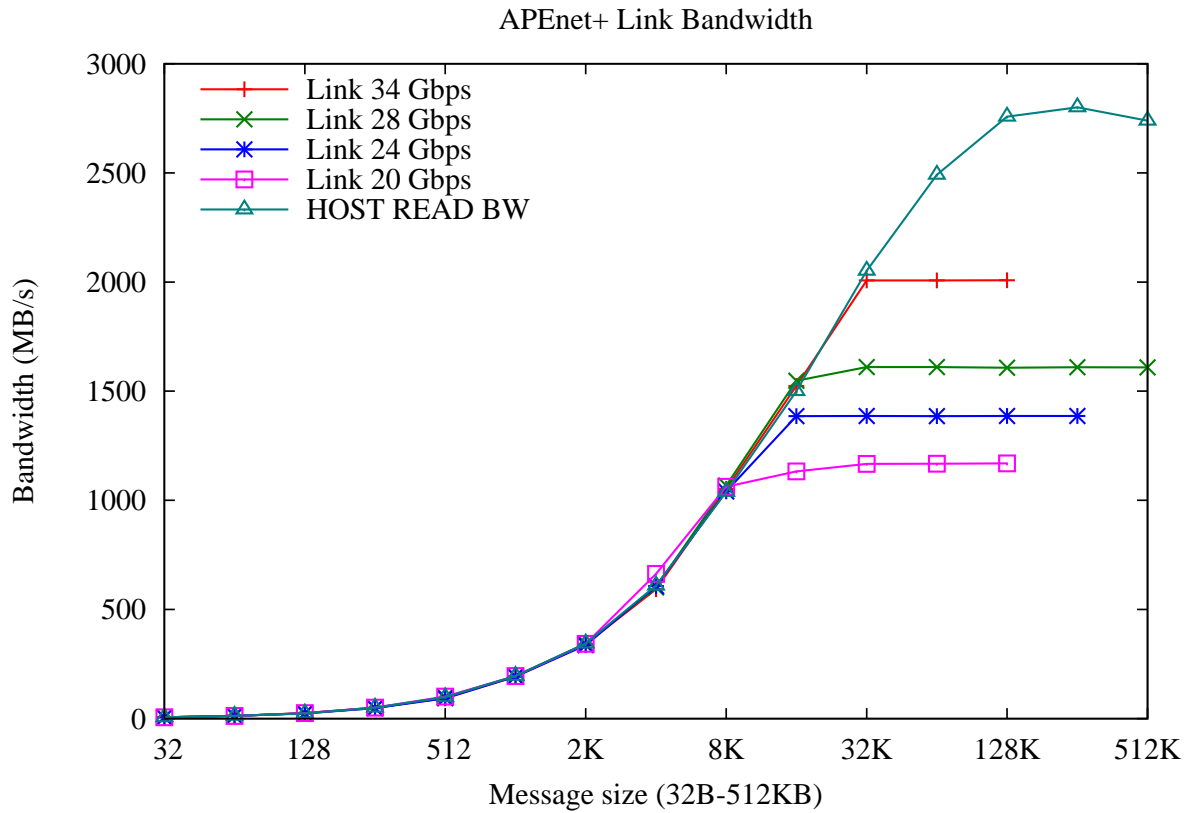}
  \caption{\apenetp Link Bandwidth. The curves were obtained by
    setting the frequency of transceivers respectively at 425 MHz
    (34Gbps), 350 MHz (28 Gbps), 300 MHz (24 Gbps) and 250 MHz (20
    Gbps). The Link Bandwidth $BW_{L}$ overlaps the Host Read
    Bandwidth $BW_{H}^{READ}$ until message size of 8KB-32KB,
    depending of the operating frequency of the transceivers.}
  \label{fig:link_bw}
\end{figure}

As shown in figure \ref{fig:link_bw}, measurements of bandwidth were
taken by varying the frequency of the transceivers to evaluate the
performance of the existing implementation of the \hw{LINK CTRL}.

In order to have a thorough evaluation of the present deployment of
\hw{LINK CTRL}, it must be taken into account that 8B10B encoder
causes a loss of 20\% for bandwidth. Then the maximum achievable link
bandwidth ($BW_{L}^{MAX}$) at the four different operating frequency
is 3.4 GB/s, 2.8 GB/s, 2.4 GB/s and 2.0 GB/s.
Our plots show an Efficiency Factor $E_{TOT}$ of approximately 60\% of
the theoretical maximum bandwidth, due to the implemented protocol
regardless of the set frequency.

A deeper analysis of the adopted transmission protocol can help to
understand the reasons for the performance degradation. The typical
\apenetp data packet is formed by 1 \header, 1 \footer (two 16-byte
words) and a \payload of maximum size equal to 4096 bytes.
The \hw{LINK CTRL} adds two additional 16-byte words (\magic and
\start) to initiate a new data transaction. Then 64 bytes of protocol
($P$) are necessary for each data packet. By focusing on maximum data
packet size ($S_{MAX}$) we can define a Efficiency Factor $E_{1}$
equal to:

$$E_{1}=\frac{S_{MAX}}{P+S_{MAX}}=\frac{4096}{64+4096}=0.985$$

The \hw{LINK CTRL} is responsible for managing the data flow between
two neighboring nodes preventing the FIFO overflow. The \hw{LINK CTRL}
keeps track of how many 16-byte data words have been sent. When it
reaches the values set in a programmable threshold ($T_{RED}$) the
data transmission is interrupted.
The \hw{LINK CTRL} ensures that the receiving FIFOs (\hw{RX LINK FIFO})
have been emptied at least up to value $T_{YELLOW}$, yet programmable
and with the constraint $T_{YELLOW} < T_{RED}$.
The information about the status of \hw{RX LINK FIFO} (how many words
are placed into the receiving FIFOs) is contained in 16-byte word
called \credit.
Referring to figure \ref{fig:link_bw} it is possible to estimate the
data-interruption cycles. \hw{LINK CTRL} is able to measure the flight
time of a 16-byte word over the Physical Layer (Remote Latency):

$$L_{R}=35\ cycles$$   

The \hw{LINK CTRL} has to wait for the flight time of the last sent
data and of \credit containing the actual status of \hw{RX LINK FIFO}
before resuming data transmission.
Furthermore, we must consider the time needed to transfer the
occupancy information from the \hw{RX LINK FIFO} to the \hw{LINK CTRL}
of the receiving node and from the \hw{RX CTRL} to the \hw{TX CTRL} of
the transmitting node. The receiving side of a \hw{LINK CTRL} clock is
reconstructed from the incoming data (Clock Data Recovery) while on
the transmitting side the clock is locally generated by PLLs. A
synchronization is essential to exchange data between receiving and
transmitting side (Local Latency):

$$L_{L}=20\ cycles$$   

Then the Total Latency of the updated status of FIFOs is:

$$L_{T}=2 \times L_{R}+2 \times L_{L}=110\ cycles$$ 

Unfortunately waiting for $L_{T}$ cycles is not enough. Indeed, the
\hw{TX CTRL} of the receiving node might be busy in transferring data
packets. Since the maximum packet size is 4096 bytes and for each
cycle a 16-byte word will be transmitted, the transmitter could be
committed over 256 cycles. To avoid this latency addiction \hw{TX
CTRL} sends a \credit every $C$ cycles.
The transmission protocol requires the submission of a 16-byte word
(\magic) every time that a \credit is submitted, to distinguish it
from the \payload (Word-Stuffing Protocol). Therefore, this mechanism
introduces a new factor of efficiency :

$$E_{2}=\frac{C}{C+2}$$

At this point we can define the ${W}$ cycles of transmission
interruption to get a \credit containing the updated status of the
\hw{RX LINK FIFO}.  In the worst case the receiving node takes $C$
cycles to submit the proper \credit so the interruption at the
transmitting node is:

$$W=L_{T} + C$$

When the value contained in the \credit is less than $T_{YELLOW}$ the
\hw{TX CTRL} resumes data transmission.

In case of normal data traffic (in a test this means just one
transmitting node and one receiving node, with data coming from a
single channel and the other channels in idle) writing and reading
speed of the \hw{RX LINK FIFO} are basically equal. Thus the first
\credit that arrives to the transmitter should contain a value below
the $T_{YELLOW}$, and with good approximation we can say that the
\hw{RX LINK FIFO} are empty and the transmitting node can resume data
transmission.
Summarizing, the transmitting node is able to send words for $T_{RED}$
cycles continuously and stops for $W$ cycles waiting for the \credit
before resuming the transmission and so on, with an additional
Efficiency Factor:

$$E_{3}=\frac{T_{RED}}{T_{RED}+W}$$

We can then define the Total Efficiency Factor:

$$E_{T}=E_{1} \times E_{2} \times E_{3}$$

$T_{RED}$ is set to 506 (to maintain a margin of safety in writing
\hw{RX LINK FIFO}) and thus we define $E_{T}$ as a function of the only
variable $C$:

$$E_{T}=0.985 \times \frac{C}{C+2} \times \frac{506}{616+C}$$

Maximizing this function for $C$ in the range $C\in[0;55]$, we obtain:

$$C=35.1$$

Then setting $C$ = 35 and substituting in formulas for $E_{2}$,
$E_{3}$, $E_{T}$

$$E_{2}=\frac{C}{C+2}=0.946$$
$$E_{3}=\frac{T_{RED}}{T_{RED}+W}=0.777$$
$$E_{T}=E_{1} \times E_{2} \times E_{3} = 0.724$$

Therefore with the current settings, we expect a bandwidth equal to
72.4\% of the theoretical maximum and instead we observe a plateau at
60\% regardless of the set frequency.
This additional loss is due to the routing algorithm adopted. We must
take into account that no transaction starts if the receiving module
does not have sufficient space to accommodate the entire message.
The information contained in the \credit is available for the
\hw{ROUTER}. The \hw{ROUTER} does not instantiates a new transaction
if there is not enough space in the \hw{RX LINK FIFO} to accommodate
the whole data-packet. Once that the sending of a data packet is
complete the \hw{FLOW CTRL} underestimates the space available in the
remote FIFO until receiving a \credit:

$$SPACE\ IN\ FIFO=T_{RED}-S_{MAX}=250<256$$

Then the \hw{ROUTER} does not allow sending a packet with payload of
4KB before the arrival of a \credit that updates the status of the
\hw{RX LINK FIFO}.
This changes the efficiency factor $E_{3}$. After sending every
packet, \hw{TX CTRL} waits for $W$ cycles:

$$E_{3}=\frac{S_{MAX}}{S_{MAX}+W}=0.638$$

With this correction the Total Efficiency Factor is in line with the
observed data:

$$E_{T} = 0.595$$

\begin{figure}[!hbt]
  \centering
  \includegraphics[trim=0mm 20mm 0mm 20mm,clip,width=\textwidth]{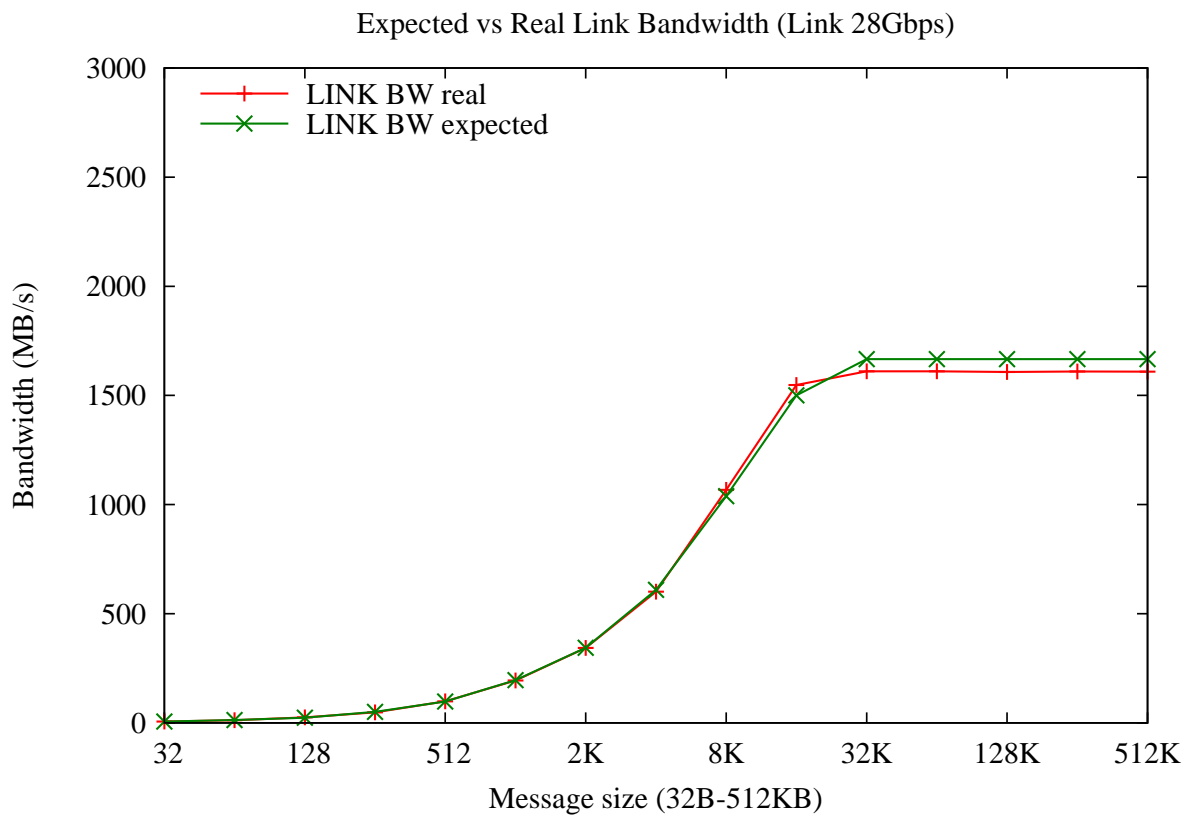}
  \caption{A comparison between the results obtained in a
  point-to-point bandwidth test and the theoretical bandwidth
  calculated considering the Total Efficiency Factor $E_{T}$.}
  \label{fig:link_theory}
\end{figure}

Observing figure \ref{fig:link_bw} we note that if the host read
bandwidth ($BW_{H}$) is less than the link maximum bandwidth
$BW_{L}^{MAX}$ then the $BW_{L}=BW_{H}$.
With this arrangement it is possible to estimate the bandwidth as a
function of message size with good results, as shown in figure
\ref{fig:link_theory}.

It is straight-forward that to improve the \hw{LINK CTRL} performance
you need to increase the Total Efficiency Factor $E_{T}$.

$E_{1}$ and $E_{2}$ are closely related to the transmission protocol
and the hardware implementation of the transceiver. Modifying these
factors leads to a major re-write of the hardware code. The factor
$E_{1}$ could benefit from a different choice for the transmission
protocol but would involve a re-write of the \hw{LINK CTRL} state
machine. $E_{2}$ is related to the latency of transceivers due to
DC-balancing, alignment and bonding of the channels. In addition, the
current values are acceptable considering that together they result in
a loss of 7\% of performance.

\begin{figure}[!hbt]
  \centering
  \includegraphics[trim=0mm 20mm 0mm 20mm,clip,width=\textwidth]{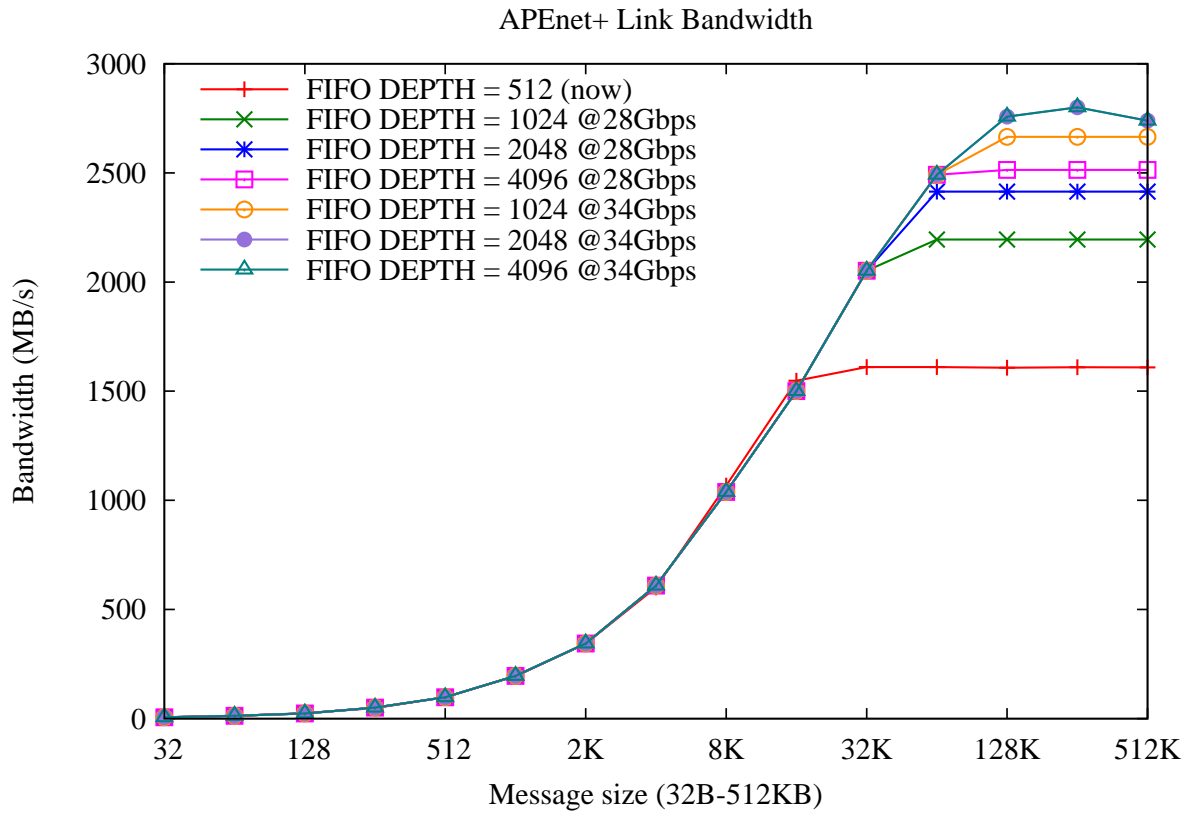}
  \caption{Expected Bandwidth at varying \hw{RX LINK FIFO} depth}
  \label{fig:link_expectation}
\end{figure}

As mentioned earlier, $E_{3}$ is related to the latency of the channel
and to the programmable threshold $T_{RED}$. With currently set value
(506) is not possible to allocate two packets of maximum size (4KB) at
the same time inside the receiving FIFOs. The only condition that
would make it possible is to set the programmable threshold equal to
the depth of \hw{RX LINK FIFO} ($T_{RED}={FIFO\_DEPTH}$). This choice
is not safe in case of transfers between remote nodes. A minimum
misalignment (due to jitter, noise, synchronization of signal at
different clock frequency) could result in the overflow of the
receiving FIFOs with consequent loss of transmitted data.
Thus the most straight way to increase $T_{RED}$ is to increase the
depth of the \hw{RX LINK FIFO}, then change $E_{3}$, resulting in
improved performance, without any major changes to the hardware code.
Table \ref{tab:efficiency} and figure \ref{fig:link_expectation} show
the benefits that would be obtained by changing the depth of the
FIFOs, at the attained clock frequency 350MHz (28Gbps) and at 425MHz
(34Gbps), which is the maximum frequency that can be achieved by
Altera Stratix IV GX transceivers.

\begin{table}[!hbt]
\centering
\setlength\extrarowheight{2pt}
\begin{tabular}{|l|cccc|}
\hline
\hline
FIFO DEPTH & $E_{3}$ & $E_{T}$ & $BW_{L}^{MAX}$@28Gbps & $BW_{L}^{MAX}$@34Gbps  \\
\hline
512        & 0.638   & 0.595   & 1666 MB/s  & 2023 MB/s     \\
1024       & 0.841   & 0.784   & 2195 MB/s  & 2665 MB/s     \\
2048       & 0.925   & 0.862   & 2414 MB/s  & 2931 MB/s     \\
4096       & 0.964   & 0.898   & 2514 MB/s  & 3060 MB/s     \\
\hline
\hline
\end{tabular}
\caption{Efficiency Factor at varying the \hw{RX LINK FIFO} depth.}
\label{tab:efficiency}
\end{table} 

Observing the values that could be obtained at 425 MHz can be taken as
immediate the choice to quadruple (or even make 8 times greater) the
\hw{RX LINK FIFO}. In this way the maximum link bandwidth becomes
larger than host read bandwidth and the transmission protocol does not
slow down the data transfer.
Unfortunately we have to deal with the resources available on the
Stratix~IV.  Increasing 4 (8) times the depth of the FIFOs implies
that each FIFO occupies 32~KB (64~KB) compared to the current 8KB. It
should also be noted that each of the 6 channels is equipped with two
receiving data FIFOs to ensure a deadlock free routing
(virtual-channel) for a total of 384~KB (768~KB). Whereas the
Stratix~IV provides (1.74~MB) of embedded memory it is necessary to
limit the amount of memory to allocate for data transfer.
For this reason, we have scheduled a new revision of the \hw{LINK
CTRL} in the near future with receiving data FIFOs 2 times larger
(16~KB).  This option seems to be the right balance between used
resources and expected performance improvement.

Indeed, with these conditions it will be possible to achieve a maximum
link bandwidth of 2650~MB/s, with a loss of \texttildelow5\% compared
to the host read bandwidth.
This result can be achieved only setting the maximum operating
frequency for the links (425~MHz). Currently the transceivers are
considered stable at a frequency of 350~MHz. In the future we will try
to optimize the operating frequency changing Equalization, 
Pre-Emphasis and DC Gain. However because of
the noise of the cables, the coupling between connector and cable and
especially the delicate coupling between daughter-card and board, it
could not be possible to achieve the maximum frequency for all
channels.  This could lead to a further loss of performance (for
example at 350~MHz the maximum achievable bandwidth is 2200~GB/s).
Obviously, this obstacle can be easily circumvented by doubling the
size of the FIFOs further (32~KB). During the next year, very
demanding memory optimizations will be completed (RX processing
accelerator) and we can assess more precisely the amount of available
memory.
Nevertheless, a further analysis factor $E_{3}$ leads to focus on
another limitation that affects performance negatively. The \hw{LINK
CTRL} is forced to interrupt the data transmission waiting for an
update on the status of the remote FIFO to avoid overflow them.
At a later stage it may be interesting to evaluate the possibility of
modifying the \hw{LINK CTRL} to avoid the interruption of data
transmission.  This change would result in a re-write of only a
portion of the link code hardware: the one responsible of the data
flow \hw{FLOW CTRL}.

%
%Also look at Table \ref{tab:degradation} and Figure \ref{fig:link_expectation}
%it is noted that the improvement of performance tends to decrease with increasing
%FIFO DEPTH.
%
%This behavior becomes clearer by analyzing Figure \ref{fig:Dfactor} which shows
%the trend of the degradation factors $D_{3}$ and $D_{T}$ at varying the fifo 
%depth (and therefore the variation of packets that can be allocated within the
%FIFO before the link interrupts the transmission of the data to wait for the 
%\texttt{Credit})
%
%\begin{figure}[!hbt]
%  \centering
%  \includegraphics[width=\textwidth]{degradation.png}
%  \caption{Degradation factor}
%  \label{fig:Dfactor}
%\end{figure}
%

\begin{figure}[!hbt]
  \centering
  \includegraphics[width=0.9\textwidth]{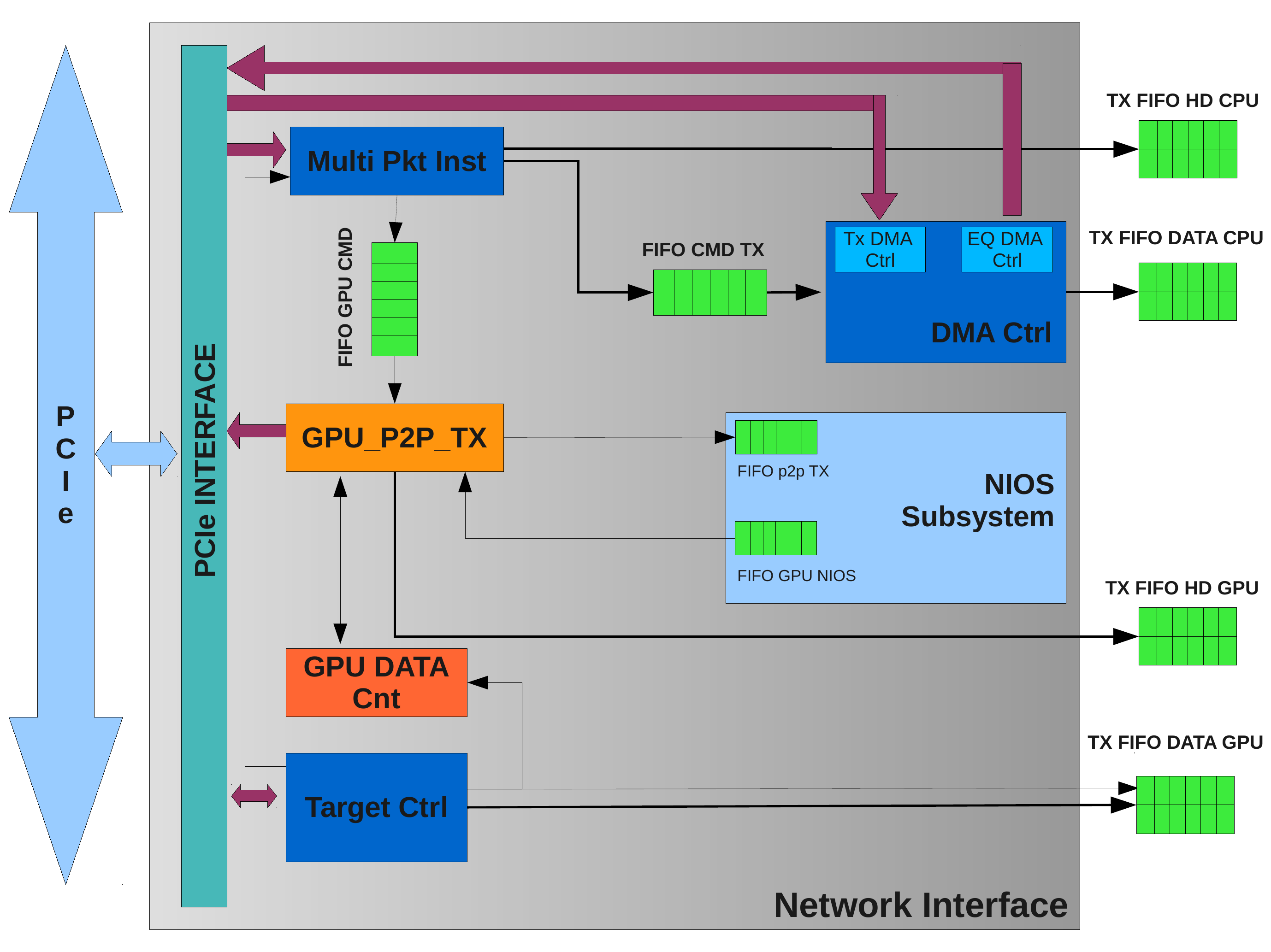}
  \caption{\apenetp's PCI read (TX) flow.}
  \label{fig:TX_datapath}
\end{figure}

\subsubsubsection{Tx Acceleration} %  (double PCI channel, ...) - Francesca
\label{sub:TxAcceleration}
On the transmit data path the DNP handles transfers from CPU/GPU
through the \PCIe port, forwarding the data stream to the TX FIFOs
(figure \ref{fig:TX_datapath}).

Referring to figure \ref{fig:TX_datapath}, PCI read transactions (TX)
are issued by command-packets, each one composed by four 128bit-words:
\header, \footer, \hw{CMD0} and \hw{CMD1}. The first two are the
\header and the \footer of the data-packet that will be sent to the 3D
network, while \hw{CMD0} and \hw{CMD1} are commands that contain
information used respectively to program the data PCI DMA read and to
communicate to the \hw{EVENT QUEUE DMA CTRL} the completion of data
reading (for more details see section \ref{sec:eventq}).

Command-packets are written by the host in a \ringbuffer, which has a
fixed size and is managed by two pointers: a \writepointer and a
\readpointer.  The \writepointer is controlled by the software main
processor and it is written in the DNP register \wrptr; on the
contrary, the \readpointer is managed by the \hw{MULTI PKT INST}, and
it is recorded in a read only register \rdptr.

A difference between \wrptr and \rdptr informs the DNP of the presence
of new command-packets to execute. 
In this case, the \hw{MULTI PKT INST} programs a PCI DMA read
transaction of $N$ command-packets, starting from the address of the
last received command-packet, where
$$N=\wrptr-\rdptr$$ 
%
%The number $N$ of the command-packets loaded depends on value of 
%registers \wrptr and \rdptr. 
As soon as PCI DMA read transaction ends, \hw{MULTI PKT INST} computes
the new \ringbuffer \readpointer and updates the \rdptr register (this
mechanism is explained in detail in D6.1 ~\cite{euretile:D6_1}).

\begin{figure}[!hbt]
  \centering
  \includegraphics[width=0.9\textwidth]{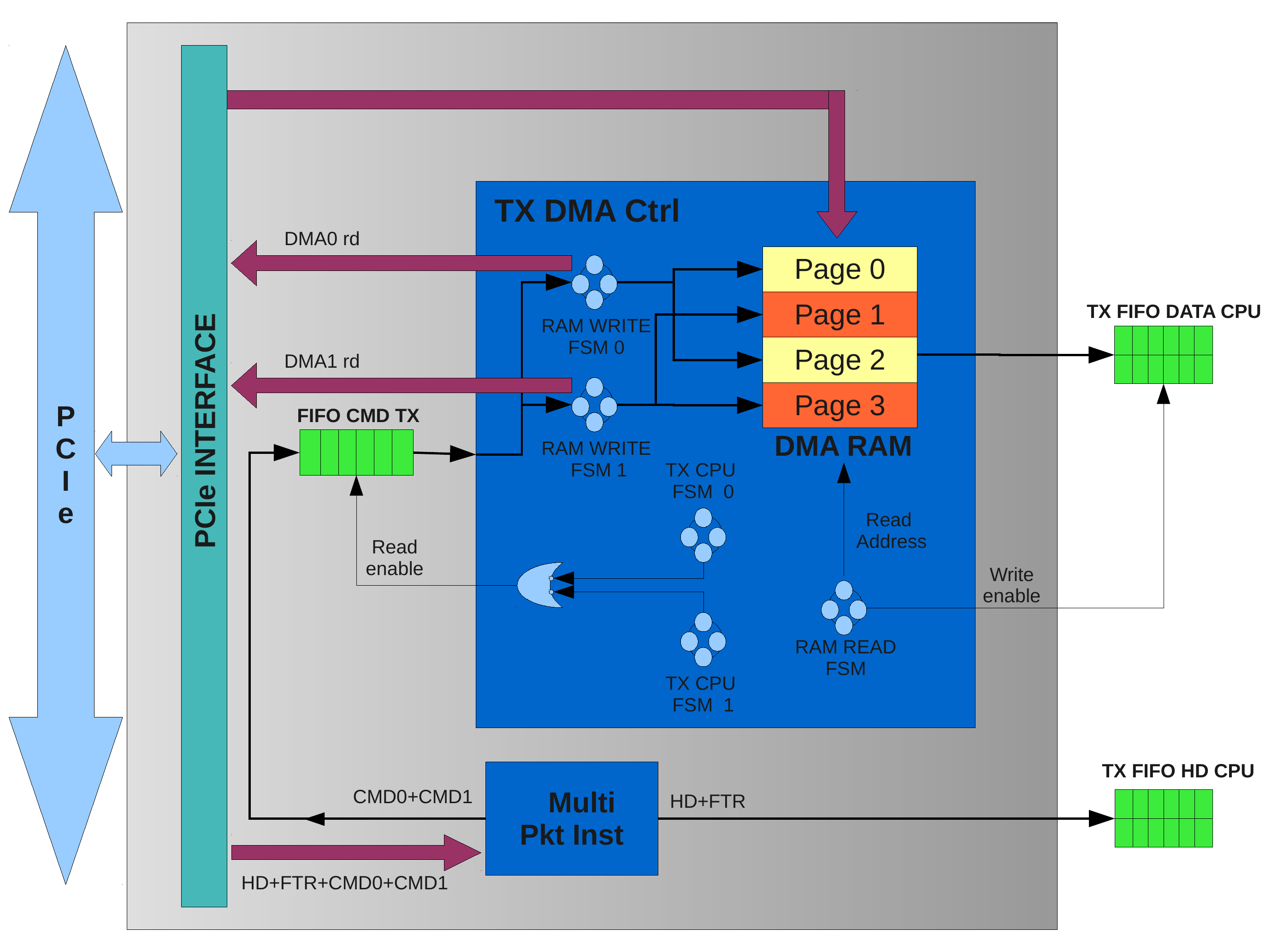}
  \caption{\apenetp's CPU TX flow.}
  \label{fig:TX_CPU}
\end{figure}

%{Host interface}
In case of CPU TX transaction, \header and \footer are directly
written in \hw{TX FIFO HD CPU}, while \hw{CMD0} and \hw{CMD1} are
pushed in \hw{FIFO CMD TX} (figure \ref{fig:TX_CPU}).

Using size and address specified in the \hw{CMD0}, \hw{TX DMA CTRL} is
able to program the requested DMA to load the packet's payload. In
order to accelerate the data transfers, the DNP implements two
separated DMA channels (DMA0 and DMA1): it instantiates transactions
on the \PCIe bus switching from one channel to the other, reducing the
latency between two consecutive data requests.

\hw{TX CPU FSM 0} and \hw{TX CPU FSM 1} pop commands from the \hw{FIFO
CMD TX} alternately; data received from channel DMA0 are written by
\hw{RAM WRITE FSM 0} in the pages 0 and 2 of the \hw{DMA RAM}, while
\hw{RAM WRITE FSM 1} is in charge of pages 1 and 3.

%At the end of payload's writing in the \hw{DMA RAM}, the \hw{CMD1} is 
%read too.
%
At the end of the \hw{DMA RAM} writing process, \hw{TX DMA CTRL}
analyzes the \hw{CMD1}.
In case of not dummy command (that is \hw{CMD1} bit 3 - 2="00"), the
\hw{TX DMA CTRL} communicates to the main processor that required
operation has been executed through \hw{EVENT QUEUE DMA CTRL} (see
section \ref{sec:eventq}); dummy \hw{CMD1} is simply dismissed.
Finally \hw{RAM READ FSM} recursively reads all written RAM pages and
pushes the \payload in \hw{TX FIFO DATA CPU}.

\begin{sidewaysfigure}
  \centering
  \includegraphics[width=\textheight]{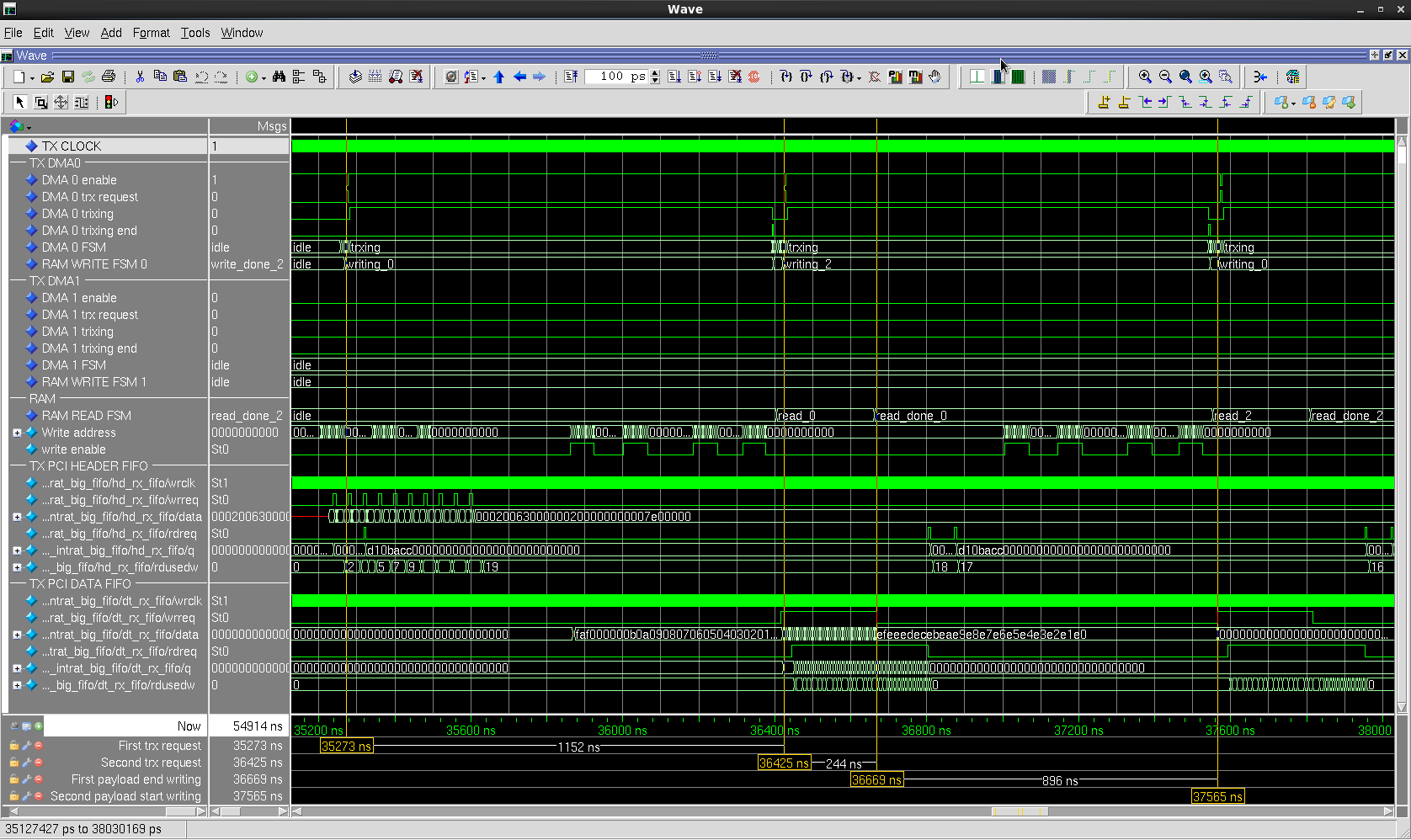}
  \caption{\apenetp's CPU TX simulation with 1 DMA channel.
Time elapsed between two consecutive DMA requests is very long
(1152~ns) and the \payload of the second packet is written in \hw{TX
FIFO DATA CPU} after 896~ns.}
  \label{fig:TX_sim_1DMA}
\end{sidewaysfigure}

\begin{sidewaysfigure}
  \centering
  \includegraphics[width=\textheight]{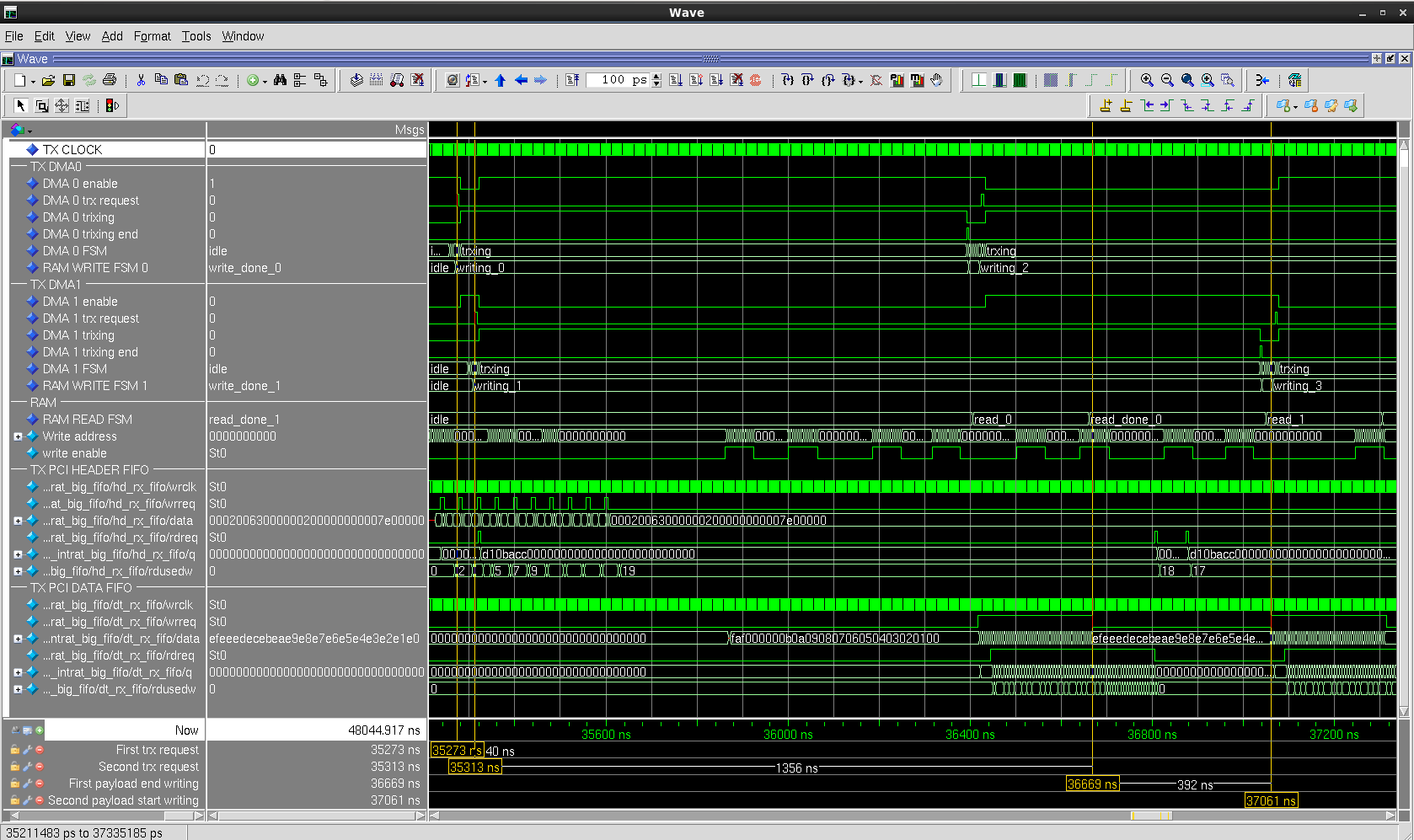}
  \caption{\apenetp's CPU TX simulation with 2 DMA channels.
The second request occurrs only 40~ns after the first one, 
and the complete \payload of the second packet is available 
only 392~ns after the first packet \payload.}
  \label{fig:TX_sim_2DMA}
\end{sidewaysfigure}

Figures \ref{fig:TX_sim_1DMA} and \ref{fig:TX_sim_2DMA} show waveforms
of a simulation of two READ commands sent to \apenetp.

In the first case only one DMA is enabled, and the \hw{DMA RAM} uses 2
pages; time elapsed between two consecutive DMA requests is very long
(1152~ns) and the \payload of the second packet is written in \hw{TX
FIFO DATA CPU} after 896~ns. In the second case the DMA 1 channel
requests data after 40~ns despite of DMA 0, and the complete \payload
of the second packet is available only 392~ns after the first one.

\begin{figure}[!hbt]
  \centering
  \includegraphics[trim=15mm 20mm 15mm 20mm,clip,width=\textwidth]{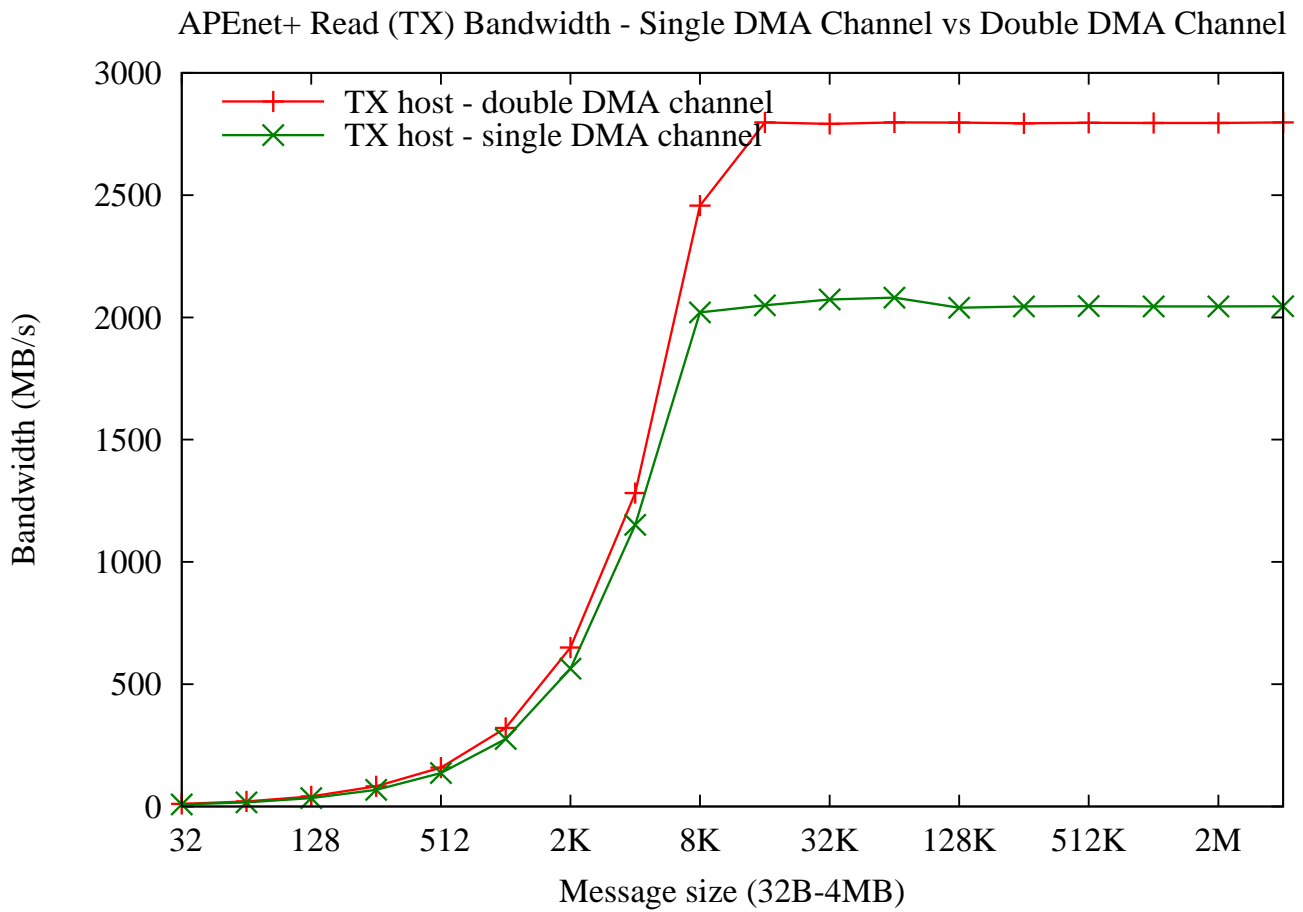}
  \caption{Bandwidth with one or two DMA channels for HOST to HOST
  transactions.}
  \label{fig:2DMA_BW}
\end{figure}

Figure \ref{fig:2DMA_BW} compares the read bandwidth measured using
one or two DMA channel implementation, showing a bandwidth improvement
of 40\%.

%\subparagraph{P2P (GPU oriented) interface}%Ottorino

 \begin{figure}[!htb]
  \centering
  \includegraphics[width=0.9\textwidth]{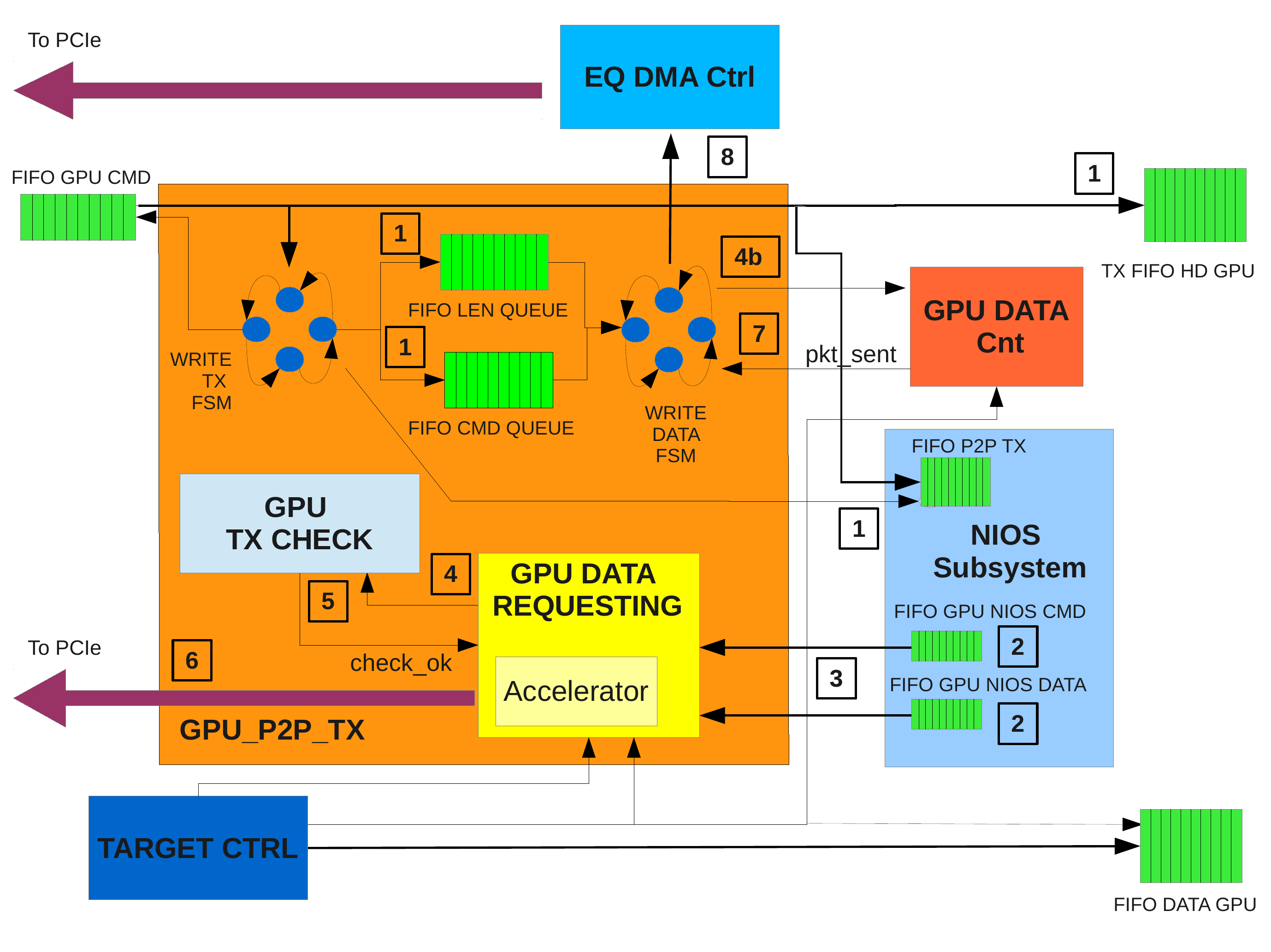}
  \caption{\apenetp's GPU p2p tx.}
  \label{fig:tx_gpu}
  
\end{figure}
Unlike host buffer transmission, which is completely handled by the
kernel driver, GPU data transmission is delegated to \apenetp.  In
particular, the \apenetp board is able to take part in the so-called
\PCIe \PtoP (P2P) transactions: it can target GPU memory by ordinary
RDMA semantics with no CPU involvement and dispensing entirely with
intermediate copies.

In case of GPU read transaction, command-packets are pushed into
\hw{FIFO GPU CMD} (figure \ref{fig:tx_gpu}). \ptoptx reads the
\hw{FIFO GPU CMD} and moves \header and \footer towards \hw{TX FIFO HD
GPU} and \hw{CMD0} in \hw{FIFO P2P TX} to request data to the
GPU. Finally it sends \hw{CMD1} to the \hw{EVENT QUEUE DMA CTRL} to
generate a completion.
%
%GPU\_P2P\_TX==gpu_target_tx!
%
 
The first version of \ptoptx ( version \textit{V1} ) reads just one command-packet at a
time, waiting for the reception of the whole packet's \payload from
the GPU before accomplishing the following command-packet.
%
% GPU\_P2P\_TX block is able to send multiple requests to the GPU.
%
The slow rate of read requests emitted by the \ptoptx towards the
GPU and the additional latency of the GPU response cause poor
performance.
%
% which is quite understandable as the GPU memory subsystem is
% optimized for throughput rather than for latency.
%

In the last year we modified \ptoptx, implementing the packet-command
multiple-read functionality, in order to collect a huge amount of data
from the GPU memory, managing many \apenetp packets simultaneously.
%
%in order to request from the GPU the data of many packets simultaneously (version V2).  
%
For this purpose we implemented a pipelining system, which requires
some additional features to manage flow of request to the GPU:
a \hw{FIFO DATA GPU} overflow control
(\hw{GPU TX CHECK}) and a request controller (\hw{GPU DATA
REQUESTING}).
%
% GPU DATA REQUESTING== DMA SERVICE!!
%
%As soon as \hw{FIFO GPU CMD} contains a command-packet
As in the previous version, \hw{WRITE TX FSM} sends \header and
\footer to \hw{TX FIFO HD GPU} and \hw{CMD0} to the \hw{FIFO P2P TX}.
Differently \ptoptx (version \textit{V2} ) pushes the \header also in \hw{FIFO LEN QUEUE}
and \hw{CMD1} in \hw{FIFO CMD QUEUE} (\textbf{1} in figure
\ref{fig:tx_gpu}).

The \nios exploits the information contained in \hw{FIFO P2P TX}
(\hw{CMD0}) to generate GPU data request: it fills \hw{FIFO GPU NIOS
CMD} with the necessary information to instantiate PCI transactions
and \hw{FIFO GPU NIOS DATA} with the messages for the GPU
(\textbf{2}).
% to communicate with the GPU device 

\hw{GPU DATA REQUESTING} pops data from NIOS FIFO (\textbf{3}) and
transmits the packet's size to \hw{GPU TX CHECK} (\textbf{4}); this
block prevents \hw{FIFO DATA GPU} overflow following the equation:
%
 %GPU TX CHECK %dma_srv_req_check %& %gpu_data_check %in

\begin{equation} 
\label{CHECK_fifo}
W_{REQ}-W_{WRT}+W_{NEW}<W_{FREE}
\end{equation}

in which $W_{REQ}$ is the number of words already requested to the
GPU, $W_{WRT}$ is the number of words already written in \hw{FIFO DATA
GPU}, $W_{NEW}$ is the number of word to request and $W_{FREE}$ is the
free space in \hw{FIFO DATA GPU}.

At the same time \hw{WRITE DATA FSM} pops the length of the processing
packet from the \hw{FIFO LEN QUEUE} (as explained above this FIFO
contains \header in \ptoptx) and sends it to the \hw{GPU DATA
Cnt} which monitors, by means of a data counter, the data writing
process in \hw{FIFO DATA GPU} (\textbf{4b}).

\hw{GPU TX CHECK} is in charge of limiting the number of requests for the
GPU; for this purpose it registers the number of sent requests
($R_{SENT}$) and served requests ($R_{DONE}$), and it compares their
difference with a programmable value written in a dedicated register
($R_{MAX}$).

\begin{equation} 
\label{CHECK_req}
R_{SENT}-R_{DONE}<R_{MAX}
\end{equation}

In case that inequalities \ref{CHECK_fifo} and \ref{CHECK_req} are
true (the space in \hw{FIFO DATA GPU} is enough to contain the entire
\payload of the requested packet and the number of pending request is
acceptable) \hw{GPU DATA CHECK} asserts \hw{check\_ok} signal
(\textbf{5}) and \hw{GPU DATA REQUESTING} starts the transaction
(\textbf{6}).
%
%packet's size specified in CMD0 
%
%
% GPU_DATA_CNT=GPU TARGET RAM CTRL
%
The end of writing operation is asserted by \hw{Pkt\_Sent} signal
(\textbf{7}), and \hw{WRITE DATA FSM} starts sending \hw{CMD1} to
\hw{EVENT QUEUE DMA CTRL} to program the TX completion (\textbf{8}).

As already explained, \hw{GPU DATA REQUESTING} reads the \hw{FIFO GPU
NIOS CMD} and programs fixed sized DMA write transaction (8 bytes)
popping GPU command from \hw{FIFO GPU NIOS DATA}, that contains the
physical address of the requested data.

The calculation of this physical address is a demanding task for the
\nios, that reduces its performance and slows the whole system. For
this reason we developed a new component inside \hw{GPU DATA
  REQUESTING}: the \hw{ACCELERATOR} (\ptoptx version \textit{V3}).

This block, enabled by a \nios \emph{special word} that acts as
template, produces a GPU command's sequence exploiting the information
contained in the template.

In this way \hw{GPU DATA REQUESTING} is able to send one request per
cycle to the GPU despite of many cycle required to \nios. This
improvement increases significantly the bandwidth of P2P GPU, from
$\sim$500~MB/s to $\sim$1600~MB/s as shown in section \ref{sec:bwlat}.

%  \begin{figure}[!hbt]
%   \centering
%   \includegraphics[width=\textwidth]{apenet_6link_gpu_prefetch_logx.png}
%   \caption{Bandwith with or without prefetch on GPU to GPU transfer}
%     \label{fig:gpu_prefetch}
% \end{figure}

%The last generation of GPU\_P2P\_TX (V3) is able to pre-fetch an unlimited amount of
%data so as to keep the GPU read request queue full, while
%at the same time back-reacting to almost-full conditions
%of the different on-board temporary buffers (TX FIFO DATA GPU, TX FIFO HD GPU, FIFO p2p TX).

\subsubsubsection{Software Interface Accelerator (Event Queue)}
\label{sec:eventq}
The \CQ (CQs) is one of the main component of The Remote Direct Memory
Access (RDMA) event-based protocol.

The \apenetp communicates to the CPU the completion of each performed
operation by generating an event, \ie writing in the \CQ.

\begin{figure}[!hbt]
  \centering \includegraphics[trim=30mm 45mm 30mm
  35mm,clip,width=\textwidth]{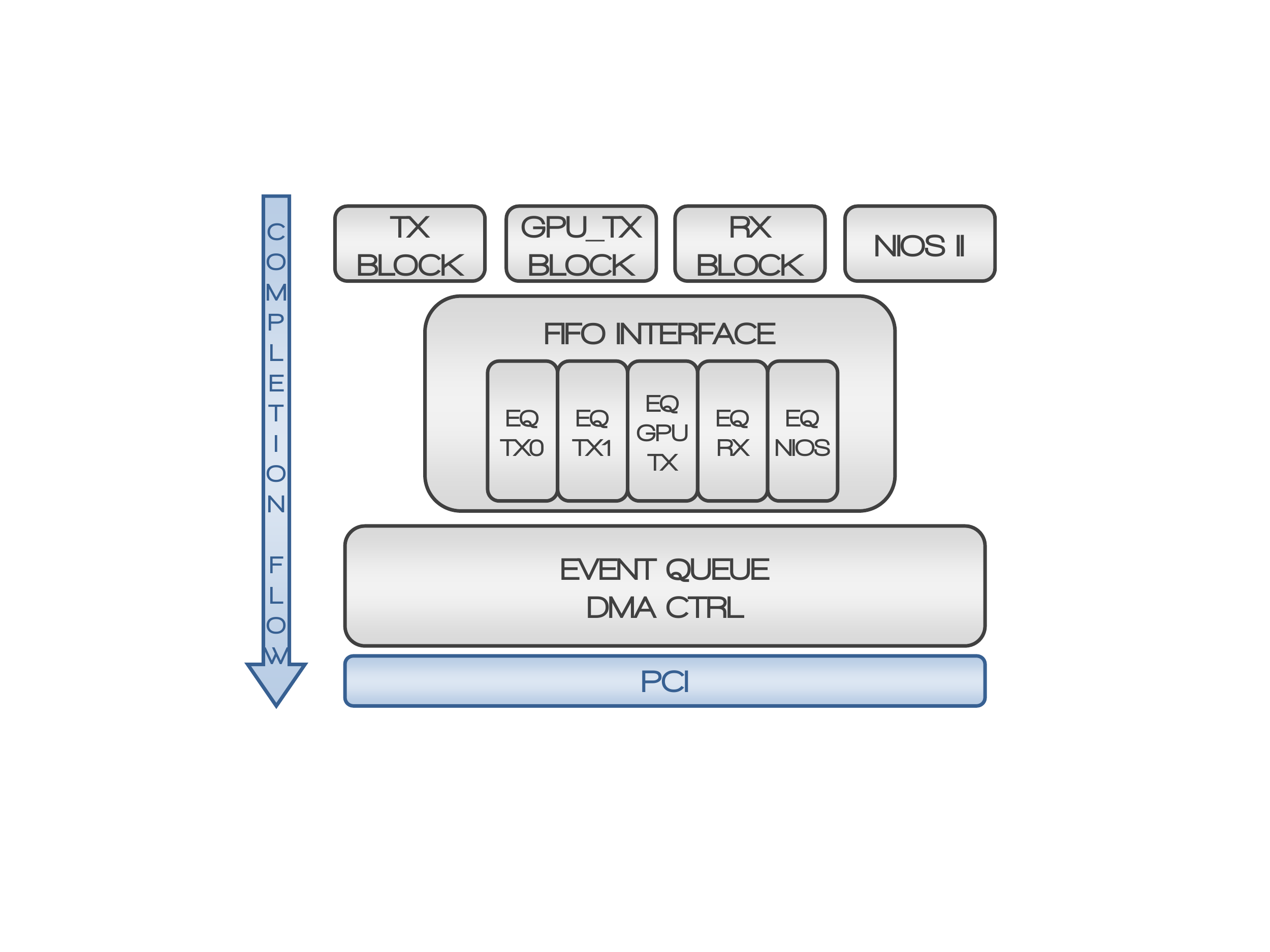}
  \caption{An overview of \hw{EVENT QUEUE DMA CTRL} and typical
  completion flow.}
  \label{fig:EQ}
\end{figure}

The \hw{EVENT QUEUE DMA CTRL} (see figure~\ref{fig:EQ}) is able to
accelerate this communication via fast PCI write transfers.

Currently we define 5 types of event:

\begin{itemize}
\item \TXaevent: \hw{TX BLOCK} completes the data read transfer from
the host memory through DMA channel 0.
\item \TXbevent: \hw{TX BLOCK} completes the data read transfer from
the host memory through DMA channel 1.
\item \GPUTXevent: \hw{GPUTX BLOCK} completes the data read transfer
from the GPU memory through DMA channel 4.
\item \RXevent: \hw{RX BLOCK} completes the data write transfer to the
host/GPU memory through the DMA channel 6.
\item \niosevent: it can be used to accelerate communication between
the CPUs in the host and the \mbox{micro-controller} in the FPGA on
board (for future development).
\end{itemize}

The completion event consist of two 128-bit words. As shown in table
~\ref{tab:eq} the \SIprotocol provides a well defined message for each
event.

\begin{table}[!htb]
\centering
\setlength\extrarowheight{2pt}
\begin{tabular}{|l|c|p{8cm}|}
\hline
\hline
\textbf{Logic Block}      & \textbf{First Completion Word} & \textbf{Second Completion Word}  \\
\hline                                                                                        
\hw{TX BLOCK DMA0}        & \hw{CMD1}                      & \hw{TX0 MAGIC WORD}          \\ 
\hw{TX BLOCK DMA1}        & \hw{CMD1}                      & \hw{TX1 MAGIC WORD}          \\            
\hw{GPU\_TX BLOCK}         & \hw{CMD1}                      & \hw{GPUTX MAGIC WORD}       \\            
\hw{RX BLOCK}             & \hw{HEADER}                    & \hw{RX MAGIC WORD}           \\            
\nios                     & \hw{NIOS CMD}                  & \hw{NIOS MAGIC WORD}         \\
\hline                    
\hline                    
\textbf{Word}             & \textbf{MSB}                   & \textbf{LSB}                \\
\hline
\hw{TX0 MAGIC WORD}       & \hw{11111111DAD0DAD0}          & \hw{11111111DAD0DAD0}       \\
\hw{TX1 MAGIC WORD}       & \hw{22222222DAD0DAD0}          & \hw{22222222DAD0DAD0}       \\
\hw{GPUTX MAGIC WORD}     & \hw{00000000DAD0DAD0}          & \hw{00000000DAD0DAD0}       \\
\hw{RX MAGIC WORD}        & \hw{PHYS. ADDRESS}             & \hw{FOOTER(63 downto 0)}    \\
\hw{NIOS MAGIC WORD}      & \hw{33333333DAD0DAD0}          & \hw{33333333DAD0DAD0}       \\
\hline
\hline
\multicolumn{3}{|c|}{\textbf{\hw{CMD1} and \hw{NIOS CMD} layout}}\\
\hline
\textbf{Field}            & \textbf{Name}                  & \textbf{Description}        \\
\hline
  1 -  0                  & pad0                           & spare bits                  \\
  3 -  2                  & tag                            & COMP\_EQ or COMP\_NONE      \\
 18 -  4                  & code                           & \nios or hardware COMP      \\
 20 - 19                  & port id                        & process id                  \\
 31 - 21                  & pad2                           & spare bits                  \\
 63 - 32                  & data                           & in case of \nios COMP contains message from the \muC \\
127 - 64                  & magic                          & TX queue entry address      \\
\hline
\hline
\end{tabular}
\caption{A detailed view of the completion events.}
\label{tab:eq}
\end{table}

Thus, once a Logic Block completes a PCI transfer, it writes the
completion words in the corresponding FIFO. The \hw{EVENT QUEUE DMA
CTRL} Finite State Machine checks the FIFO Interface
continuously. When anyone of the FIFOs contains the completion words,
the \hw{EVENT QUEUE DMA CTRL} instantiates a fixed size PCI write
transfer (32 bytes). In this way the completion process adds just few
overhead cycles to initialize the PCI transaction (about 10~cycles,
40~ns). Furthermore we reserved a DMA channel (DMA7) for \hw{EVENT
QUEUE DMA CTRL} then the completion process is totally parallelized to
the process of the corresponding Logic Block. Summarizing the typical
completion-flow is:

\begin{itemize}
\item Logic Blocks perform the entrusted tasks at the same time.
\item One of the Logic Blocks completes its task and it writes the two
words of completion in the corresponding \hw{EVENT QUEUE FIFO}.
\item The Logic Block continues performing its tasks while the
\hw{EVENT QUEUE DMA CTRL} instantiates a 32~bytes PCI write transfer
through DMA7 to communicate the event completion.
\end{itemize}

In this way no latency is introduced with a relevant performance gain.

\subsubsubsection{Logic for Sensors Interface} %% Roberto did it!
\label{sec:sensor}
\apenetp card carries a number of on-board sensors for power and
temperature monitoring. They are organized in two groups, one for on
board temperature, FPGA temperature and 12V power monitoring, the
second group for all others on board voltages (\ie 3.3V used by DDR3,
1.1V and 1.4V needed by Stratix transceivers, \dots).  The two groups
have separated serial interface directly connected to Altera's MAX2 
CPLD (which is responsible for configuration and
setting of all on board devices).

Temperature is monitored with IC Maxim MAX1619, which features an
internal temperature sensor and a connection to an external temperature
diode, which in our case is connected to Altera Stratix internal sensor.
12V voltage and current is sampled with a single IC Linear LTC4151. 
Both ICs are interfaced to the programmable configuration device 
through the same I2C line at different addresses.

Dedicated circuitry has been implemented to sample all others power
rails, using two Linears LTC2418 -- 8 lines differential analog
to digital converters suited for sensing applications --. The ICs
communicate with the programmable configuration device through a
4-wire digital protocol (SPI-like).

On the MAX2 device is implemented all the logic needed to gather and
decode data from the sensors and to configure them at start up time,
in particular I2C and SPI protocols. A number of internal 16-bit
registers are used to store the data from sensors upon request. These
registers are read and controlled from the main FPGA, through a 16-bit
wide bus, and are memory mapped in the Altera \nios embedded
microprocessor address space through the Avalon bus. Hence, a
configuration or a data read from a sensor is performed with a
specific \nios instruction in the system firmware.

%\subparagraph{Microcontroller Custom Instructions (check if last year...) - Andrea}
\subsubsection{Software Development}
\label{sec:swstack}
%%laura
For what regards 2012, the objectives of the Software Development
activities were the support to the HW innovations on the \apenetp IP
and to open the path towards the support of the \euretile complete
toolchain. A lot of effort is employed to achieve the best
performances and usability for the hardware platform. The software
layers have to match the same requisites. As a result the software
development done during 2012 consisted in the implementation in the
\apenetp Kernel Driver of the necessary support to the new hardware
blocks and features (TX accelerator, Event Queue, NIOS completion) as
well as a continuous tuning to reduce the latency due to the driver
itself, and the support to a newer version of the Linux Kernel
(v3). To make easier the exploitation of the 3D network and the
advantages that come, from the application point of view, from the use
of the \apenetp device, more standardized and user friendly APIs have
been ported and developed, including an OpenMPI first
implementation. To address the project objective of a final and
complete integration between all the \euretile HW and SW layers the
development of a preliminary version of the DNA-OS driver for \apenetp
has been started together with TIMA.

\begin{figure}[!htbp]
\centering
%\hspace{-12pt}
\includegraphics[trim=20mm 10mm 20mm 10mm,clip,width=0.75\textwidth]{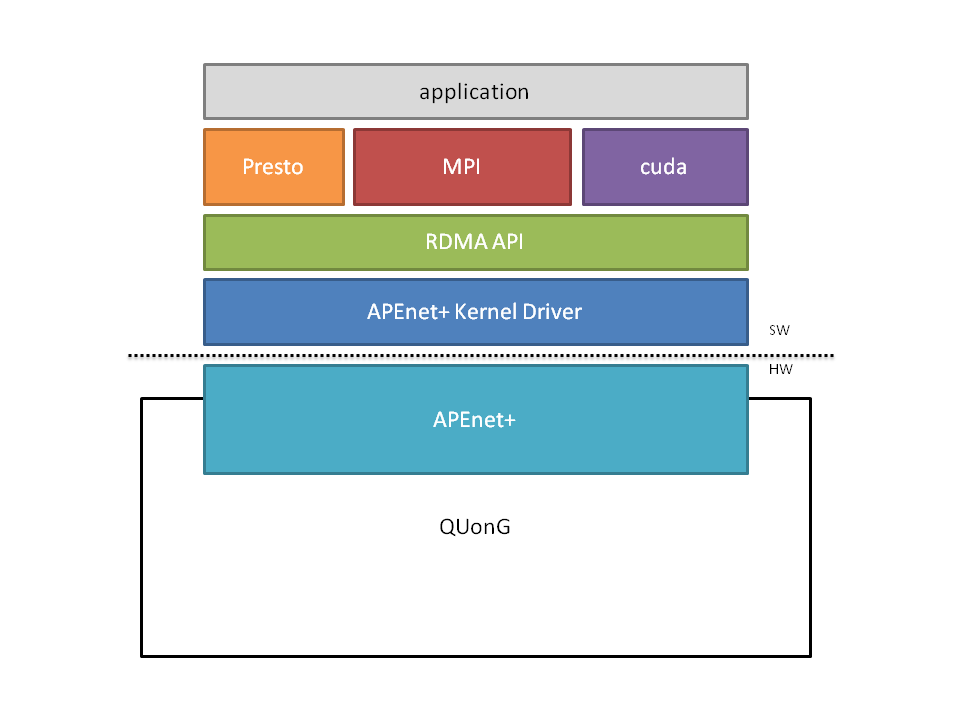}
\caption{Communication software stack for the \apenetp card in the
\quong platform.}
\label{fig:apenetsw}
\end{figure}

\begin{figure}[!htbp]
%\hspace{-12pt}
\centering
\includegraphics[trim=20mm 10mm 20mm 20mm,clip,width=0.75\textwidth]{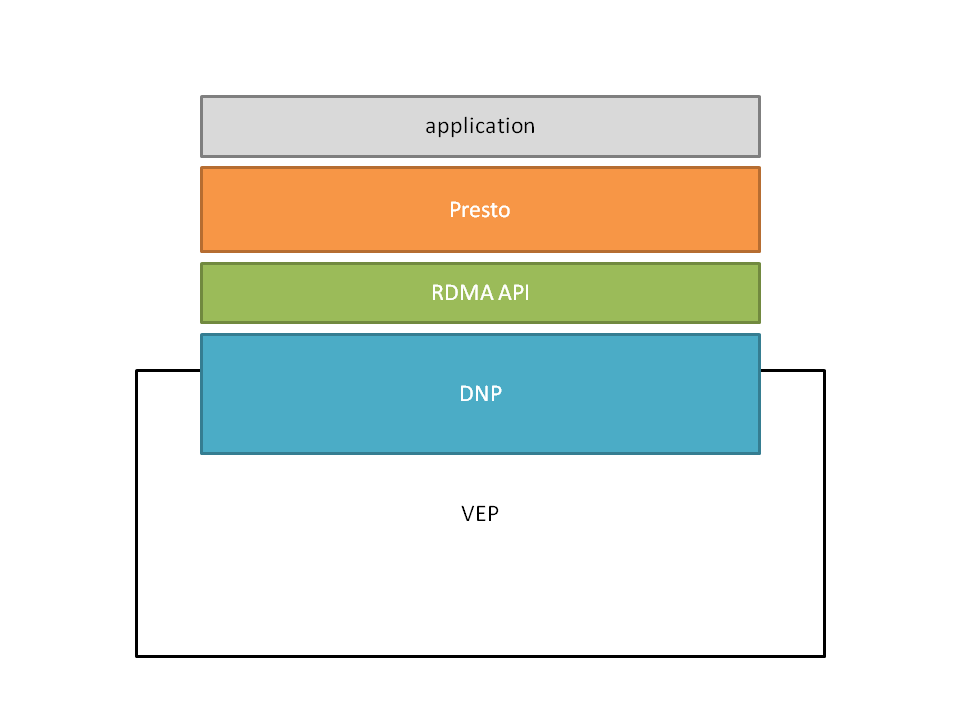}
\caption{Communication software stack for the DNP (DNP SystemC model)
in the Virtual \euretile Platform.}
\label{fig:dnpsw}
\end{figure}

The software stack for DNP/\apenetp is outlined, for what regards
communication, in figures \ref{fig:apenetsw} and \ref{fig:dnpsw}. In
the following paragraphs we describe the components of this stack
relevant to highlight the work done during 2012. The RDMA API has been 
described in D2.1~\cite{euretile:D2_1} and some more can be found in 
section \ref{sec:testsuite}.
\subsubsubsection{PCI Kernel Driver}
All the RDMA primitives included in the RDMA API are actually
implemented in a Linux kernel driver properly developed for
\apenetp. The kernel driver is a run-time loadable dynamic object
which abstracts the \apenetp card as a simple character device, a
simple abstraction of an I/O unformatted data stream. Most of the RDMA
APIs are channeled towards the kernel driver through the IOCTL system
call. The most complex APIs are those involving buffer registration,
which is mandatory for receive buffers, involves memory-pinning and
virtual-to-physical translation, and persists until an explicit buffer
deregistration. Pinning and translation are also implicitly done for
communication primitives, like RDMA PUT and GET.  As memory pinning is
a time-consuming operation, involving the manipulation (locking,
traversal, \etc) of the data structures related to the OS memory
management (MM) subsystem, we developed a Pin-Down Cache (PDC), a
caching subsystem which keeps pinned memory areas in a proper data
structure and delays the un-pinning to the closing (program tear-down)
phase.  The MM traversal related to pinning of memory buffers is also
used for the population of the virtual-to-physical (V2P) translation
tables. The physical address of buffer memory is the address of that
memory as it is seen from the peripheral bus, \PCIe. As
the concept of virtual/physical addressing scheme is also present for
the GPU of our choice, a similar technique is implemented for the GPU
as well. V2P tables are produced during buffer pinning and
communicated to the \apenetp firmware, which uses them, for example,
in the receive path to translate virtual addresses carried by packets
into physical addresses, suitable to be used on the \PCIe bus.

During 2012 a number of new features have been implemented in the
\apenetp driver:
\begin{itemize}
\item Support to the novel hardware features in the GPU TX side,
  described in section \ref{sub:TxAcceleration}.
\item Support to the NIOS completions (see \ref{sec:eventq})
\item Management of overlapping entries in the PDC; user buffers may
  overlap, the driver is able to manage both the case of intersecting
  and inclusive memory ranges by merging the entries and maintaining a
  list of collisions.
\item Support to Linux kernels version 3.X.X. %%Verificare!
\end{itemize}

\subsubsubsection{MPI for \apenetp}
%%laura
\label{sec:MPI}
MPI is a standard syntax and semantics for message-passing for
parallel computing platforms \cite{MPIstd}. The MPI primitives
integrate with commonly used programming languages in a number of
efficient implementations. OpenMPI \cite{OpenMPI} is the one we chose
to port MPI for the \apenetp interconnection card: it is open source,
it is portable and it is used by many TOP500 supercomputers
\cite{top500}, which testify its value for the HPC community and the
necessity for our project to put effort in this activity.

The MPI API basically provides a virtual topology, synchronization and
communication functionalities between \emph{processes} that are mapped
onto computing nodes instances.  The MPI library functions include
(but are not limited to) point-to-point rendezvous-type send/receive
operations, choosing of logical process topology, combining partial
results of computations (gather and reduce operations), synchronizing
nodes (barrier operation) as well as obtaining network-related
information (number of processes, process id, \etc). Point-to-point
operations come in synchronous, asynchronous, buffered, and ready
forms, to allow both relatively stronger and weaker semantics for the
synchronization aspects of a rendezvous-send.

\begin{figure}[htbp]
\centering
%\hspace{-12pt}
\centering
\includegraphics[trim=0mm 30mm 0mm 20mm,clip,width=0.9\textwidth]{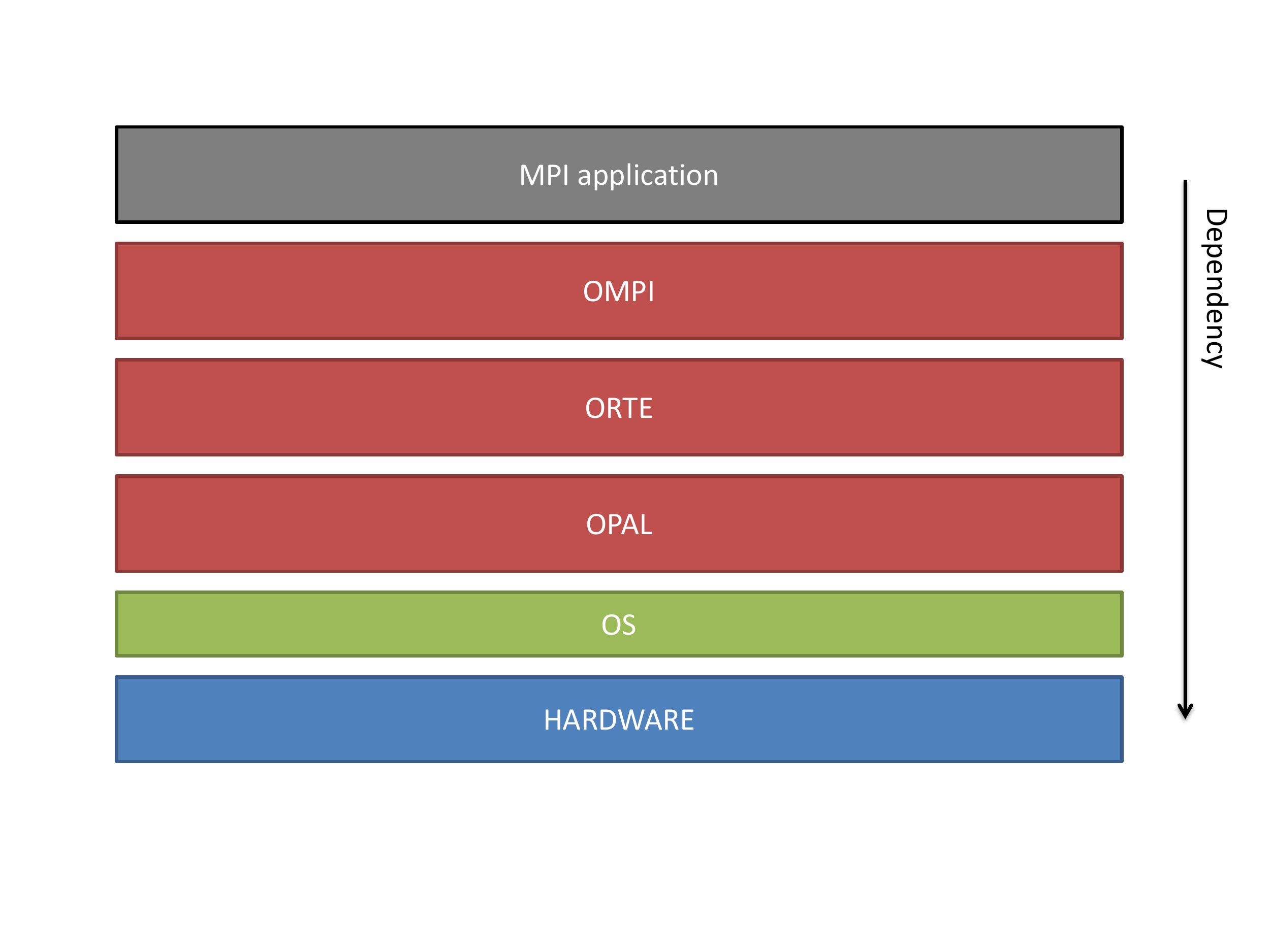}
\caption{OpenMPI layered structure. The OMPI layer contains the
  implementation of the MPI API, ORTE is the runtime environment that
  launches processes and manages resources, OPAL contains utility
  code. Each layer is implemented as a library.}
\label{fig:MPIframws}
\end{figure}

The OpenMPI library has a layered structure in terms of
dependency (figure \ref{fig:MPIframws}):
\begin{itemize}
\item \textbf{Open MPI (OMPI)} - The uppermost layer, contains the
  proper implementation of the MPI API.
\item \textbf{Open Run-time Environment (ORTE)} - The middle layer: it
  is a common interface to the runtime system, it is responsible of
  launching processes, out of bound communication, resource
  management.
\item \textbf{Open Portability Access Layer (OPAL)} - The bottom
  layer, it mostly contains utility code.
\end{itemize}

The substructure is based on the concepts of \emph{frameworks} and
\emph{components}: the \emph{components} are runtime loadable plugins,
each of them is included in a \emph{framework}. Components of the same
type are in the same framework.  This modular structure allows for
example to specify at run-time which type of communication device to
use when running an MPI program.  Each component can have one or more
instances, called \emph{modules}.

\begin{figure}
%\hspace{-12pt}
\centering
\includegraphics[trim=0mm 10mm 0mm 20mm,clip,width=0.9\textwidth]{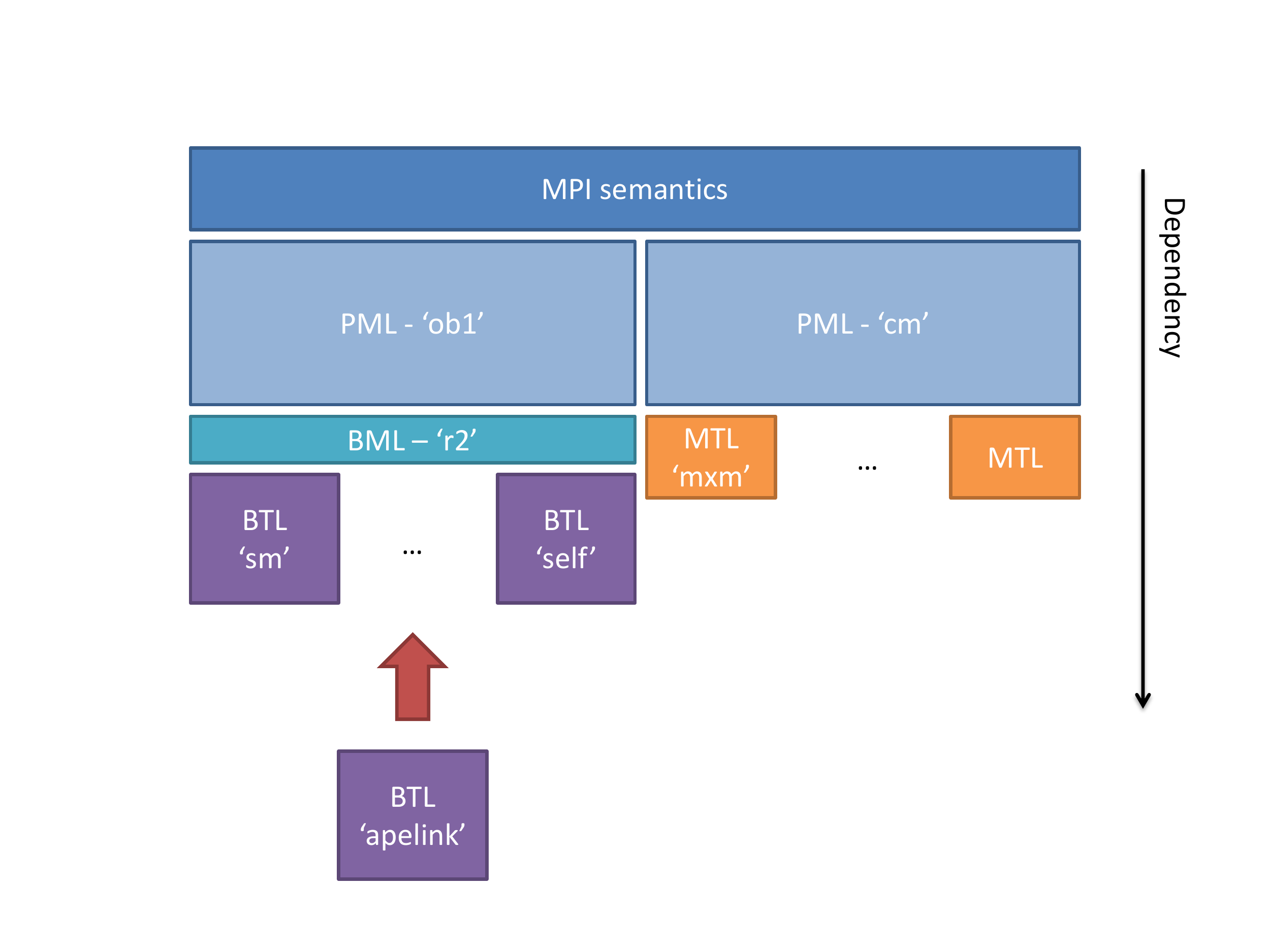}
\caption{The OpenMPI structure is based on \emph{frameworks} and
  \emph{components}. Each framework can have multiple
  components, for example the PML framework has the 'ob1' or the
  'cm' components. Each of them supports a different low level
  communication framework, BTL for 'ob1' and MTL for 'cm'. Each BTL
  component allows a different communication means or device, for
  example the BTL 'apelink' is the one written to use the \apenetp
  card.}
\label{fig:MPIstruct}
\end{figure}

Here we restrict our description to the frameworks shown in figure
\ref{fig:MPIstruct}, organized hierarchically:
\begin{itemize}
\item \textbf{PML - Point-to-point Message Layer}, it implements the
  high level MPI point-to-point semantics and manages messages
  scheduling and matching as well as the progress of the different
  communication protocols (for long or short messages). The most
  important component in the PML framework is \texttt{ob1}, which is
  specifically targeted to BTLs, while the PML \texttt{cm} supports
  MTLs.
\item \textbf{BTL - Byte Transfer Layer}, it is the framework that
  provides a uniform method for raw data transfer for numerous
  interconnects, both send/receive and RDMA based. BTLs components of
  common use are: \texttt{sm}, for intra-node communication between
  processes; \texttt{self}, for intra-process communication (a
  process that sends data to itself, useful to implement the
  collective semantics); \texttt{tcp}, for inter-node communication
  via sockets; \etc
\item \textbf{MTL - Matching Transport Layer}, it's the framework
  designed for networks that are able to implement message matching
  inside the communication library.
\item \textbf{BML - Byte Management Layer}, in this framework the
  \texttt{r2} component is responsible of opening and multiplexing
  between the BTLs.
\end{itemize}

In this context, the work of porting MPI communication library for the
\apenetp card can be seen as the implementation of a new BTL component
called \texttt{apelink} that relies on the \apenetp RDMA API (see
\ref{sec:testsuite}).


Writing a new BTL basically consists in defining a series of functions,
which are used by the upper PML component to manage communication.
For the \texttt{apelink} BTL these handles can be summarized in table
\ref{tab:btlapelink}. These are necessary for a minimal support to the
MPI API, a few more can be defined to enable advanced features, like
efficiency-tuned RDMA protocols, fault-tolerance, \etc

\begin{table}[!htb]
\centering
\setlength\extrarowheight{3pt}
\begin{tabular}{|l|p{9cm}|}
\hline
Function name & Description\\
\hline
\texttt{mca\_btl\_apelink\_add\_procs} & Discovers which processes are reachable by this module and create endpoints structures.\\
\texttt{mca\_btl\_apelink\_del\_procs} & Releases the resources held by the endpoint structures.\\
\texttt{mca\_btl\_apelink\_register} & Registers callback functions to support send/recv semantics.\\ 
\texttt{mca\_btl\_apelink\_finalize} & Releases the resources held by the module.\\
\texttt{mca\_btl\_apelink\_alloc} & Allocates a BTL specific descriptor for actual data.\\
\texttt{mca\_btl\_apelink\_free} &  Releases a BTL descriptor.\\
\texttt{mca\_btl\_apelink\_prepare\_src} & Registers user buffer or pack data into pre-registered buffer and return a descriptor that can be used for send/put.\\
\texttt{mca\_btl\_apelink\_prepare\_dst} &  Prepares a descriptor for send/rdma using the user buffer if contiguous or allocating buffer space and packing.\\
\texttt{mca\_btl\_apelink\_send} & Initiates a send.\\
\texttt{mca\_btl\_apelink\_put} & Initiates a RDMA WRITE from a local buffer to a remote buffer address.\\
\texttt{mca\_btl\_apelink\_get} & Initiates a RDMA READ from a remote buffer to a local buffer address.\\
\texttt{mca\_btl\_apelink\_register\_error\_cb} & Registers callback function for error handling.\\
\hline       
\end{tabular}
\caption{Functions defined by the BTL apelink.}
\label{tab:btlapelink}
\end{table}

A minimal and non-optimized implementation of the \texttt{apelink} BTL
is available and aligned with the OpenMPI repository trunk at July
2012.  This implementation has not been completely tested yet, but has
been proven to work with simple MPI applications, like DPSNN as
reported in section \ref{sec:DPSNN}.

The \texttt{apelink} BTL can be
selected at launch time instead of the usual ones (infiniband,
tcp,...) for example:
\begin{verbatim}
mpirun -mca btl apelink,sm,self -n PROCESSES -host HOSTLIST program_name
\end{verbatim}

Obviously the MPI frameworks add a substantial overhead
to the communication latency so a consistent optimization work on the
BTL is necessary to obtain interesting performances.
%%laura: the following is from CASS13 submitted paper
We are now working on implementing the necessary logic inside the
\apenetp OpenMPI module to support GPU-aware point-to-point primitives
implemented via the GPU \PtoP technique.  The problem with GPU-aware
MPI is the one with small message size, where the eager protocol,
which is often used in MPI implementations, cannot be used
effectively. In the eager case, the matching between the MPI Send and
the MPI Recv is handled by the receiving node and so it is arbitrarily
delayed. This means that the sender does not know whether the
destination buffer will end up on either host or GPU, and, more
importantly, the APEnet+ firmware, on processing the incoming packet
on the receiver side, does not know which kind of buffer to put the
data onto. Of course, this will be resolved as soon as the application
code on the receiving node posts a matching request, but it will be
too late to avoid memory copies; more importantly, this does not fit
with the RDMA model.  One option we would like to experiment with is
to embed additional GPU-related meta-data, necessary for the Send/Recv
matching, in the eager packet, and to delay the processing of the
packet, probably buffering it on the card, until the MPI Recv is
actually executed by the application and its information are uploaded
to the card; this would give the receiving card the opportunity to put
the data in the correct final destination. With eager intensive
traffic, this would probably be limited by on-board buffer space, and
the additional processing could easily become counter-effective.  The
easiest solution would be to disable eager for GPU buffers altogether
and to always employ rendez-vous. In this case, when a GPU buffer is
used on the send side, the sender initiates the rendez-vous protocol,
but the receiver side code has to be properly modified to handle this
case.  

\subsubsubsection{Presto}
%laura
\label{sec:presto}
\emph{Presto} is an MPI-like communication library that implements
RDMA data transfers using an \emph{eager} or \emph{rendez-vous}
protocol on top of the DNP-\apenetp RDMA API.  It's implemented in C
and it's available both for the DNP in VEP and for the \apenet+
card. Presto doesn't offer the same variety of primitives of MPI, but
because of this simplicity it is actually a thin and light-weight
software layer.  Table \ref{tab:prestoapi} shows a list of the Presto
communication primitives and the corresponding MPI primitives.

\begin{table}[!htb]
\centering
\setlength\extrarowheight{3pt}
\begin{tabular}{|p{10cm}|p{8cm}|}
\hline
\multicolumn{2}{|l|}{\emph{Basic functions}} \\
\hline
\texttt{pr\_init(pr\_mgr\_t \*\*m)} & like \texttt{MPI\_Init()}\\
\texttt{pr\_fini(pr\_mgr\_t \*m)}  & like \texttt{MPI\_Finalize()}\\
\texttt{pr\_get\_num\_procs(pr\_mgr\_t \*m, int \*n\_procs)}    & like \texttt{MPI\_Comm\_size()}\\
\texttt{pr\_get\_self\_rank(pr\_mgr\_t \*m, pr\_rank\_t \*rank)} & like \texttt{MPI\_Comm\_rank()}\\
\texttt{pr\_get\_clock(pr\_mgr\_t \*m, pr\_clock\_t \*clk)}     & fast cycle counter, in usec\\
\texttt{pr\_get\_clock\_res(pr\_mgr\_t \*m, int \*clk\_res)}    & get clock resolution\\
\texttt{pr\_get\_time(pr\_mgr\_t \*m, pr\_time\_t \*tm)}        & get wall-clock time in msec like \texttt{gettimeofday(), MPI\_Wtime()}\\
\hline
\hline
\multicolumn{2}{|l|}{\emph{\PtoP primitives}}\\
\hline
\multicolumn{2}{|l|}{\emph{blocking}}\\
\hline
\texttt{pr\_send(pr\_mgr\_t \*m, pr\_rank\_t dest, pr\_word\_t\* buf, size\_t nwords)} & blocking send, like \texttt{MPI\_Send()}\\
\texttt{pr\_recv(pr\_mgr\_t \*m, pr\_rank\_t src,  pr\_word\_t\* buf, size\_t nwords)} & non blocking receive, like \texttt{MPI\_Recv()}\\
\texttt{pr\_bcst(pr\_mgr\_t \*m, pr\_rank\_t orig, pr\_word\_t\* buf, size\_t nwords)} & broadcast, like \texttt{MPI\_Bcast()}\\
\hline
\multicolumn{2}{|l|}{\emph{non-blocking}}\\
\hline
\texttt{pr\_isend(pr\_mgr\_t \*m, pr\_rank\_t dest, pr\_word\_t\* buf, size\_t nwords, pr\_req\_t \*r)} & non-blocking send, like \texttt{MPI\_Isend()}\\
\texttt{pr\_irecv(pr\_mgr\_t \*m, pr\_rank\_t src,  pr\_word\_t\* buf, size\_t nwords, pr\_req\_t \*r)} & non-blocking receive, like \texttt{MPI\_Irecv()}\\
\texttt{pr\_wait(pr\_mgr\_t \*m, pr\_req\_t r)} & wait for a request to complete, like \texttt{MPI\_Wait()}\\
\texttt{pr\_waitall(pr\_mgr\_t \*m, pr\_req\_t \*r, size\_t n)} & wait for all the requests to complete, like \texttt{MPI\_Waitall()}\\
\texttt{pr\_test(pr\_mgr\_t \*m, pr\_req\_t r)} & test for the completion of a request, like \texttt{MPI\_Test()}\\
\texttt{pr\_req\_free(pr\_mgr\_t \*m, pr\_req\_t r)} & free a request, \texttt{MPI\_Request\_free()}\\
\hline
\hline
\multicolumn{2}{|l|}{\emph{Collective primitives}}\\
\hline
\texttt{pr\_barrier()} & like \texttt{MPI\_Barrier()}\\
\hline       
\end{tabular}
\caption{Functions defined by the Presto API.}
\label{tab:prestoapi}
\end{table}

The Presto library have been available to the EURETILE partners since
the early stages of the project, allowing the early integration and
use of the DNP SystemC model in the Virtual EURETILE Platform (see
D5.1~\cite{euretile:D5_1} and D5.2~\cite{euretile:D5_2}).

More effort is still necessary to make the Presto library optimized in
order to extract the best performances from the \apenetp card.  Even
if MPI is a more standard tool for application development, Presto can
be seen as the simplest way to reach the best performance and
usability for the \apenetp card even with legacy MPI applications,
that in principle would require just a renaming of the communication
primitives to run in a Presto+\apenetp environment. Also the
similarities in the implementation of MPI and Presto and the lack of
complexity of the latter, make our library a good testbench to
experiment a GPU-RDMA integration and find solutions to fix the
communication protocol in case of small GPU packets as mentioned in
section \ref{sec:MPI}.

\subsubsubsection{DNA-OS Driver}
DNA-OS was developed targeting
MP-SoC's and embedded systems; devices using the \PCIe are not usually
represented in such an IT ecosystem.
The target of adapting DNA-OS to run within an x86 platform and drive
a \PCIe device like \apenetp required significant work:
\begin{itemize}
\item accommodating the basic requirements of a generic \PCIe device
  needed modifications to the DNA-OS Hardware Abstraction Layer (HAL)
  and the creation of an API for a generic bus driver (described in
  detail in section 6.1.2 and 6.1.3 of D4.2~\cite{euretile:D4_2});
\item the addition of services in DNA-OS to manage the sources of
  interrupts within an x86 platform --- peripherals like timers and
  Ethernet/\apenetp cards --- and interaction with a user (described
  in detail in section 5.1 and 5.3 of D4.2~\cite{euretile:D4_2});
\item \textit{ex novo} development of a prototype driver specific to
  the \apenetp card (described in detail in section 2.6.1 of D8.1~\cite{euretile:D8_1}).
\end{itemize}

First two items are structural modifications which were mandatory for
having the chance of running the DNA-OS run on x86 platform as a
'lightweight' OS while the third item is specific to the \apenetp
card.
With these additions in place, the modified infrastructure was tested
and validated by a simple program produced by the DNA-OS toolchain and
running on x86 that queries and sets some board internal registers
through the PCIe Base Address Register 0 (BAR0).
This program --- more complete details on its operation are available
in section 6.1.5 of D4.2~\cite{euretile:D4_2}  --- is the first stage of what will become the complete
\apenetp driver for DNA-OS.

\subsubsection{Test, Debugging \& Measures}
\subsubsubsection{Setup of Test \& Debug (T\&D) environment}
\label{sec:IBinstall}
The environment where the \apenetp card is being developed --- from
now on the T(est) \& D(ebug) platform --- which is actually also a
prototype for the \euretile HPC platform, the \quong cluster, consists
of a mostly homogeneous x86 cluster where the nodes are 1U units
equipped with dual Xeon CPUs.
More precisely, the T\&D is composed of (not mentioning off-the-shelf
components like the Gbit Ethernet switch, ECC DRAM banks, \etc):
\begin{itemize}
\item 10 SuperMicro X8DTG-D dual Xeon SuperServers;
\item 9 Mellanox ConnectX VPI MT26428 InfiniBand Host Channel Adapters (HCAs)
\item 1 SuperMicro X8DAH+ dual Xeon SuperWorkstation;
\item 1 SuperMicro X8DTG-QF dual Xeon SuperWorkstation;
\item 1 Mellanox MTS3600/U1 QDR 36-port switch;
\item 13 \nvidia GPUs systems:
\begin{itemize}
\item 2 Tesla arch. discrete cards (2xC1060);
\item 7 Fermi arch. discrete cards (2xC2050, 3xM2050, 2xM2070);
\item 4 Fermi arch. cards housed into 1 S2070 multi-GPU rack.
\end{itemize}
\end{itemize}

The diversity of the nodes is partly due to different procurement
epochs but also to cater to different needs; \eg, the two
SuperWorkstation units are 4U tower systems which, although more bulky
than the 1U SuperServer ones, allow for easy 'open-heart' inspection
of the \apenetp board when it is attached to a protocol analyzer ---
\eg the Teledyne LeCroy Summit T2-16 --- to peruse the low-level
workings of the \PCIe bus.
The S2070 multi-GPU rack of the T\&D was the first assembly of the
elementary building block which is the keystone for the \quong
cluster, made of two 1U SuperServers that 'sandwich' the 1U S2070
rack; the result is a unit with the optimal ratios of 4 GPUs vs. 4
CPUs and 4 GPUs vs. 2 \apenetp cards compacted into a 3U volume.

The CPUs are quad-core (X5570/E5620) and hexa-core (X5650/X5660)
variants of Sandy-Bridge dual-core Xeons with frequencies in the
2.40\textdiv2.93 GHz range and ECC RAM from 12 to 48 GB per node.
Differently populated topologies can be tested on the T\&D rearranging
the connections of the \apenetp cards: closed 1-dim loops, 2-dim and
3-dim torus meshes (4x2 is actually in test), \etc
The InfiniBand HCAs and switch are necessary for comparison purposes
to a network fabric fat-tree topology having the high bandwidth and
low latency currently considered state-of-the-art for an HPC cluster.

A worth mentioning feature of the SuperMicro SuperServers is the
compliance with the Intelligent Platform Management Interface (IPMI)
for out-of-band management.
This means that every T\&D node has onboard a subsystem (called a
Baseboard Management Controller or BMC) which operates independently
of the host OS and lets an administrator perform a number of tasks
that would usually require physical proximity with the device --- as
accessing BIOS settings, temperature/voltage probing or even
power-cycling for the unit --- through an out-of-band connection, in
this case conveyed over an auxiliary Ethernet plug.
The value of IPMI is twofold:
\begin{itemize}
\item the debug phase of low level, potentially system-disrupting
  software is made considerably easier --- a machine locked hard by an
  unstable, under development kernel device driver can be power-cycled
  without compelling the programmer to physically reach the cluster
  and flip the switch;
\item it provides essentially for free a number of machine stats ---
  like voltages, fan speeds or system temperatures --- that can be
  logged in real-time and eventually used to assess the health status
  of the machine.
\end{itemize}

The operating system is a stock CentOS 5.X GNU/Linux, with standard
2.6.18 kernel line; we are currently evaluating migration to the newer
CentOS 6.X with 2.6.32 kernel line, which is to be the target OS for
the \quong cluster.
The software stack for the GPUs --- driver and development kit --- is
standard \nvidia CUDA in both versions 4.2 and 5.0.

\subsubsubsection{Software Testsuite (Tools, Synthetic Benchmarks)}
\label{sec:testsuite}
The development of the \apenetp card required writing a great deal of
software; besides the fundamental middleware needed to implement a
working API for communication with the board --- see \ref{sec:swstack}
for more details ---, a large number of small test codes (mostly in C
supplemented by scripts in Bash/TCSH) were written.
They are necessary for automation of the procedure that 'kickstarts'
the board to a usable state at system startup and, more importantly,
to stimulate in the most punctual way all the device's subsystems, in
order either to timely verify that each modification to the firmware
was not causing regressions or to help in debugging them when they
appeared.
Moreover, especially when the APIs are under development so that
resorting to extensive rewrites of a more complete application is
unfeasible, these synthetic tests are invaluable to immediately gauge
the gains or losses in performance for every optimization that was
added to the board design along the course of its evolution.

\begin{figure}[!hbt]
  \centering
  %\hspace{-12pt}
  \centering
  \includegraphics[width=\textwidth]{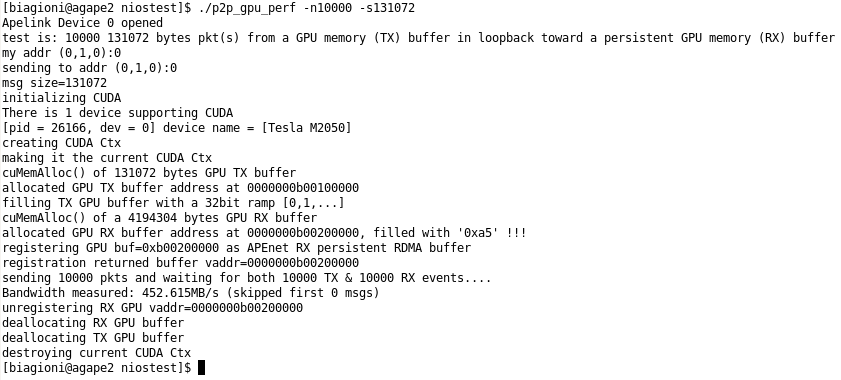}
  \caption{Host console output with \textit{p2p\_gpu\_perf} results.}
  \label{fig:p2p_gpu_perf_host}
\end{figure}

An exhaustive list of these tests is not very informational and a
recap of the relevant data --- aggregated bandwidth and latency ---
are given in \ref{sec:bwlat} together with a comparison to results
gathered by the equivalent OSU benchmarks~\cite{Traff:2012:OMB-GPU}
over the InfiniBand fabric of the T\&D platform described in
\ref{sec:IBinstall}.
We just show here an example of the data gauged by one of these tests,
\emph{p2p\_gpu\_perf}, that measures the bandwidth of a loopback
GPU-to-GPU data transfer; it exercises one of the most complex path
inside the \apenetp.
In figure \ref{fig:p2p_gpu_perf_host} we have the aggregated data of
the test as seen by the host together with some debugging output.
In this test, data packets in transit through the RX path of the
transfer are processed in a number of steps performed by the \nios
processor.

\begin{figure}[!hbt]
  \centering
  %\hspace{-12pt}
  \centering
  \includegraphics[width=0.45\textwidth]{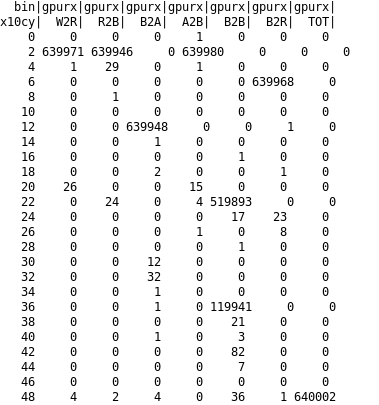}
  \caption{\nios console output with snapshot of profiling histogram.}
  \label{fig:p2p_gpu_perf_NIOS}
\end{figure}

In figure \ref{fig:p2p_gpu_perf_NIOS} we see a binning histogram with
a run-down of these steps --- \eg \emph{B2A} is the time taken by
looking up a buffer in a virtual-to-physical address conversion plus
the initialization of the memory write, while \emph{B2B} is the time
taken by the actual virtual-to-physical memory address translation ---
and their duration in cycles, averaged over a large number of runs and
output directly by the \nios.

This highly granular profiling allowed us to focus the activity on the
areas that benefited mostly of tight optimization or were worth the
extra effort of an hardware implementation.

\subsubsubsection{Bandwidth \& Latency Measures}
\label{sec:bwlat}
The \apenetp benchmarks were performed on the T\&D platform ---
hardware and software details are in section \ref{sec:IBinstall}.
The \apenetp cards used were preliminary with a reduced link speed of
28~Gbps.
%
% sigle node test
%

\begin{figure}[!htb]
\centering
\includegraphics[trim=0mm 20mm 0mm 20mm,clip,width=.75\textwidth]{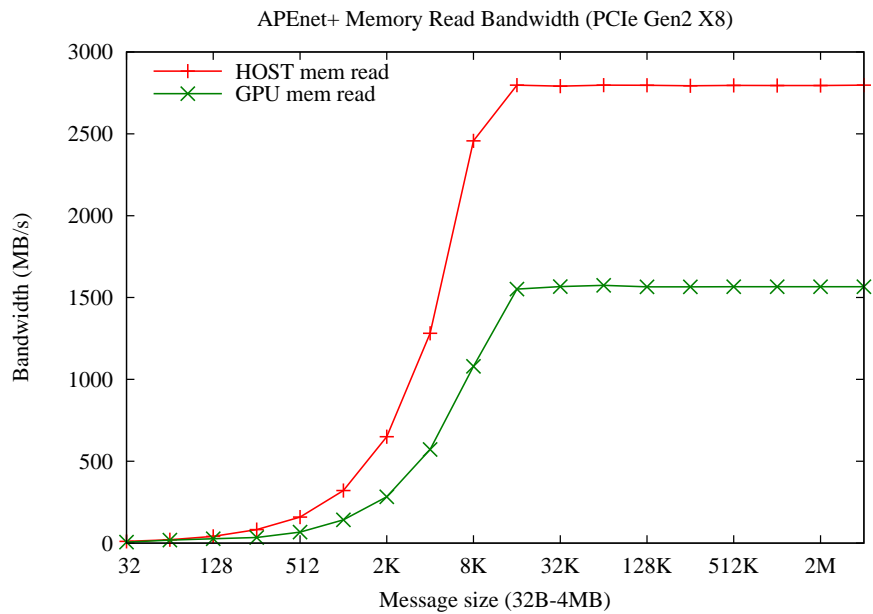}
\caption{A comparison between HOST mem read and GPU mem read. The
plots are obtained in a loop-back test flushing the TX injection
FIFOs.}
\label{fig:apenet_tx_bw}
\end{figure}

\begin{figure}[t]

 \begin{minipage}[t]{0.48\textwidth}
   \centering
   \includegraphics[trim=15mm 20mm 15mm 20mm,clip,width=\textwidth]{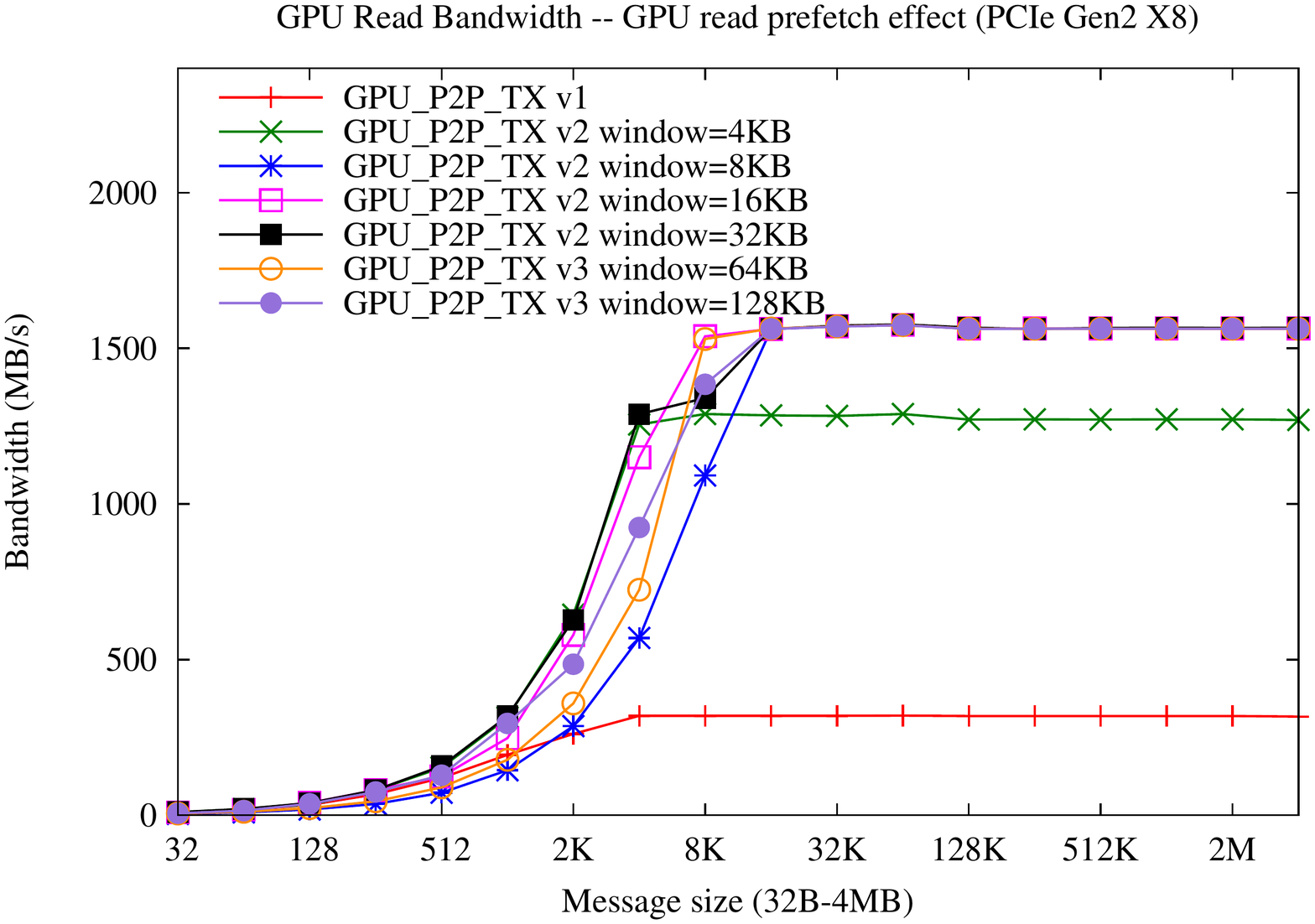}
   \caption{Single-node GPU memory reading bandwidth, showing the
  performance at varying message size, obtained by flushing TX
  injection FIFOs. Different curves corresponds to the three \ptoptx
  implementations and to different pre-fetch window sizes, where
  appropriate. Figures are highly unstable for small message sizes,
  mainly due to software related issues under queue-full conditions.}
 \label{fig:gpu_tx_prefetch_bw}  
 \end{minipage}
 \hspace{3mm}
 \begin{minipage}[t]{0.48\textwidth}
 \centering
 \includegraphics[trim=15mm 20mm 15mm 20mm,clip,width=\textwidth]{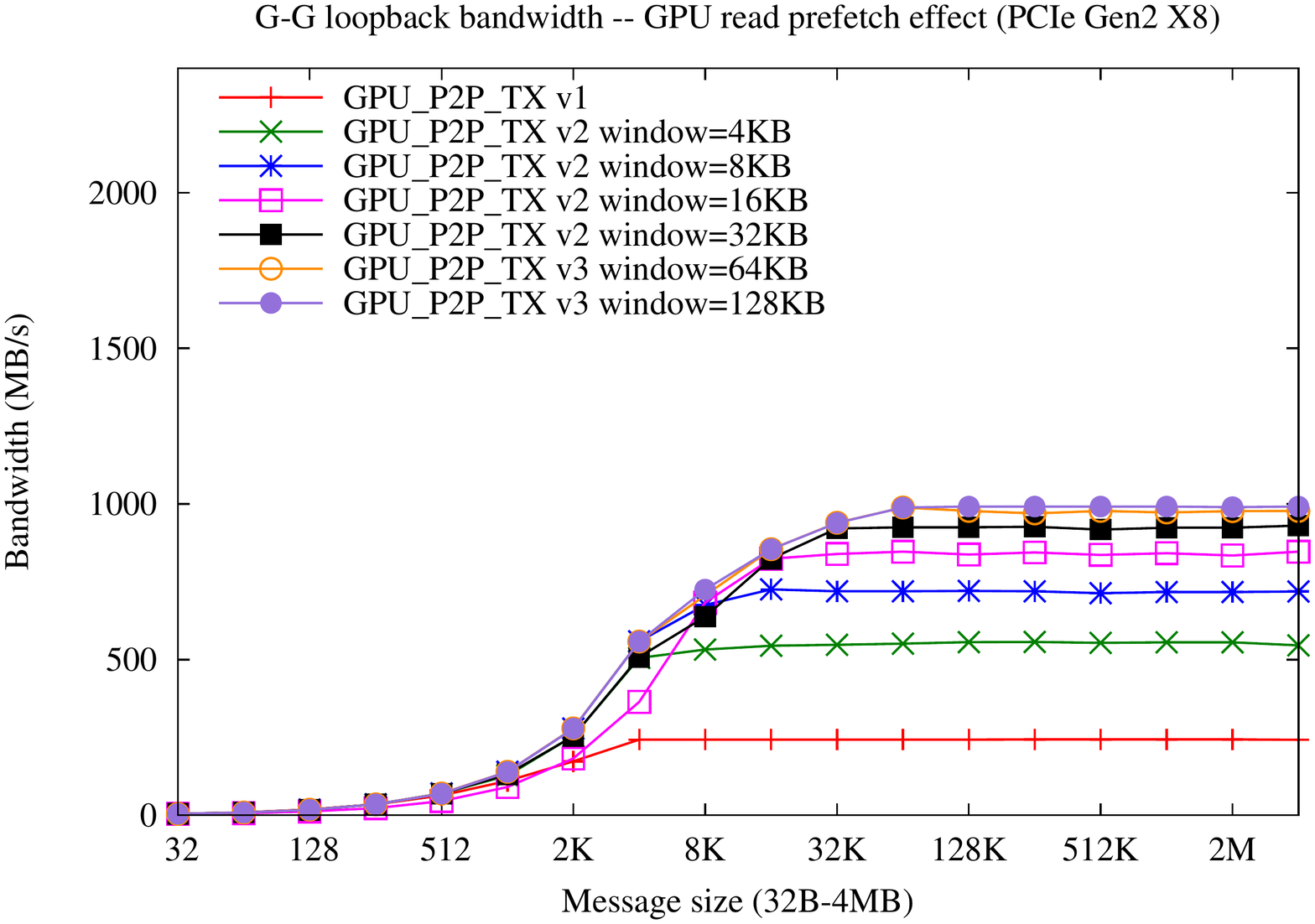}
 \caption{Single-node GPU memory loop-back bandwidth, at varying
  pre-fetch threshold size. Different curves are as in the the
  previous plot. The full loop-back send-and-receive bandwidth is
  plotted and it is limited by the \nios \mbox{micro-controller}
  processing capabilities. }
 \label{fig:gpu_prefetch_bw}
 \end{minipage}
\end{figure}

To give an idea of the performance and limitations of the current
implementation, in table~\ref{tab:lowlevel} we collected the memory
read performance, as measured by the \apenetp device, for buffers
located on either host or GPU memory and figure~\ref{fig:apenet_tx_bw}
shows a comparison between HOST and GPU memory read bandwidth at
varying of the message size.
\begin{small}
\begin{table}[htbp]
\centering
\setlength\extrarowheight{2pt}
\begin{tabular}{|l|l|l|l|}
\hline
\textbf{Test}  & \textbf{Bandwidth} & \textbf{GPU/method} & \textbf{\nios active tasks}\\
\hline
Host mem read  & 2.8~GB/s &             & none \\
GPU mem read   & 1.5~GB/s & Fermi/P2P   & \ptoptx \\
%GPU mem read   & 150MB/s & Fermi/BAR1  & \ptoptx \\
GPU mem read   & 1.6~GB/s & Kepler/P2P  & \ptoptx \\
%GPU mem read   & 1.6GB/s & Kepler/BAR1 & \ptoptx \\
\hline
GPU-to-GPU loop-back   & 1.1~GB/s  & Fermi/P2P  & \ptoptx + RX\\
\hline
Host-to-Host loop-back & 1.2~GB/s  &            & RX \\
\hline
%\hline
\end{tabular}
\caption{\apenetp \mbox{low-level} bandwidths, as measured with a
  \mbox{single-board} loop-back test. The memory read figures have
  been obtained by flushing the packets while traversing \apenetp
  internal switch logic.  Kepler results are for a pre-release K20
  (GK110), with ECC enabled. Fermi results are with ECC off.  GPU and
  \apenetp linked by a PLX \PCIe switch.}
\label{tab:lowlevel}
\end{table}
\end{small}
The complexity of the GPU \PtoP read protocol and the limitations of
our implementation set a limit of 1.5~GB/s to the Fermi GPU memory
read bandwidth, which is roughly half that obtained for host memory
read (2.8~GB/s).
For reference, the GPU-to-host reading bandwidth, as obtained by
\texttt{cudaMemcpy}, which uses the GPU DMA engines, peaks at about
5.5~GB/s on the same platform (note \nvidia GPU is an x16 I/O slot wide
PCI Express card, while \apenetp is only x8 wide).
We underline that this is the reading bandwidth as measured from
\apenetp through the GPU \PtoP protocol, neither the internal device
bandwidth, which is instead available to kernels running on the GPU,
nor the GPU DMA engine bandwidth, \eg \texttt{cudaMemcpy()}.
The last two lines of table~\ref{tab:lowlevel} and
figure~\ref{fig:apenet_tx_bw} show that, when the packet RX processing
is taken into account by doing a loop-back test, the peak bandwidth
decreases from 2.8~GB/s to 1.2~GB/s in the host-to-host case, and from
1.5~GB/s to 1.1~GB/s in the GPU-to-GPU case, \ie an additional 10\%
price to pay in the latter case.
The last column in the table shows that the \nios
\mbox{micro-controller} is the main performance bottleneck.
We are currently working on adding more hardware blocks to accelerate
the RX task.
The values reported in table~\ref{tab:lowlevel} are obtained as the
peak values in a loop-back performance test, coded against the
\apenetp RDMA API.
The test allocates a single receive buffer (host or GPU), then it
enters a tight loop, enqueuing as many RDMA PUT as possible to keep
the transmission queue constantly full.
Figure~\ref{fig:gpu_tx_prefetch_bw} is a plot of GPU reading bandwidth
at varying message size, estimated by using the test above and by
flushing TX injection FIFOs, effectively simulating a zero-latency
infinitely fast switch.
The original \textit{V1} \ptoptx implementation without pre-fetching (for more
details see \ref{sub:TxAcceleration}) shows its limits.
\ptoptx \textit{V2} (HW acceleration of read requests and limited pre-fetching)
shows a 20\% improvement while increasing the pre-fetch window size
from 4KB to 8KB.
Unlimited pre-fetching and more sophisticated flow-control in \ptoptx
\textit{V3} partially shows its potential only in the full loop-back plot of
figure~\ref{fig:gpu_prefetch_bw}.
Here the \nios handles both the \ptoptx and the RX tasks, therefore
any processing time spared thanks to a more sophisticated GPU TX
flow-control logic reflects to an higher bandwidth.
\begin{figure}[t]
 \begin{minipage}[t]{0.48\textwidth}
   \centering
  \includegraphics[trim=15mm 20mm 15mm 20mm,clip,width=\textwidth]{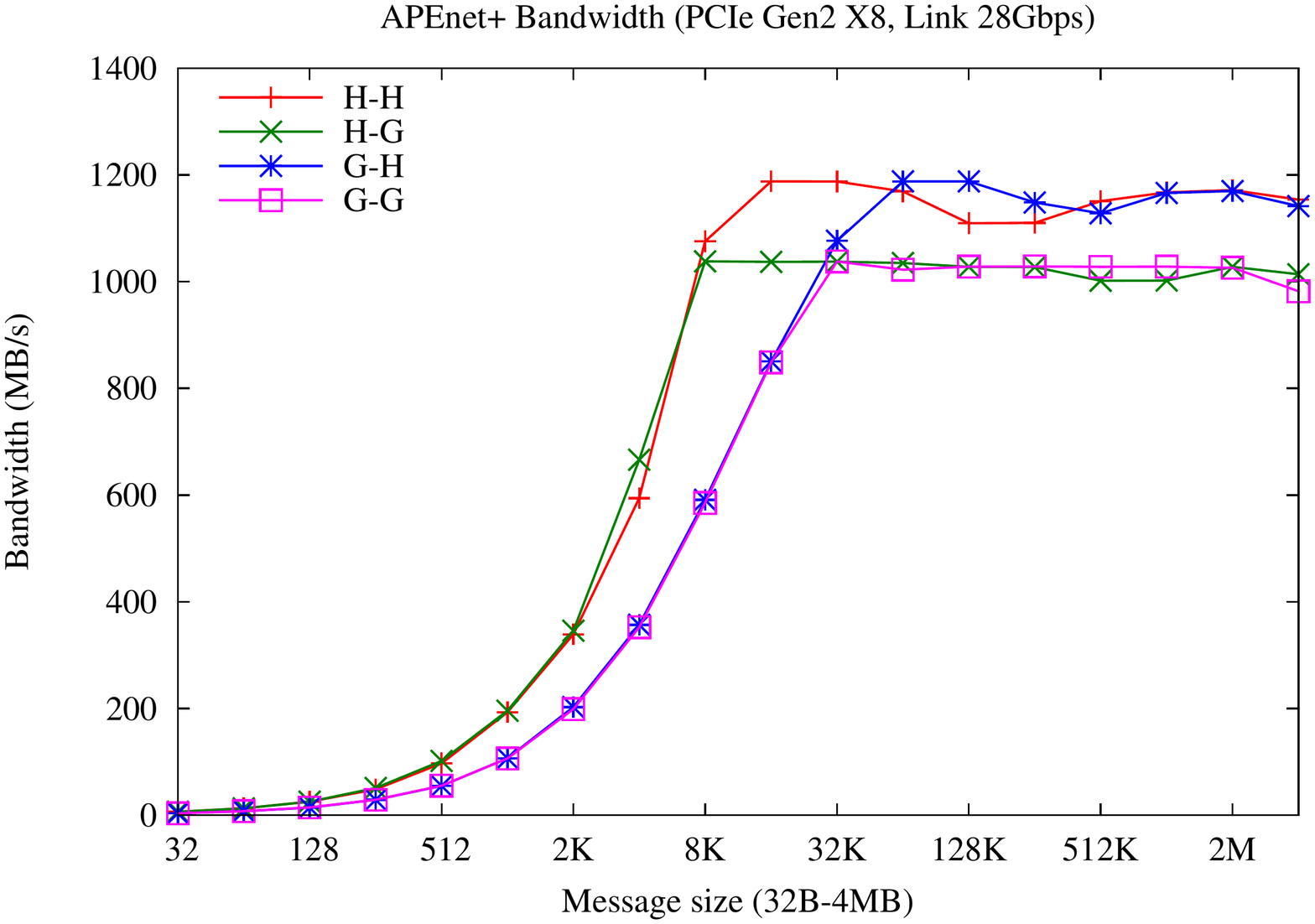}
  \caption{\mbox{Two-nodes} \mbox{uni-directional} bandwidth test, for
  different combinations of both the source and the destination buffer
  types. When source is in GPU memory, the overhead is visible; at
  8KB, the bandwidth is almost half that in the host memory case. The
  bandwidth cap is related to the limited processing capabilities of
  the \nios \mbox{micro-controller}. }
  \label{fig:apenet_bw} 
 \end{minipage}
 \hspace{3mm}
 \begin{minipage}[t]{0.48\textwidth}
 \centering
 \includegraphics[trim=15mm 20mm 15mm 20mm,clip,width=\textwidth]{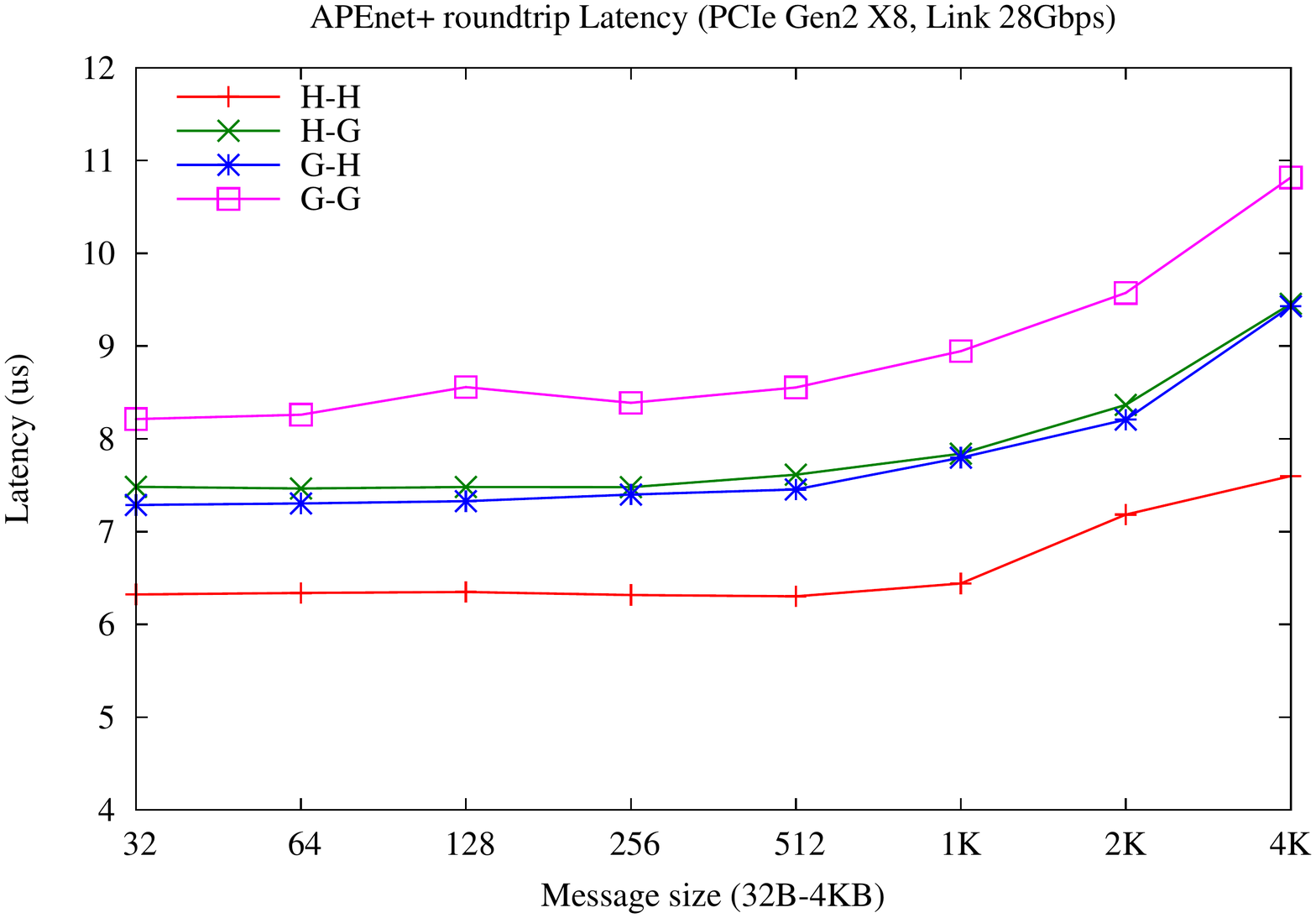}
 \caption{\apenetp latency, estimated as half the round-trip
  latency. Different combinations of both the source and the
  destination buffer types.}
 \label{fig:apenet_lat}
 \end{minipage}
\end{figure}

As shown above, reading bandwidth from GPU memory and RX processing
are the two keys limiting factors of the current \apenetp
implementation.
Therefore, it can be expected that they influence the communication
bandwidth between two nodes in different ways, depending of the type
of the buffers used.

To measure the effect of those factors independently, we run a two
node bandwidth test on \apenetp, in principle similar to the MPI
OSU~\cite{Traff:2012:OMB-GPU} uni-directional bandwidth test, although
this one is coded in terms of the \apenet RDMA APIs.
The benchmark is basically a one-way point-to-point test involving two
nodes. The receiver node allocates a buffer, on either host or GPU
memory, registers it for RDMA, sends its address to the transmitter
node, starts a loop waiting for N buffer received events and ends by
sending back an acknowledgment (ACK) packet.  The transmitter node
waits for an initialization packet containing the receiver node buffer
(virtual) memory address, writes that buffer N times in a loop with
RDMA PUT, then waits for a final ACK packet.

The plot in figure~\ref{fig:apenet_bw} shows the bandwidth of \apenetp
for the four different possible combinations of source and destination
buffer types:
for source buffers located in host memory, the best performance of
1.2~GB/s is reached, with a 10\% penalty paid when receive buffers are
on the GPU, probably related to the additional actions involved, \ie
switching GPU \PtoP window before writing to it.
For GPU source buffers, the GPU \PtoP reading bandwidth is the
limiting factor, so the curves are less steep and only for larger
buffer sizes, \ie beyond 32KB, the plateau is reached.
Clearly, the asymptotic bandwidth is limited by the RX processing, but
the overall performance is affected by the transmission of GPU
buffers.
Interestingly, the Host-to-GPU performance seems to be a very good
compromise bandwidth-wise, \eg for 8KB message size the bandwidth is
twice that of the GPU-to-GPU case.
Of course this plot is good for analyzing the quality of the \apenetp
implementation, but it says nothing about which method is the best for
exchanging data between GPU buffers, \ie in which ranges GPU \PtoP is
better than staging on host memory.

\begin{figure}[t]
 \begin{minipage}[t]{0.48\textwidth}
   \centering
   \includegraphics[trim=15mm 20mm 15mm 20mm,clip,width=\textwidth]{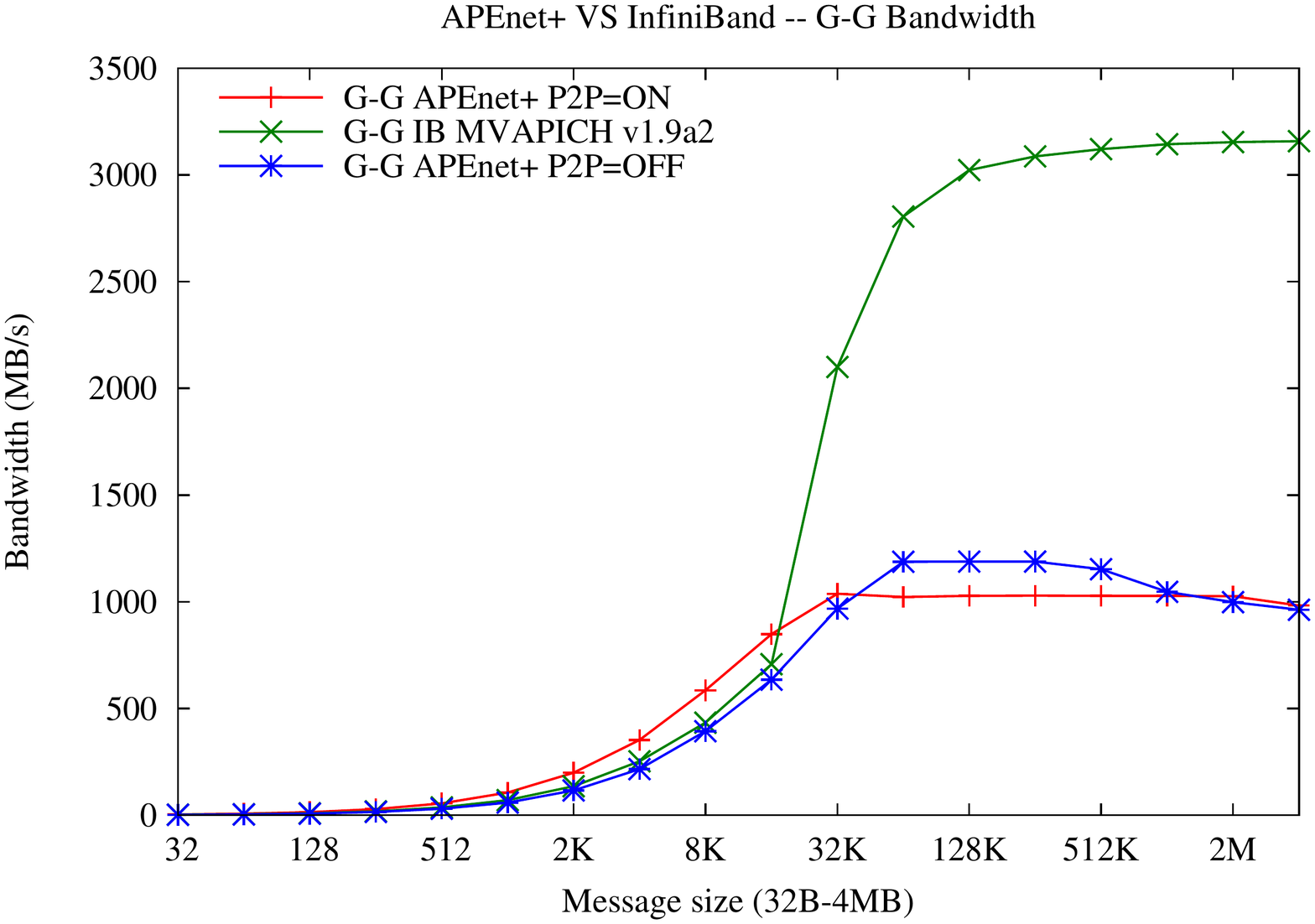}
   \caption{\mbox{Two-nodes} \mbox{uni-directional} bandwidth test,
  GPU-to-GPU. P2P=OFF case corresponds to the use of staging in host
  memory. MVAPICH2 result on OSU MPI bandwidth test is for reference.}
   \label{fig:apenet_vs_ib_bw}
 \end{minipage}
 \ \hspace{2mm} \hspace{3mm} \
 \begin{minipage}[t]{0.48\textwidth}
 \centering
\includegraphics[trim=15mm 20mm 15mm 20mm,clip,width=\textwidth]{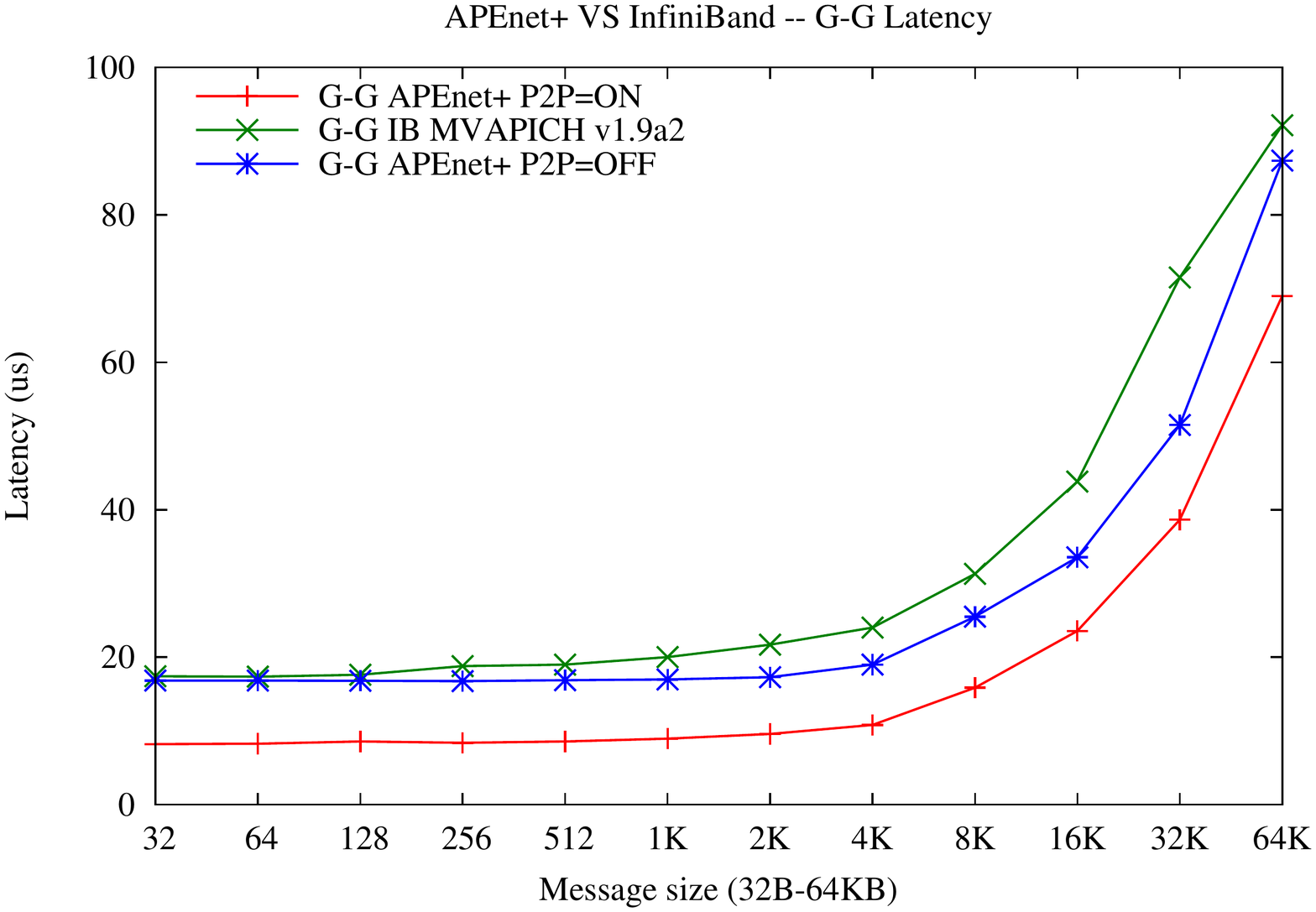}
\caption{\apenetp latency, GPU-to-GPU case. \PtoP has 50\% less
  latency than staging. The MVAPICH2 plot is the GPU OSU latency test
  on Infiniband.}
\label{fig:apenet_vs_ib_lat}
 \end{minipage}
\end{figure}

To this end, figure~\ref{fig:apenet_vs_ib_bw} is a plot of the
GPU-to-GPU communication bandwidth, with three different methods:
\apenetp using GPU \PtoP; \apenetp with staging of GPU data to host
memory; OSU bandwidth test, using MVAPICH2~\cite{MVAPICH2} over
Infiniband, which uses a pipelining protocol above a certain
threshold, used for reference.
The GPU \PtoP technique is definitively effective for small buffer
sizes, \ie up to 32KB; after that limit, staging seems a better
approach.
\begin{figure}[!hbt]
  \centering \includegraphics[trim=0mm 20mm 0mm
20mm,clip,width=.75\textwidth]{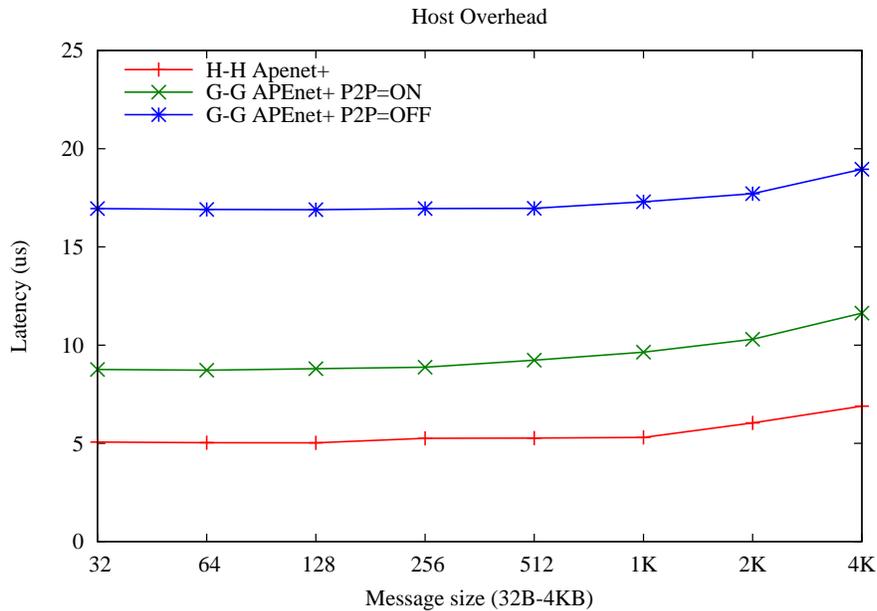}
\caption{\apenetp \textit{host overhead}, estimated via bandwidth
test.}
\label{fig:host_overhead}
\end{figure}

Figure~\ref{fig:apenet_vs_ib_lat} is more useful to explore the
behaviour of GPU \PtoP on small buffer size.
Here the latency, estimated as half the round-trip time in a ping-pong
test, shows a clear advantage of the \PtoP implementation with respect
to staging (P2P=OFF in the figure), even on a very low-latency network
as Infiniband.
Indeed, the \apenetp \PtoP latency is 8.2\us, while for \apenetp with
staging and MVAPICH2/IB it is respectively 16.8\us and 17.4\us.
In the latter case, most of the additional latency comes from the
overhead of the two CUDA memory copy (\texttt{cudaMemcpy}) calls
necessary to move GPU data between temporary transmission buffers.
By subtracting the \apenetp H-H latency (6.3\us in
figure~\ref{fig:apenet_lat}) from the \apenetp latency with staging
(16.8\us), the single \texttt{cudaMemcpy} overhead can be estimated
around 10\us, which was confirmed by doing simple CUDA tests on the
same hosts.

%---------------------------------------------------------------
% host overhead
The run times of the bandwidth test, for short message size, are plot
in figure~\ref{fig:host_overhead}.
In the LogP model~\cite{Culler:1993:LogP}, this is the host overhead,
\ie the fraction of the whole message send-to-receive time which does
not overlap with subsequent transmissions.
Of those 5\us in the Host-to-Host case, at least a fraction can be
accounted to the RX processing time (3\us estimated by cycle counters
on the \nios firmware).
%
%%The additional 3\us in the GPU-to-GPU (P2P=ON) case should be quite
%%related to \PtoP protocol as implemented by \apenetp, \eg the
%%3+1.8 \us \ptoptx overhead in figure~\ref{fig:pcie_timings}.
% and the are at least in part responsible for those 8\us.
%
When staging is used instead (P2P=OFF), out of the additional 12\us
(17\us - 5\us of the host-to-host case), at least 10\us are due to the
\texttt{cudaMemcpy} device-to-host, which is fully synchronous with
respect to the host, therefore it does not overlap.

In conclusion, the GPU \PtoP, as implemented in \apenetp, shows a
bandwidth advantage for message sizes up to 32KB.
Beyond that threshold, at least on \apenetp it is convenient to give
up on \PtoP by switching to the staging approach.
Eventually that could have been expected, as architecturally GPU \PtoP
cannot provide any additional bandwidth, which is really constrained
by the underlying \PCIe link widths (X8 Gen2 for both \apenetp and
Infiniband) and bus topology.

\begin{figure}[t]
 \begin{minipage}[b]{0.48\textwidth}
   \centering
   \includegraphics[trim=15mm 20mm 10mm 15mm,clip,width=\textwidth]{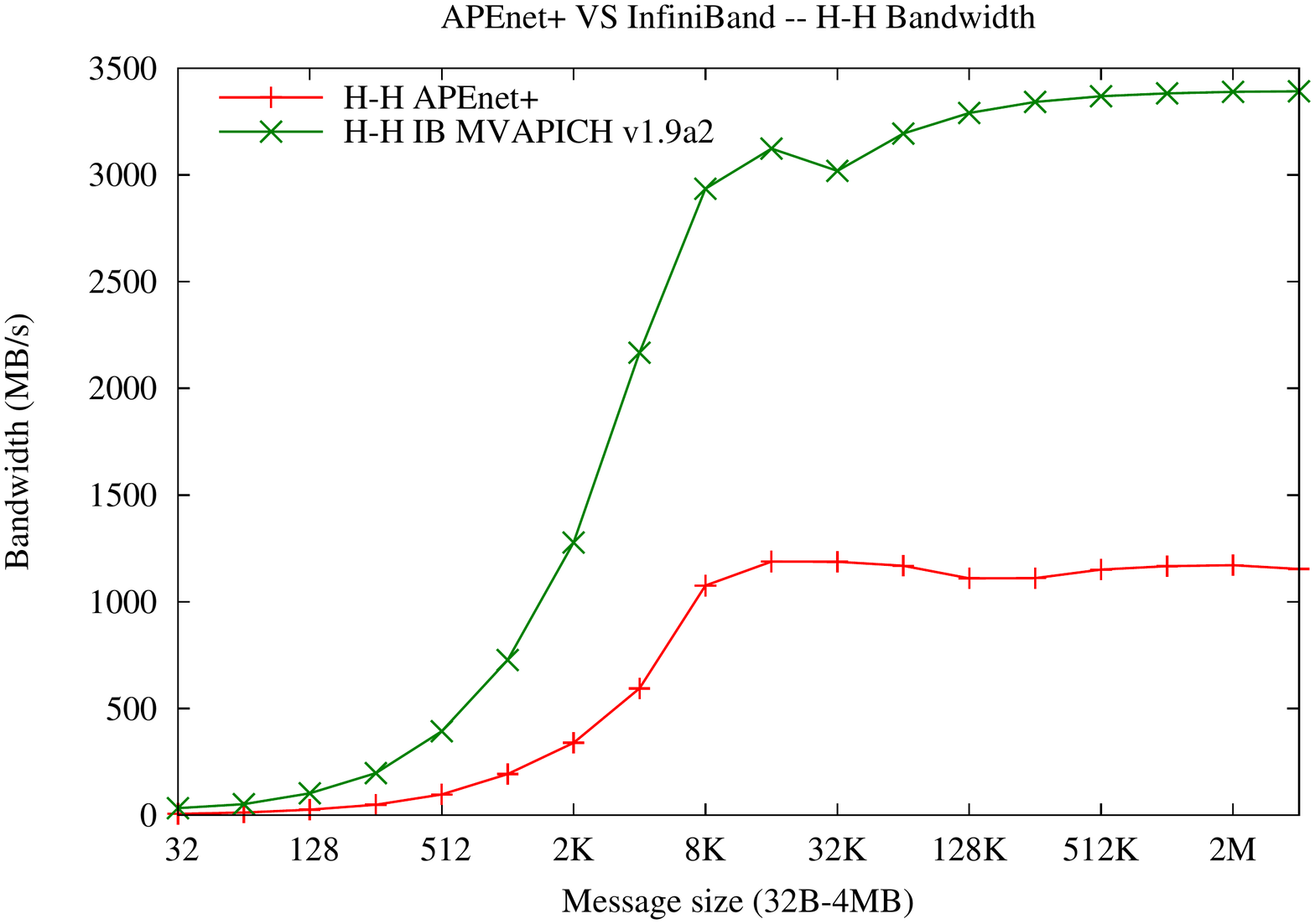}
   \caption{Host-to-Host Bandwidth comparison between \apenetp and
   Infiniband.}
   \label{fig:apenet_vs_ib_h2h_bw}
 \end{minipage}
 \ \hspace{2mm} \hspace{3mm} \
 \begin{minipage}[b]{0.48\textwidth}
  \centering
   \includegraphics[trim=15mm 20mm 15mm 20mm,clip,width=\textwidth]{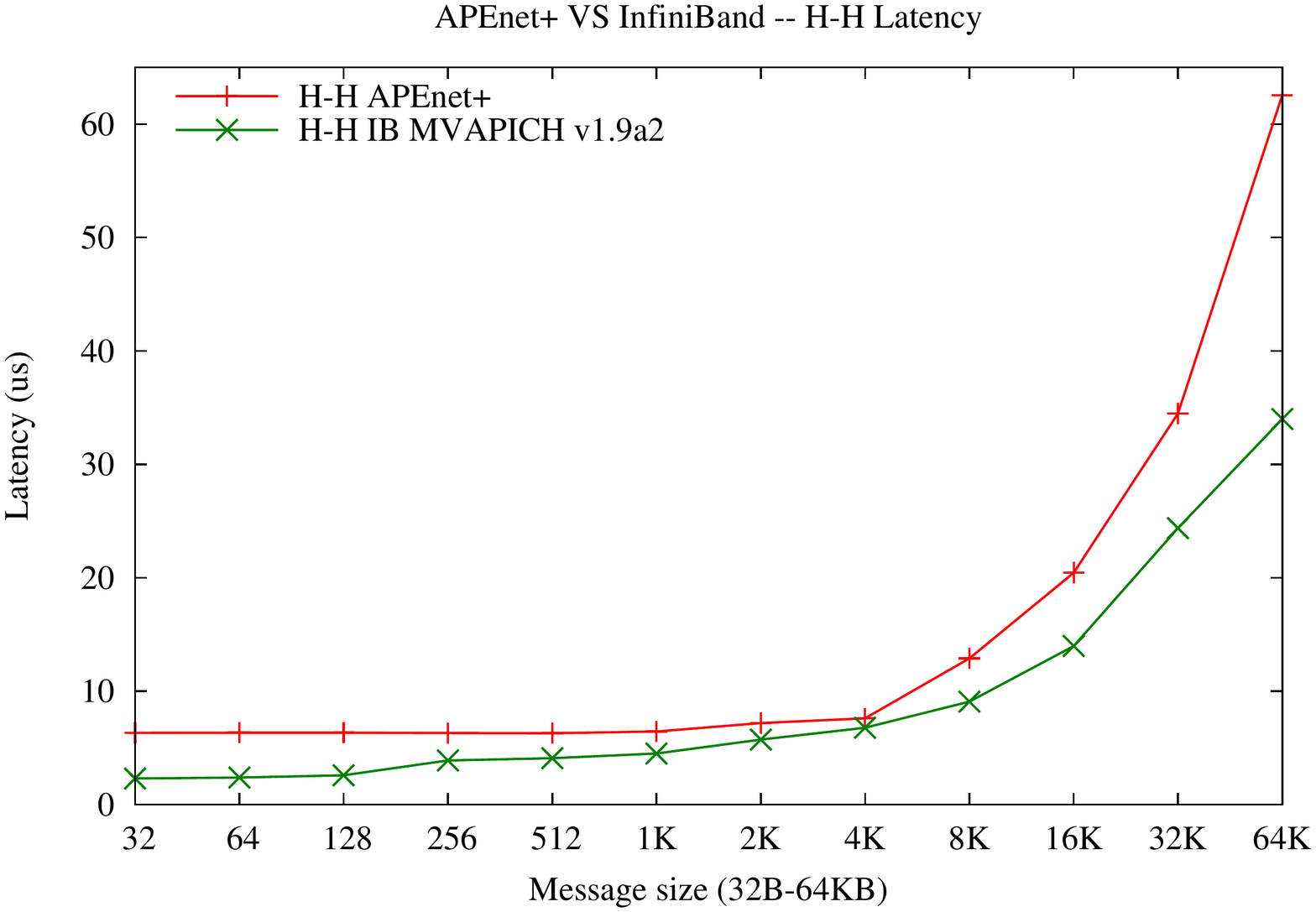}
   \caption{Host-to-Host Latency comparison between \apenetp and
   Infiniband.}
   \label{fig:apenet_vs_ib_h2h_lat}
 \end{minipage}
\end{figure}

Figure~\ref{fig:apenet_vs_ib_h2h_bw} and
figure~\ref{fig:apenet_vs_ib_h2h_lat} show a comparison between
\apenetp and Infiniband in case of Host-to-Host transaction.
In this case \apenetp can not exploit any optimization for small
message size, as \PtoP in case of GPU-to-GPU transaction. On the other
hand Infiniband takes advantage of dedicated path for different
message sizes to optimize the latency.
When the message size increases (greater than 8~KB) the \apenetp RX
processing is limited by the reduced performance of the \nios. This
limitation explains the huge gap in bandwidth for great size messages.
In section~\ref{sec:link_bw_lat} we define a roadmap to increase the
performance of the link equalizing the HOST memory read bandwidth.
Thus the only critical part remaining is the RX processing one.
At the moment we are defining a re-working strategy of the hardware
code to remove several tasks from \nios, trying to relieve it,
optimizing the performance of the RX path.
Referring to table~\ref{tab:lowlevel} and
figure~\ref{fig:apenet_tx_bw} the maximum bandwidth that \apenetp can
achieve is the HOST memory read one (2.8~GB/s) not much less than the
Infiniband Host-to-Host bandwidth (3.3~GB/s).

\subsubsubsection{Effective Frequency: theoretical approach to \apenetp 
optimization}

The results of benchmarks reported in section~\ref{sec:bwlat}, allowed us
to select hardware/software blocks to be optimized, in order to further 
speed-up the \apenetp design. 
An alternative approach to the physical benchmarking leading to similar
results is the analysis sketched in this paragraph, where a new blocks
evaluation index is introduced.  
The amount of available memory is one of the FPGA most critical resources
The on-chip memory available in Stratix IV EP4SGX290NF45C2, the 
\apenetp card device, is limited to 1.74~MB .
Thus, the architectural choices performed during the hardware development
phase aims to minimize the amount of used memory and to achieve the target
performance. The right balance between these two aspects is one of the 
factors behind the success of a project.

To perform a deeper analysis about the connection between memory
consumption and achieved bandwidth, we introduce the parameter
Effective Frequency ($f_{eff}$) defined as the ratio of used memory to
the achieved peak bandwidth with the implemented architecture.
$$f_{eff}=\frac{PEAK\ BANDWIDTH}{USED\ MEMORY}$$

Then we introduce the ratio $O$ of effective bandwidth to the
operating frequency $f_{real}$.

$$O=\frac{f_{eff}}{f_{real}}$$

The aim of these parameters is to detect the limitations of the
architectural decisions and to assist in re-allocating the available
memory resources.
$O$ is defined in a range $O\in[0;1]$.
The closer to 1 the value approaches, the higher the level of memory
optimization.
The proper use of parameter $O$ is clarified by the following points.

\begin{itemize}
\item First of all, implement and optimize the architecture to achieve
the desired performance.
\item Once the desired performance is reached, analyze the memory
consumption and calculate the $f_{eff}$ and $O$.
\item Optimize the memory consumption without affecting the achieved
performance.
\item Uniform memory optimization level of all logic blocks of the
implementation, eliminating the presence of bottlenecks.
\end{itemize}
\begin{small}
\begin{table}[htbp]
\centering
\setlength\extrarowheight{2pt}
\begin{tabular}{|l|cccccc|}
\hline
\hline
\textit{Logic Block}  & \textit{Used Mem} & \textit{\% of Mem} & \textit{Peak Bandwidth} & $f_{eff}$ & $f_{real}$ & $O$    \\
\hline                                                                                  
\hw{TX BLOCK}         & 0.105 MB          & 6.0\%              & 2.8 GB/s                & 26.7  MHz & 250 MHz    & 0.107  \\ 
\hw{GPUTX BLOCK}      & 0.088 MB          & 5.1\%              & 1.5 GB/s                & 17.0  MHz & 250 MHz    & 0.068  \\ 
\hw{RX BLOCK}         & 0.070 MB          & 4.0\%              & >2.0 GB/s               & >28.6 MHz & 250 MHz    & >0.114 \\  
\hw{TORUS LINK}       & 0.167 MB          & 9.6\%              & 9.6 GB/s                & 57.5  MHz & 175 MHz    & 0.328  \\                                                                                   
\hline                                                                                     
\nios                 & 0.402 MB          & 23.1\%             & 1.2 GB/s                &  3.0  Mhz & 200 MHz    & 0.015  \\
\hline
\hline
%\hline
\end{tabular}
\caption{An overview of memory consumption and its connection to
the achieved performance.}
\label{tab:memory}
\end{table}
\end{small}
We identified 5 key logic blocks to evaluate the performance of the
\apenetp card:

\begin{itemize}
\item the \hw{TX BLOCK} is responsible of reading from the host
memory.
\item the \hw{GPUTX BLOCK} is responsible of reading from the GPU
memory, implementing the \PtoP capabilities to optimize the latency of
the reading process.
\item the \hw{RX BLOCK} is responsible of writing to the host/GPU
memory.
\item the \hw{TORUS LINK} is responsible of data-transmission between
neighbouring nodes.
\item the \nios the on-board \mbox{micro-controller} simplifies the
 tasks of \hw{GPUTX BLOCK} and mainly of \hw{RX BLOCK}.
\end{itemize}

In table~\ref{tab:memory} we collect the amount of memory used in the
current implementation of the main logic blocks.

The first line of the table~\ref{tab:memory} reports the memory
consumption due to the implementation of the \hw{TX BLOCK}.  \hw{TX
BLOCK} is the more performing logic block and it satisfies our
expectation. It defines a good balance between performance and memory
consumption (only 6\%) then $O_{TX}$ can be taken as reference to
evaluate the memory optimization level of the remaining logic blocks.

The second line collects the \hw{GPUTX BLOCK} status. In this case
$O_{GPUTX}$ is lower than the \hw{TX BLOCK} one.
We have to consider that \PtoP is a convenient optimization for small
size messages. The size limitation prevents to reach the maximum
performance. The effective frequency is an evaluating parameter in
case of peak performance. Then it confirms that for greater size
messages a different GPU memory reading protocol is necessary.
As explained above the staging approach is a good alternative to
increase the peak bandwidth of the GPU memory read process considering
also the high level of optimization of the \hw{TX BLOCK}.

To analyze the RX block by itself we modified the one-way
point-to-point test involving two nodes described before, to bypass
the \nios during the memory writing process.
Basically the receiver node communicates to the transmitter node the
physical memory address of the allocated buffer, on host memory,
instead of the virtual one. In this way when the \hw{RX BLOCK}
receives the data-packet can use the physical memory address contained
in the \header to perform the PCI transaction without questioning the
\nios. This make it possible to eliminate the bottleneck introduced by
the \mbox{micro-controller}.
At the moment the result of these tests is afflicted by the reduced
capabilities of \hw{TORUS LINK} (as shown in
section~\ref{sec:link_bw_lat}) but it allows to establish that \hw{RX
BLOCK} can sustain a bandwidth equal to or greater than 2.0~GB/s. In
this case $O_{RX}$ is in line with $O_{TX}$ but it could be better
once we discover the real architecture limitation.

The last line shows the memory resources reserved for \nios.
Obviously a \mbox{micro-controller} needs a huge amount of memory
compared to normal hardware blocks. In this case we want to evaluate
the choice to use the \nios in the RX chain (\nios plus \hw{RX
BLOCK}).  $O_{NIOS}$ is an order of magnitude lower of $O_{RX}$. It
results in a dramatical loss of performance. This consideration may
lead to produce an effort to move some RX functionalities from \nios
to \hw{RX BLOCK}, re-allocating memory resources, in order to maximize
the performance.

Finally, the \hw{TORUS LINK} has the higher value of $O$ ratio (three
times higher than $O_{TX}$). The value reported is the total memory
used by 6 channels and we consider the aggregated bandwidth. The
optimization of the memory resources is too high and clearly it limits
the \hw{TORUS LINK} performance.
The modification proposed in section~\ref{sec:link_bw_lat} (doubling
\hw{RX LINK FIFO}) is strengthened by this observation even if it
should cause a decrease in the effective frequency of the \hw{TORUS
LINK}. In this case a reduction of memory optimization homogenizes the
resources usage and increases the performance.
   
\subsubsubsection{HW instrumentation for debugging}
%In APEnet+ we implemented some Hardware instumentation for debugging. We are able to monitor all data path by means of a CRC (Cyclic redundancy Check) and a data parity checker. 

\begin{figure}[!hbt]
  \centering
  \includegraphics[width=\textwidth]{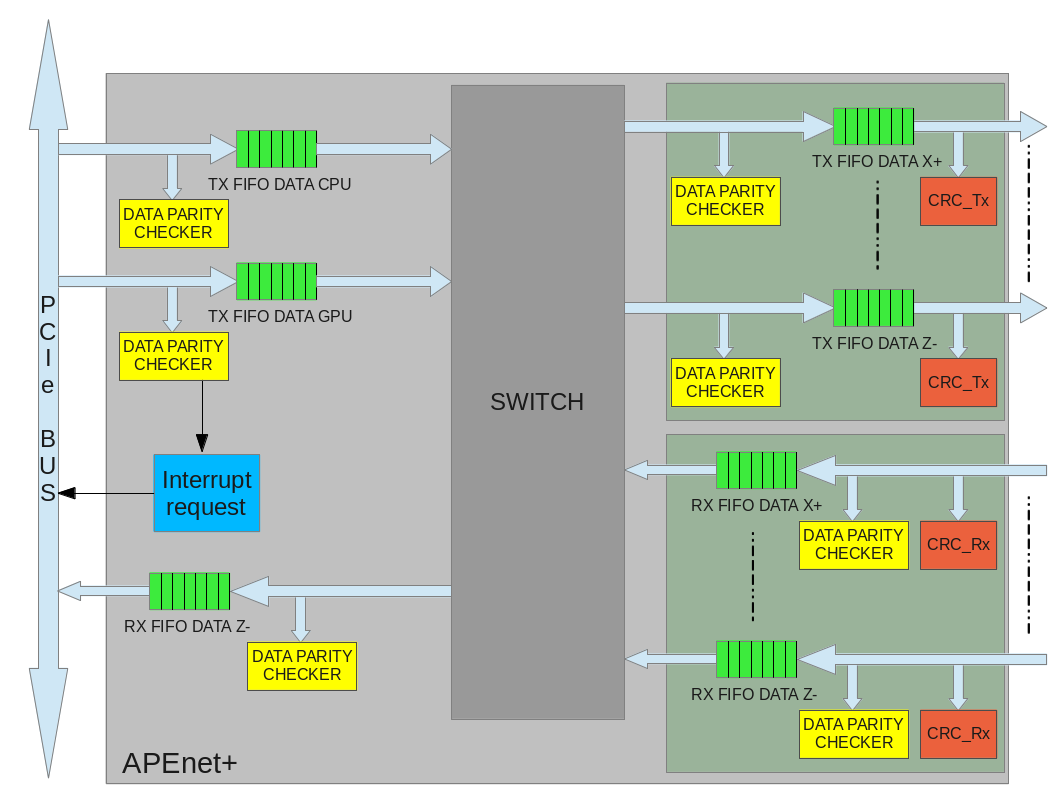}
  \caption{Hardware instrumentations for debugging on \apenetp.
%Data parity checker distribution on APEnet+
}
  \label{fig:par_check_1}
\end{figure}

%In \apenetp we implemented some hardware instrumentations for
%debugging: we monitor data transferred between two \apenetp cards by
%means CRC (Cyclic redundancy Check) blocks, and internal data path by
%means of a data parity checker (Figure \ref{fig:par_check_1}).
In \apenetp we implemented two main components devoted to debugging:
we monitor data transferred between two \apenetp cards by means
\hw{CRC} (Cyclic Redundancy Check) blocks, and internal data path by
means of a \hw{DATA PARITY CHECKER} (figure \ref{fig:par_check_1}).

\hw{CRC\_TX} calculates the check value for the whole data-packet to
be sent (\hw{header}+\hw{payload}+\hw{footer}) using the CRC-32 IEEE
standard polynomial. The check value is appended to the \footer.
The error detection is performed by \hw{CRC\_RX}, that calculates the
check value and compares it with the received one. The chosen
polynomial (CRC-32 IEEE) is able to detect a maximum number of errors
that ranges from 2 to 6.
%
%The CRC check data from their input in transmitter Torus Link until their input in Receiver Torus link. CRC we use is CRC-32 IEEE standard polynomial,
%this CRC %The error detection is performed by the receiving DNP card.
 
A \hw{TORUS LINK} is considered sick when the ratio between the number
of errors and the number of packets received overruns a given
(programmable) threshold.

%The \hw{DATA PARITY CHECKER} block, enabled by writing one user's
%register, detects error on data transferred inside \apenetp.  It 
%needs software support: the 128th bit of each payload's word have 
%to contain the parity bit of the first 127 bits. 

The \hw{DATA PARITY CHECKER} block detects error on each 128-bit data
word transferred inside \apenetp. This error-detection technique
requires a software support: the 128th bit of each payload's word have
to contain the \hw{parity bit} of the first 127 bits. The \hw{DATA
PARITY CHECKER} checks the \hw{parity bit}.

\begin{figure}[!hbt]
  \centering
  \includegraphics[width=\textwidth]{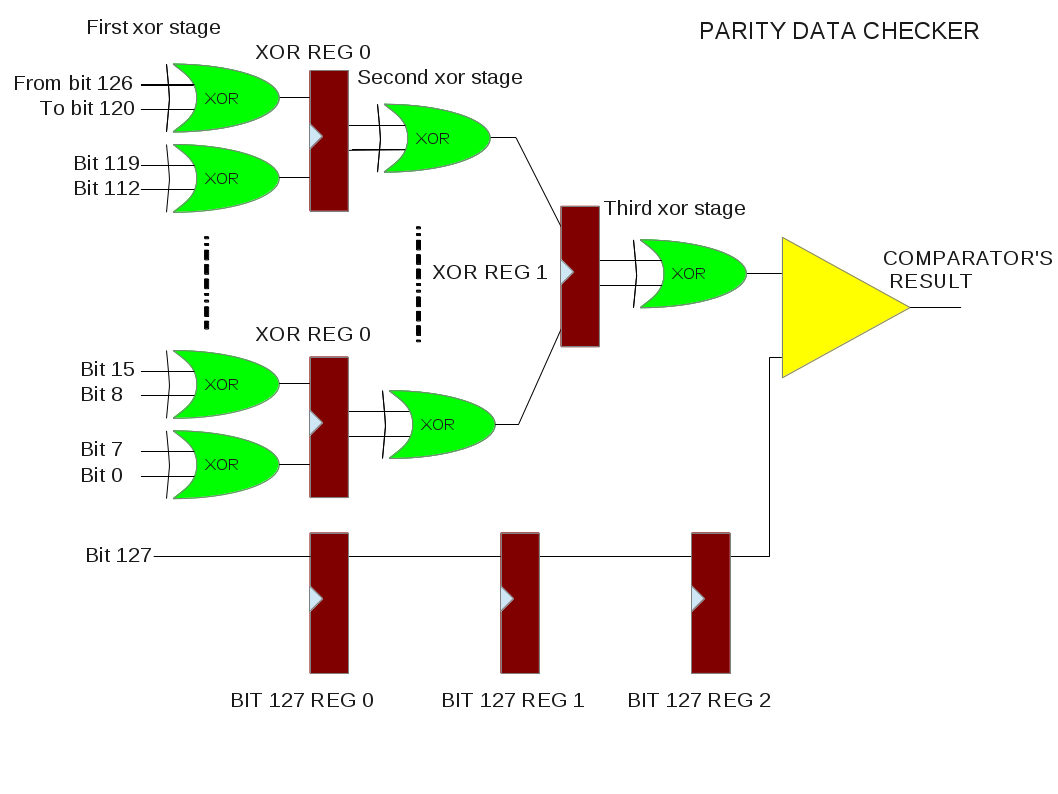}
  \caption{Architecture of \hw{DATA PARITY CHECKER} Block.}
  \label{fig:par_check_2}
\end{figure}

We implement a \hw{DATA PARITY CHECKER} in the input of each data FIFO
(as shown in figure \ref{fig:par_check_1}); it calculates for each
128-bit data word pushed in FIFO the \hw{parity bit} according to the
first 127 bits, and then it compares the result with the bit 128.  If
the comparison fails, indicating that a parity error occurred, the
block sends an interrupt request and writes in a dedicated register
the incorrect word. Parity calculation is made by a XOR chain, as show
in picture \ref{fig:par_check_2}.

Once got interrupt, we can investigate and then modify the path
between data FIFO connected at the \hw{DATA PARITY CHECKER} that
founds error. Usually these problems are due to a long path between a
signal's source and its destination.
%
%
%There is a DATA PARITY CHECKER block in front of each fifo data as show in picture(\ref{fig:par_check_1}).
%When DATA PARITY CHECKER is enabled, software has in charge of sends only data word which bit 128 report parity of the other bits of the word.
%DATA PARITY CHECKER calculate for each data word parity of its first 127 bits and then compares the result with the bit 128, sending an interrupt if the compare is not succesfully.
%
%To monitor data path between each APEnet+ input data fifo to each APEnet+ output data fifo we implement a block called
%DATA PARITY CHECKER. Despite CRC functionality DATA PARITY CHECKER block is enabled by meanwrite of one user's register and in addition it needs software support.
%When DATA PARITY CHECKER is enabled, software has in charge of sends only data word which bit 128 report parity of the other bits of the word.
%DATA PARITY CHECKER calculate for each data word parity of its first 127 bits and then compares the result with the bit 128, sending an interrupt if the compare is not succesfully.
%There is a DATA PARITY CHECKER block in front of each fifo data as show in picture(\ref{fig:par_check_1}).
%
% Once got interrupt, software reads relative registers to verify which DATA PARITY CHECKER found error. 
% Once found an error we can investigate and then modify the path between that fifo data and the previous to resolve the problem. 
%
%
% Once found an error we can investigate and then modify the path between that fifo data and the previous to resolve the problem. 

\subsubsubsection{Resource's utilization, power measurements and link characterization} % (occhio,...) - Francesca
\apenetp is developed on an EP4SGX290 FPGA, which is part of the
Altera 40 nm Stratix IV device family.

% and comes equipped with 48 fullduplex CDR-based transceivers, 
% supporting data  rates up to 8.5 Gbps each. It also provides a PCI-Express X8 Gen2 interface, which is complemented by 
% a commercial PCIExpress core to allow communication between the host processor and the network. 
% Moreover, an Ethernet port is foreseen in order to build an additional secondary network,with an offload engine for collective 
% communication tasks. 

In Table \ref{tab:synt_summary} and \ref{tab:synt_block} is shown an
outline of the FPGA logic usage measured with the synthesis software
Quartus II.

\begin{table}[!htb]
\centering
\begin{tabular}{|l|l|}
\hline
\hline
Logic utilization                 & 42\%                          \\
Total Thermal Power Dissipation   & 13,191 mW                     \\
Combinational ALUTs               & 70,673 / 232,960 (30\%)       \\
Memory ALUTs                      & 228 / 116,480 (< 1\%)         \\
Dedicated logic registers         & 61,712 / 232,960 (26\%)       \\
Total pins                        & 242 / 1,112 (22\%)            \\
Total block memory bits           & 7,533,432 / 13,934,592 (54\%) \\
DSP block 18-bit elements         & 4 / 832 (< 1\%)               \\
Total GXB Receiver Channel PCS    & 32 / 32 (100\%)               \\
Total GXB Receiver Channel PMA    & 32 / 48 (67\%)                \\
Total GXB Transmitter Channel PCS & 32 / 32 (100\%)               \\
Total GXB Transmitter Channel PMA & 32 / 48 (67\%)                \\
Total PLLs                        & 3 / 12 (25\%)                 \\
\hline
\hline
\end{tabular}
\caption{Summary of \apenetp logic usage on Stratix IV EP4SGX290 FPGA.}
\label{tab:synt_summary}
\end{table}
\begin{table}[htbp]
\centering
\begin{tabular}{|l|c|c|c|}
\hline
\hline
 & Network Interface Block & Switch Block & Single Link Block\\
\hline 
Combinational ALUTs       & 32,858    & 12,879    & 4,216   \\
Memory ALUTs              & 228       & 0         & 0       \\
ALMs                      & 25,893    & 12,810    & 4,413   \\
Dedicated logic registers & 29,622    & 12,321    & 3,842   \\
Block memory bits         & 4,609,848 & 2,875,392 &  8,032  \\
Power Dissipation         & 914 mW    &  3,464 mW & 1,134 mW\\
\hline
\hline
\end{tabular}
\caption{Logic usage per logic block on Stratix IV.}
\label{tab:synt_block}
\end{table}

The thermal power dissipation measurements are obtained from the power
analyzer tool by Altera, with a value for the power input I/O toggle
rate set to 25\%. The power dissipation associated to a single channel
(4 lanes and related logic) amounts to slightly more than 1
Watt.

\begin{figure}[!hbt]
  \centering
  \includegraphics[width=\textwidth]{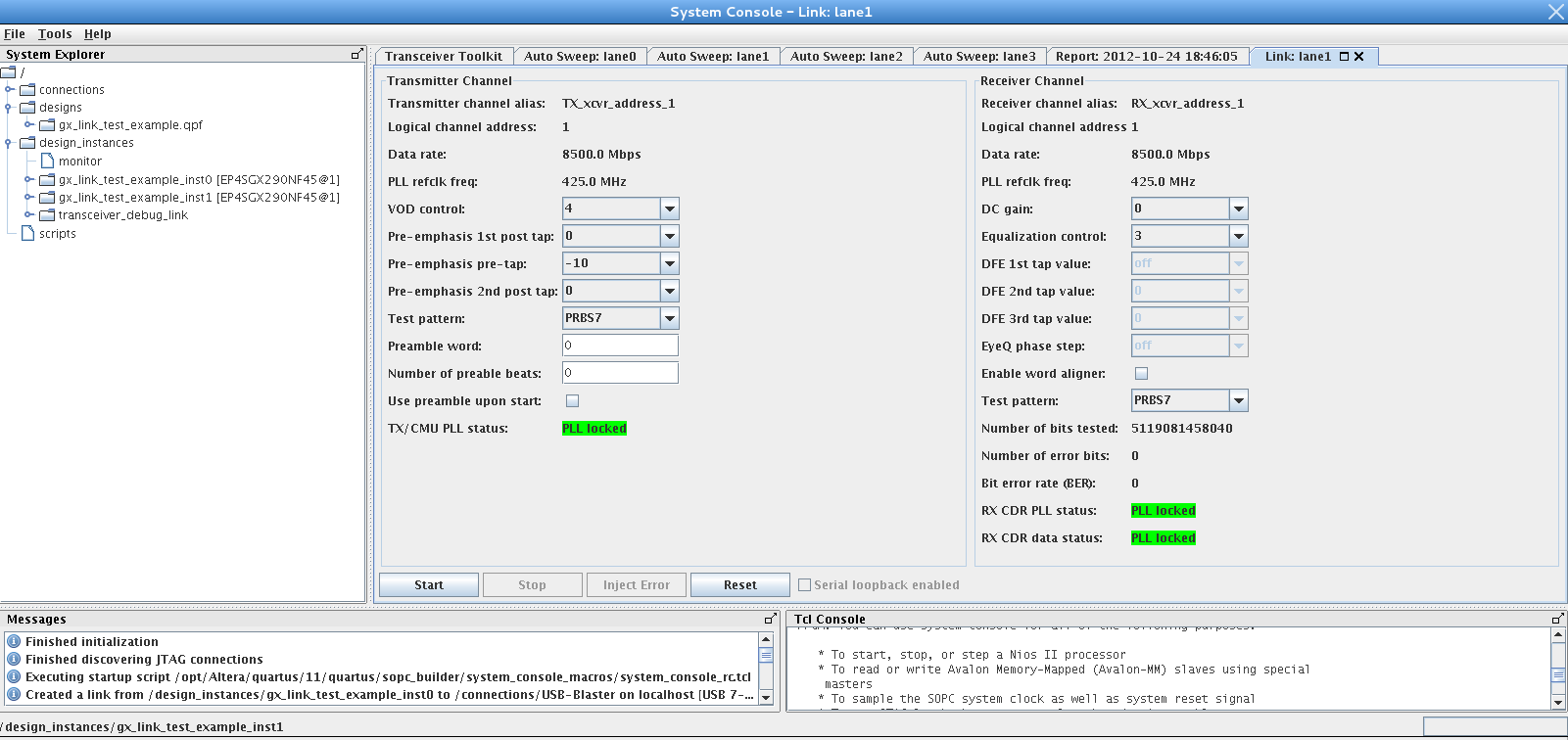}
  \caption{Transceiver Toolkit GUI: analog settings for transmitter
  and receive channels of X+ channel's lane 0.}
  \label{fig:trans_tk}
\end{figure}

The Stratix IV GX provides up to 32 full-duplex
Clock-Data-Recovery-based transceivers with Physical Coding Sublayer
(PCS) and Physical Medium Attachment (PMA). In order to produce the
clearest signal and thus being able to increase signal clock frequency
on cable, a fine tuning of PMAs analog settings is required. Since the
transmitted signal is sensitive to the physical route that it must
travel toward the receiver (cable, connectors, PCB lanes), a number of
analog settings must be appropriately tuned to offset for distortions.
%
%Manual tuning of these settings involve trials and errors, then locking
%at compile time the optimal values found in this way. 
%
Altera provides the Transceiver Toolkit as a specific tool to choose
the appropriate values for these analog controls. The Transceiver
Toolkit performs its job by tuning the Reconfig block; it can
dynamically reconfigure analog settings of the PMAs such as
Pre-emphasis and DC Gain on the transmitter side and Equalization and
VOD on the receiver side; furthermore, it offers a simple GUI to tweak
these settings and to see immediate feedback on receiver side. By
means of this tool we found the optimal settings for each of the four
lanes in all six channels (figure \ref{fig:trans_tk}).

Once the tuning phase is done, a following step is the testing of
these settings on our board with our custom logic. Our testing
firmware is made of a random signal generator (PRBS @32 bits) on the
transmitting side and a data checker on the receiving side.
Transmitter and Receiver are synchronized by a dedicated protocol
implemented in the \hw{TORUS LINK}.
%
% started after synchronization performed by a dedicated protocol, 
%connected atop the \apenetp link block.  
%
The numbers of correctly and incorrectly transferred 128-bits words
are stored in registers. In this way we can either measure the bit
error rate of the channels (number of incorrectly transferred
words/number of transferred words) and verify our alignment custom logic.
In the table \ref{tab:BER_table} we report some of the measures performed,
taken with different cable lengths, channel ports, and data rate.

\begin{table}[!t]
\renewcommand{\arraystretch}{1.3}
\centering
\begin{tabular}{|l||l|l|l|}
\hline
\hline
BER           & Direction & Cable length & Data Rate \\
\hline
$<~$7.04 E-11 & X+        & 5m           & 8.5~Gbps  \\
\hline
$<~$8.12 E-11 & Y+        & 5m           & 8.5~Gbps  \\
\hline
$<~$7.04 E-10 & Z+        & 5m           & 8.5~Gbps  \\
\hline
$<~$9.8 E-13  & X+        & 5m           & 7~Gbps    \\
\hline
$<~$2.46 E-14 & X+        & 2m           & 7~Gbps    \\
\hline
\hline
\end{tabular}
\caption{BER measurements on \apenetp's channel, taken with different 
cable lengths and data rate.}
\label{tab:BER_table}
\end{table}

We verified successful data transfer up to the maximum declared rate
per single lane (8.5~Gbps) in a defined range of the analog
settings. Anyway at this moment, for reliable operations and upper
level firmware and software validation, we use a more conservative
setting for all transceivers lanes at 7.0~Gbps, which makes the
channels raw aggregated bandwidth at 28~Gbps.

\begin{figure}
\centering
\subfloat[][\emph{5 Gbps}.]
{\includegraphics[width=.45\textwidth]{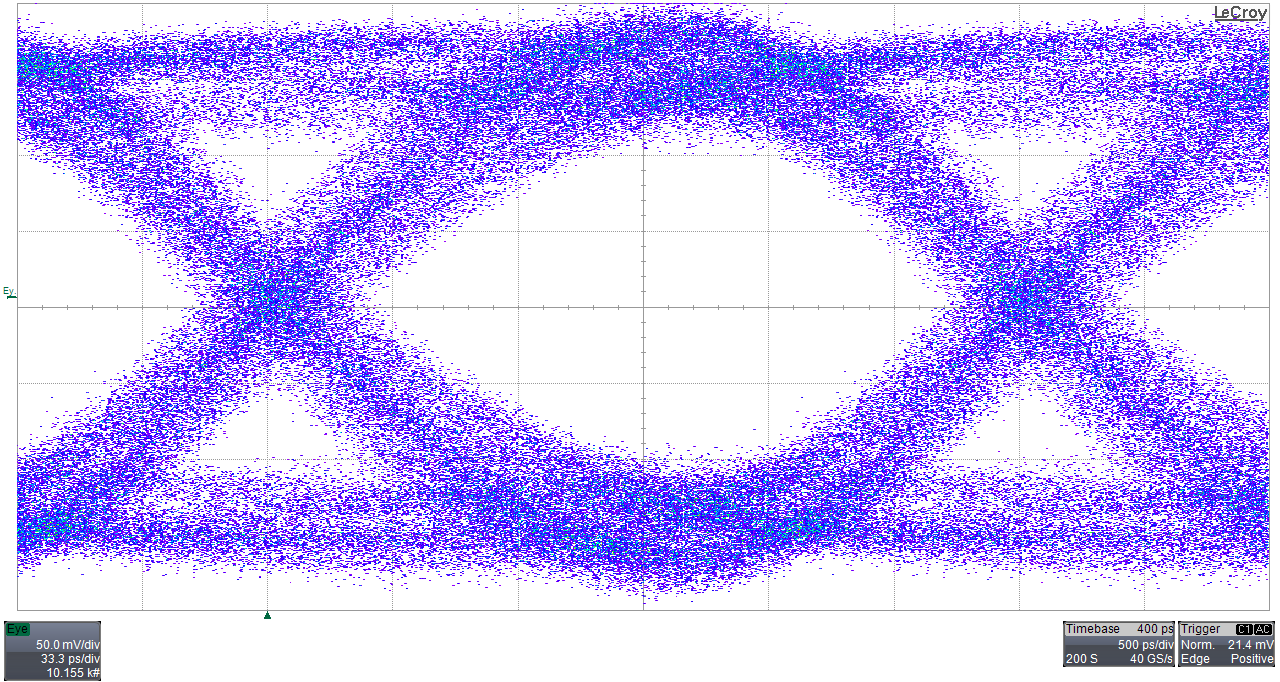}} \quad
\subfloat[][\emph{6 Gbps}.]
{\includegraphics[width=.45\textwidth]{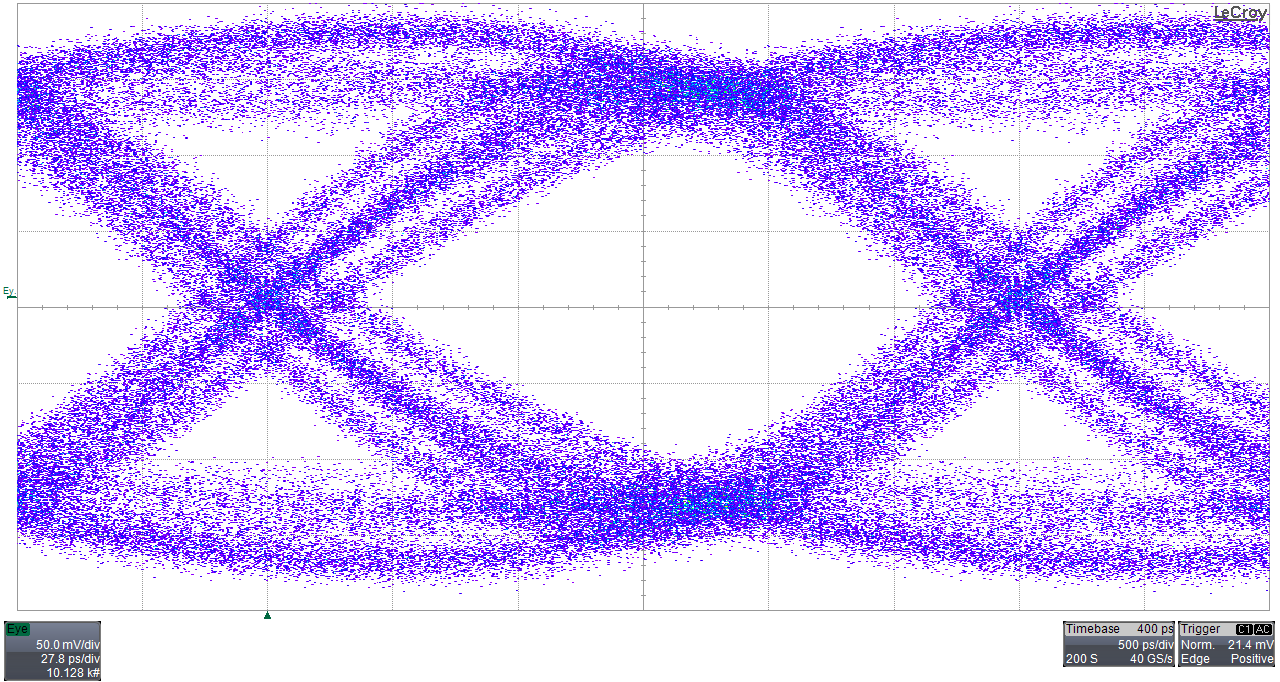}} \\
\subfloat[][\emph{8.5 Gbps}.]
{\includegraphics[width=.45\textwidth]{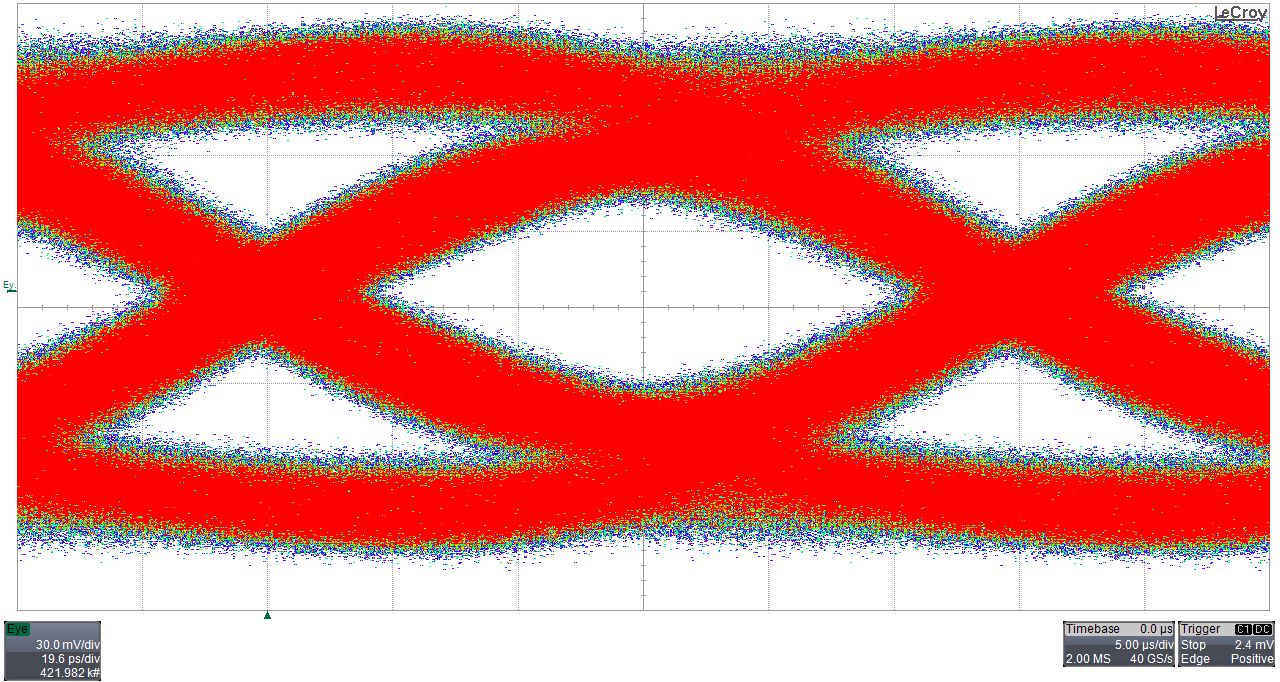}} \quad
\caption{Eye Pattern.}
\label{fig:subfig}
\end{figure}

In order to verify link quality, we prepared a testbed for measuring
signal integrity for the received data. Testbed is composed by an
\apenetp board which acts as a sender, a QSFP+ cable, and a QSFP+ to
SMA breakout custom card.
We prepared a special firmware for \apenetp, in which all the
transmitter part of the transceivers are sending a random data
pattern, with no waiting for the receive side to be aligned (as it is
a one way measurement). Measurements were performed with a 20~GHz,
40~GS/s high bandwidth sampling scope (a LeCroy WaveMaster 820Zi-A)
connected to the SMA test points on the break-out card.
In figure \ref{fig:subfig} we show the Eye Pattern at 5~Gbps, 6~Gbps
and 8.5~Gbps otteined using a cable of 1~m. length.
Signal integrity could be improved by receiver side optimizations
(like VOD or equalization) and by the insertion of a 100~$\Omega$
differential termination (as required by QSFP+ standard).

%Signal integrity is good for data rate up to 6 Gbps, after
%which signal degradation occurs. This is due to the fact that
%in this testbed no receiver side optimizations can be used
%(like VOD or equalization), and no 100 Ohm differential
%termination is inserted (as instead QSFP+ standard requires).

\subsubsection{Board Production} %% Roberto did it!
A production batch of 15 boards has been processed during 2012.  Some
minor changes were introduced in this batch, with respect to the
prototypal batch (4 boards) produced during 2011. These changes were
including fixes of previous design mistakes.

A request for tender has been initiated at the beginning of the year,
and the selected vendor resulted to be Seco S.r.l. with the lowest bid
and satisfactory technical proposal.

Vendor was encharged with complete board production, component
procurement (excluding the Altera Stratix IV component, which were
provided directly by an Altera re-seller) and complete board
assembly. Basic electrical tests have been performed at this stage.

Before validation, boards need to be initialized with a list of tasks:
\begin{itemize}
 \item Altera EPM240 component programming --- in order to have
 on-board Altera USB blaster capabilities on mini-USB connector.
 \item FTDI component initialization --- for USB connection to JTAG
 properly working.
 \item Altera EPM2210 component programming --- in order to have all
 on-board controllers working (sensors, clock management, JTAG
 capabilities, \dots).
 \item Altera Stratix IV programming.
\end{itemize}

Thus, production batch has been validated with a test suite on latest
available software and firmware.

%% was divided into:
%%\paragraph{HW provider selection - Piero
%%\paragraph{Specification for HW provider - Roberto }
%%\paragraph{Components Procurement - Roberto}
%%\paragraph{16 boards validation - Roberto}

%%Alessandro
\subsection{Assembly of the \quong Hybrid Cluster}

\begin{figure}[!hbt]
  \centering
  \includegraphics[width=0.70\textwidth]{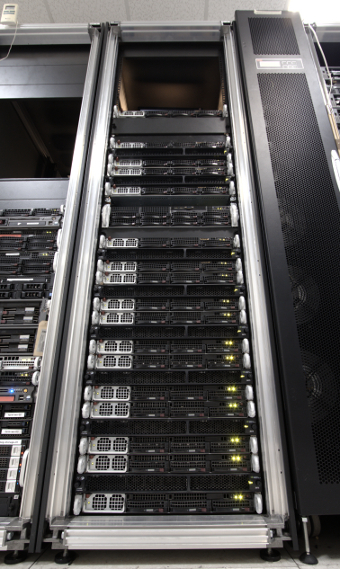}
  \caption{The \quong Hybrid Cluster}
  \label{fig:quong}
\end{figure}

During 2012 a 16 nodes \quong system has been deployed.  The hardware
procurement procedure carried out during 2011 has been further
extended in order to reach the 16 nodes count of the currently
available system (figure \ref{fig:quong}).

The \quong hybrid computing node integrates an Intel Xeon E5620 double
processor host with 48~GB of system memory, two S2075 \nvidia Fermi
class GPUs and an \apenetp 3D network card.
A 40~Gb/s Infiniband Host Controller Adapter is also available, in
order to provide a \emph{de facto} standard network interface to users
and to perform comparative benchmarks with our custom 3D network
interface.

\begin{figure}[!hbt]
  \centering
  \includegraphics[width=0.70\textwidth]{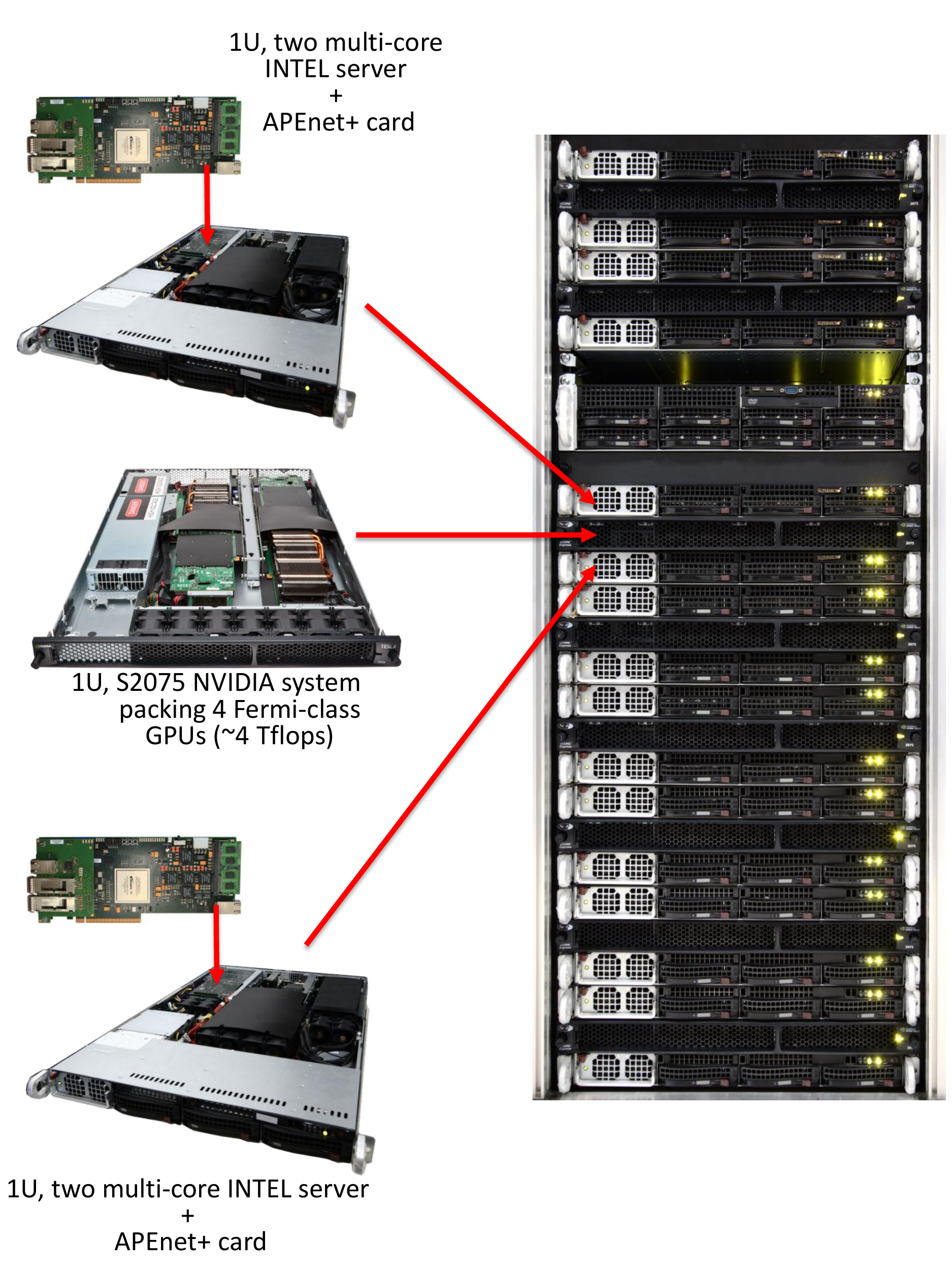}
  \caption{How to assemble a \quong sandwich.}
  \label{fig:quongsandwich}
\end{figure}

%%qui si potrebbe parlare dell'architettura del pci
The \quong elementary mechanical unit is a 3U "sandwich" built
assembling two Intel dual Xeon servers with a Next/IO vCORE Express
2075, a 1U \PCIe bus extender hosting 4 \nvidia Tesla M2075 GPU (figure \ref{fig:quongsandwich}).
Topologically it corresponds to 2 vertexes on the \apenetp 3D network.

Currently, only 4 out of 16 nodes are actually equipped with an
\apenetp board, enabling a bidimensional 2x2x1 network topology on
these nodes.
By Q2 2013 the whole system will be connected by the \apenetp 3D
network, with a 4x2x2 topology.

The software environment of the cluster is based on CentOS 6.3, with a
diskless setup where the computing nodes are bootstrapped via the GbE
network interface, and employs the \nvidia CUDA 4.2 driver and
development kit for the GPUs. Both OpenMPI and MVAPICH2 MPI libraries
are available.

Users can submit parallel jobs using the SLURM batch system, with full
support for task launch for OpenMPI and MVAPICH2 applications. 
Programs linking  the OpenMPI library can perform communications both
over \apenetp and Infiniband fabrics, those linking the MVAPICH2 can 
use only the Infiniband.

\FloatBarrier

\subsection{Applications}
\label{sec:applic}
Since its first deployment, \quong cluster is available to users from
different research areas like Lattice Quantum
Chromo-Dynamics~\cite{D'Elia:2012:zw}, Laser-Plasma
Interaction~\cite{Rossi:pic}, Complex Systems~\cite{Berganza:2012:xy}
and N-Body simulation for astrophysics~\cite{2012arXiv1207.2367C}.
Herebelow we give an outlook of the codes that were more thoroughly
examined and customized to exploit the \PtoP capabilities of the
\apenetp boards equipping the \quong cluster.

\subsubsection{OSU Micro Benchmarks}
A cluster like \quong must be graded against what is currently
considered standard reference for network performance gauging, \ie the
synthetic tests \texttt{MPI\_latency} and \texttt{MPI\_bandwidth} from
the Ohio State University Micro Benchmarks (OSU-MB)
suite~\cite{Traff:2012:OMB-GPU}.
Starting from version 3.5, they are also able to exercise any
GPU-aware directives that the MPI infrastructure should offer.
The pipelining strategies employed by different MPI stacks can be
tested over the InfiniBand network fabric on \quong and compared to
the more advanced \PtoP capabilities employed by \apenetp; a detailed
run-down of these runs and the results of the comparison can be found
in section~\ref{sec:bwlat}.

\subsubsection{Heisenberg Spin Glass}
One application that was used as testbench of the \apenetp
capabilities comes from the statistical mechanics field; it is the
simulation of a prototypical complex system called Heisenberg Spin
Glass (HSG)~\cite{Bernaschi20121416}.
This system models a cubic lattice of magnetic classical 3D spins
$\sigma_{i}$ subject to an Hamiltonian function $H=-\sum_{i \neq j}
J_{ij} \sigma_{i} \cdot \sigma_{j}$ with the $J_{ji}$ couplings
distributed with a normal Gaussian distribution law restricted to
next-neighbours interaction ($J_{ij}=0$ if $|i-j|>1$).

Scanning the lattice, site after site, is needed for every computation
performed onto the system, like in the heat-bath and over-relaxation
algorithms for lattice dynamics or computation of observables like the
internal energy.
Standard techniques of domain decomposition are applied to this scan;
the lattice is partitioned along one or more dimensions and different
processes on different nodes scan a partition each, synchronizing with
each other by exchange of lattice sites along the partition edges.
The criticality of this exchange is apparent as soon as the processes
are scaled up to make the ratio of computation time vs. exchange time
smaller and smaller, all the more so when the computing processes are
on GPUs and the exchange must take into account the staging needed to
bounce back and forth the lattice partition from GPU memories.

This is exactly the scenario where the \PtoP capabilities of the
\apenetp board can be made to shine.
After code rearrangement --- replacing a small number of key MPI calls
with equivalent ones to the RDMA API --- a thorough study of the
performance improvements was performed on the HSG application; results
are published here~\cite{Bernaschi2013250} and here~\cite{CASS2013}.

\subsubsection{Lattice Quantum Chromo-Dynamics}
Lattice Quantum Chromo-Dynamics is a challenging application and a
staple of INFN research work.
The LQCD field has greatly benefited of the strong boost given by GPU
acceleration to stencil computations: in this regard, the QUDA
library~\cite{Clark:2009wm} is a renowned, advanced software package
for multi-GPU Lattice QCD simulation that, in its standard form,
employs MPI for communication but has not yet been modified to exploit
the GPU-aware directives~\cite{Babich:2010}.

\begin{figure}[!hbt]
  \centering
  \includegraphics[trim=18mm 20mm 12mm 20mm,clip,width=0.9\textwidth]{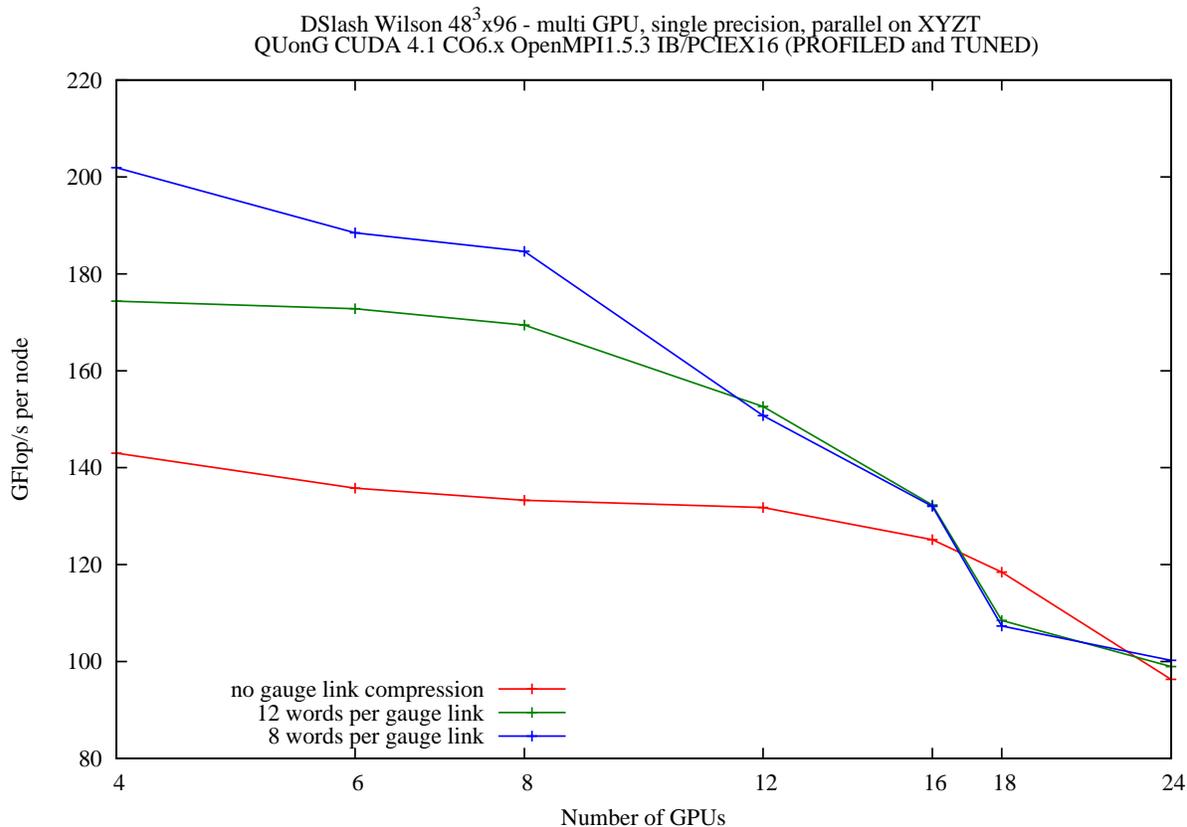}
  \caption{Strong-scaling plot of GFlops per node for a QUDA run on
    \quong.}
  \label{fig:qudagflops}
\end{figure}

Moreover, since the domain decomposition techniques employed for the
LQCD algorithms are exactly the same as in the HSG, the scaling issues
hindering the HSG application belong to the LQCD application as
well~\cite{Babich2:2011}.
Out-of-the-box QUDA was installed and run on the \quong cluster to
verify those issues; in figure~\ref{fig:qudagflops} we see the
decrease of efficiency when strong-scaling the application of the
Wilson-Dirac operator over a $48^{3}\times96$ lattice up to 24 GPUs
--- all twelve \quong nodes (2 GPUs per node) which were available
when the test was run.

\begin{figure}[ht]
  \begin{minipage}[b]{0.45\linewidth}
    \centering
    \includegraphics[trim=22mm 22mm 25mm 25mm,clip,width=\textwidth]{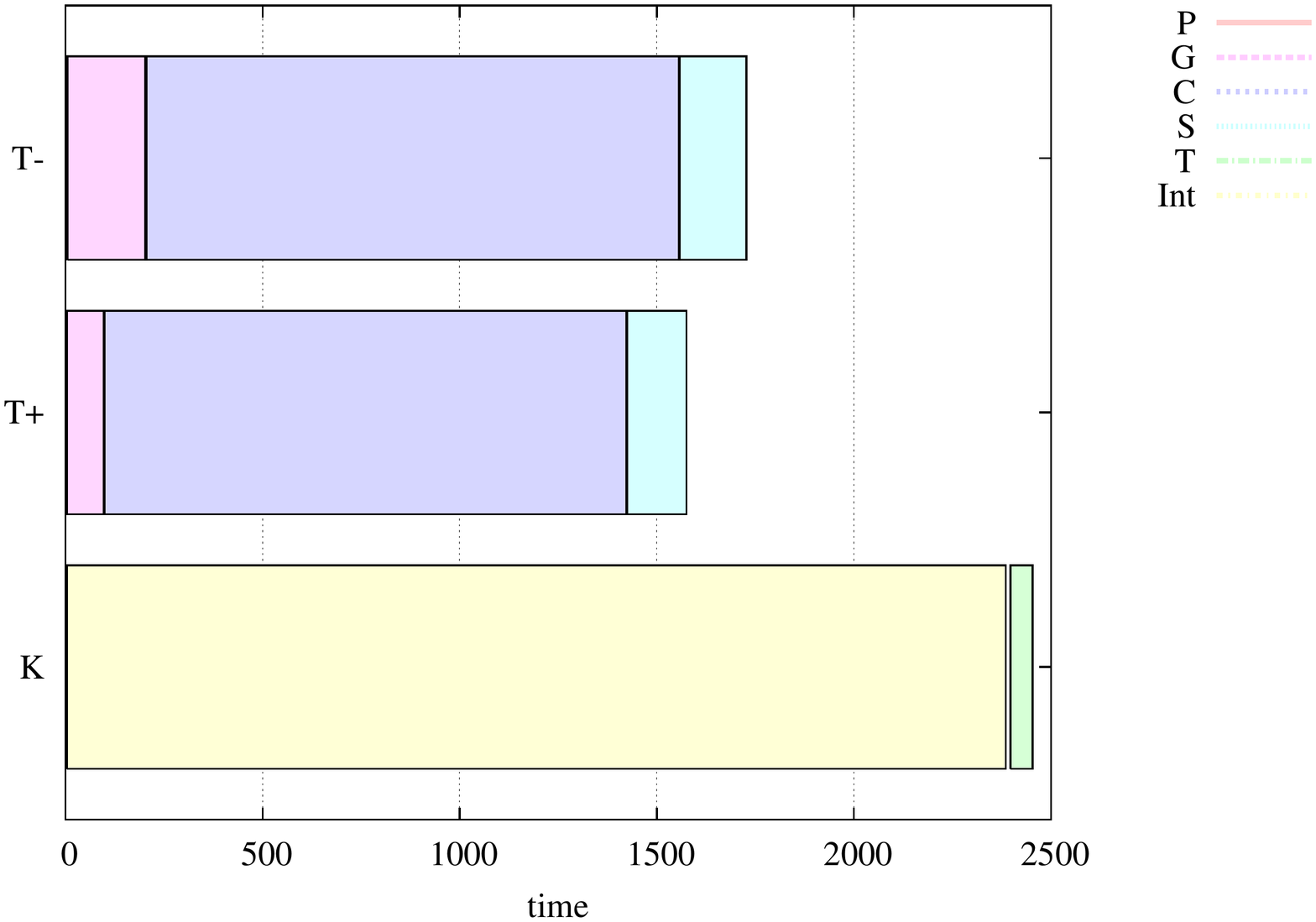}
    \caption*{Lattice partitioned along T axis, 4 GPUs.}
  \end{minipage}
  \hspace{0.5cm}
  \begin{minipage}[b]{0.45\linewidth}
    \centering
    \includegraphics[trim=22mm 22mm 25mm 25mm,clip,width=\textwidth]{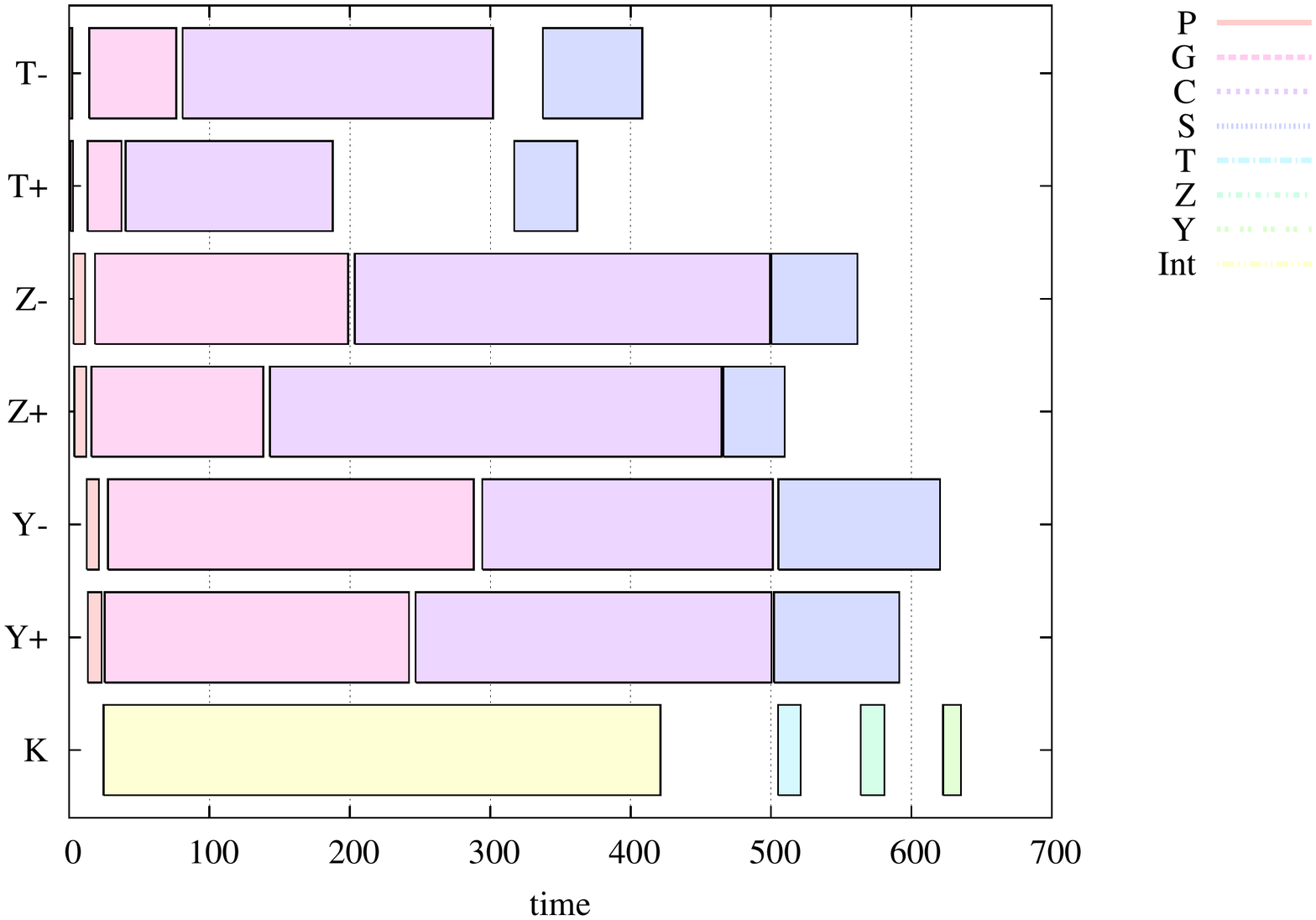}
    \caption*{Lattice partitioned along T, Z and Y axes, 24 GPUs.}
  \end{minipage}
  \caption{GANTT diagrams of a QUDA run --- $48^{3}\times96$ lattice.}
  \label{fig:qudagantt}
\end{figure}

The Gantt diagrams in figure~\ref{fig:qudagantt} clarify the problem.
On the 'K' row, the bars show the span of the computation in the bulk
('Int') and the frame of the lattice along the dimensions ('T', 'Z' and
'Y') for one GPU.
On the 'T+/-', 'Z+/-' and 'Y+/-' rows, the bars show the span of
consecutive 'P'repare, 'G'ather, 'C'ommunication and 'S'catter phases
for the frame exchange between GPUs.

On the left side of figure~\ref{fig:qudagantt}, it is seen that on a
$48^{3}\times96$ lattice split over 4 GPUs, the span of 'P', 'G', 'C'
and 'S' phases is completely overlapped to the 'Int' one, so that
computation is not delayed by communication.
On the contrary, the GPU computation kernel is throttled on 24 GPUs by
the wait for completion of the frames exchange between neighbouring
nodes (along the 'T', 'Z' and 'Y' axes in the '+' and '-' directions),
especially so since the various 'G' and 'S' phases imply a number of
GPU-to-host and host-to-GPU memory transfers that must be scheduled
one after the other, interfering with the scheduling of the CUDA
stream performing the computation kernel and adding an overall
significant latency to the computing over the frames --- see the blank
areas of the 'K' row in the right side of the figure.

Since usage of \PtoP primitives should do not require the 'G' and 'S'
phases, porting of QUDA to \apenetp is foreseen to alleviate this
degradation.
Work for adding this support to the QUDA library is underway; it is a
large and complex endeavour made even harder due to the fact that the
application was not designed from the start with GPU-awareness for the
communication primitives.

In the mean time, parts of the \euretile toolchain were put to the
test with LQCD on \quong --- especially the DAL functional simulator
and the AED of the VEP simulator --- with a reduced kernel of the main
LQCD computation that was rewritten from scratch using either the DAL
syntax and the RDMA low-level API.
A more complete description of the various forms of this LQCD
reference application for \euretile is in section 4 of
\textbf{D7.2}~\cite{euretile:D7_2}.

\subsubsection{Distributed Polychronous Spiking Neural Networks (DPSNN)}
\label{sec:DPSNN}
The investigation on \textit{"brain-inspired"} techniques for the
design of hardware and software components of future many-tile
computing systems is among \euretile key topics.
In the proposal, we stated a set of intuitions about the potentiality
of such architectural concepts to many tile software and hardware
architectures. Actually, we expect to exploit hints coming from the
cortical architecture, a system capable of extreme parallelism and low
power operation.
We identified the development of a dedicated benchmark as a key asset
to get quantitative and qualitative way-points along the project.

Such a cortical simulation benchmark should be used:
\begin{itemize}
 \item as a source of requirements and architectural inspiration
towards the extreme parallelism which will be exhibited by future
many-tile systems;
 \item as a parallel/distributed coding challenge;
 \item as a scientific grand challenge. A well designed code could
produce interesting scientific results.
\end{itemize}

During the previous reporting period (2011), in the framework of
\textbf{WP7: Challenging Tiled Applications} we defined the
characteristics of the DPSNN-STDP application benchmark (Distributed
Polychronous Spiking Neural Network with synaptic Spiking Time
Dependent Plasticity), as described by \textbf{D7.1} ~\cite{euretile:D7_1}.

This benchmark will be coded during the temporal framework of the
\euretile project both using a standard C/\cpp plus MPI environment, as
well as using the C/\cpp plus XML DAL environment.
This will allow to compare the features of the new environment in
comparison to a classical environment on a benchmark which will be
coded "from scratch" using an explicit description of parallelism.

Here, we highlight only a few results directly related to the
development of the \quong platform and \apenetp interconnection board
that are fully described by \textbf{D7.2} ~\cite{euretile:D7_2} .

In summary, during 2012, we designed and implemented the \cpp code of a
complete prototype of the DPSNN-STDP simulator. Then we run the full
simulation cycle (initialization and dynamic phases of a network
including $10^{5}$ synapses) in the \cpp plus MPI environment on a
\quong prototype, using the standard InfiniBand interconnections, and
obtaining the same behavior with different number of processes. Then
we ported and run the initialization phase of the DPSNN-STDP code
under the DAL plus \cpp environment on a prototype environment where
two \quong nodes were interconnected by two \apenetp card. We used the
same \cpp classes as building blocks to construct the DAL and MPI
network of processes.

This will permit, during 2013, to start a comparison activity between
the \apenetp interconnect system and the InfiniBand system, and among
the performances offered by the DNA-OS, MPI and Presto message passing
software layers, when applied to the DPSNN-STDP benchmark. Then we
will identify the middleware and hardware bottlenecks, and derive
architectural improvements for future generations of interconnect
systems.

\subsubsection{Breadth-First Search}
Breadth-First Search (BFS) is a fundamental graph algorithm that
systematically explores the nodes in a graph.
BFS is typically considered one of the most important graph algorithms
because it serves as a building block for many others, including
betweenness centrality calculation, connected component
identification, community structure detection and max-flow
computation.
Benchmark suites targeting graph applications perennially include BFS
as a primary element.

Most of graph algorithms have low arithmetic intensity and irregular
memory access patterns.
In this context, those irregular computation and communication
patterns are very interesting; the typical traffic among nodes that
BFS traversal is able to generate is non-deterministic all-to-all,
depending on the edge partitioning.
The buffer sizes vary as well, so that the performance of the
networking compartment is exercised in different regions of the
bandwidth plot.

Recent works~\cite{Hong:2011:BFS} have shown that, by using a Level
Synchronous BFS, a single-GPU implementation can exceed in performance
high-end multi-core CPU systems.
To overcome the GPU memory limitation, some authors
proposed~\cite{Graph400:2012:BFS} a multi-GPU code that is able to
explore very large graphs (up to 8 billion edges) by using a cluster
of GPU connected by InfiniBand.

The communication code of this BFS application was modified to use the
\PtoP calls from the RDMA API of \apenetp; although the trials were
conducted with \apenetp still in a development and testing stage, the
results obtained~\cite{Graph400:2012:IA3,CASS2013}, albeit
preliminary, show an advantage of \apenetp with respect to InfiniBand.

\subsubsection{TRIGGU (RDMA API)}
A very different endeavour for the \apenetp card is within
TRIGGU~\cite{TRIGGU:2012:NSS}, a simulation of the data flow and
processing --- the \emph{track fitting} --- of a real time event
selection system in a High Energy Physics experiment.
Data produced by the silicon detector at the Collider Detector at
Fermilab is sent from a transmitter node (simulating the detector) to
a receiver node (the selection system) where it is processed on the
GPU to look for tracks left by charged particles.

The system allows measuring both data transfer and processing latency.
The time to copy memory into and out of the GPU represents a
significant overhead for the total latency.

A number of strategies were tested in the software infrastructure to
improve upon this overhead, exploiting the GPUDirect functionalities
of latest CUDA by \nvidia and the RDMA API of the \apenetp card.
Using GPU-Aware MPI leads to a decrease in the memory transfer time
for large data sizes, while usage of \PtoP communication strategies by
means of the RDMA API over \apenetp shows a strong boost to
performance (10 \us on average) for even small data sizes.

\subsection{IP exploitation}
\subsubsection{NaNet Project}
The NA62 experiment at CERN aims at measuring an ultra-rare decay of
the charged kaon ($K^{+} \rightarrow \pi^{+} \nu\overline{\nu}$ );
signal has to be extracted from a huge background which is ten orders
of magnitude more frequent.

With an input particle rate of 10 MHz, some tens of thousands detector
channels and the requirement of avoiding zero suppression as much as
possible, triggerless readout into PCs is not feasible.

Very promising results in increasing selection efficiency of
interesting events come from a pilot project within NA62 that aims at
integrating GPUs into the central L0 trigger processor, exploiting
their computing power to implement more complex trigger primitives
\cite{2012NIMPA.662...49C}.

Independently, in the framework of the \euretile project, we were
developing the low latency, scalable, GPU-accelerated PC cluster
interconnect \apenetp, which is the first non-\nvidia device to 
exploit the \nvidia GPUDirect P2P protocol to enable zero-copy 
transfers to and from GPU memory.

From the joint effort of these two projects comes the next step of
GPUs integration for the NA62 L0 trigger: the NaNet project.

The NaNet board is a custom FPGA-based NIC featuring a standard GbE
interface and GPUDirect P2P capabilities, able to inject the UDP input
data stream from the detector front-end directly into the Fermi/Kepler
GPU(s) memory, with rates compatible with the low latency real-time
requirements of the trigger system.

This makes for a modular and customizable hybrid (CPU+FPGA+GPU) system
able to sustain low response time and real-time features and
potentially re-usable in analogous contexts of strict timing
requirements.

As main design rule, we partitioned the system in a way such that we
could offload the CPU from any data communication or computing task,
leaving to it only system configuration and GPU kernel launch tasks.

This choice was really a step in the right direction for the real-time
requirement, eliminating the unavoidable OS Jitter effects that
usually hinder system response time stability.

Data communication tasks have been entirely offloaded to NaNet by
implementing a dedicated UDP protocol handling block that communicates
with the P2P logic: this allows a direct (no data coalescing or
staging is performed) data transfer with low and predictable latency
on the GbE link-GPU data path.

\begin{figure}[!htb]
  \centering
  \includegraphics[trim=30mm 0mm 15mm 0mm,clip,width=0.8\textwidth]{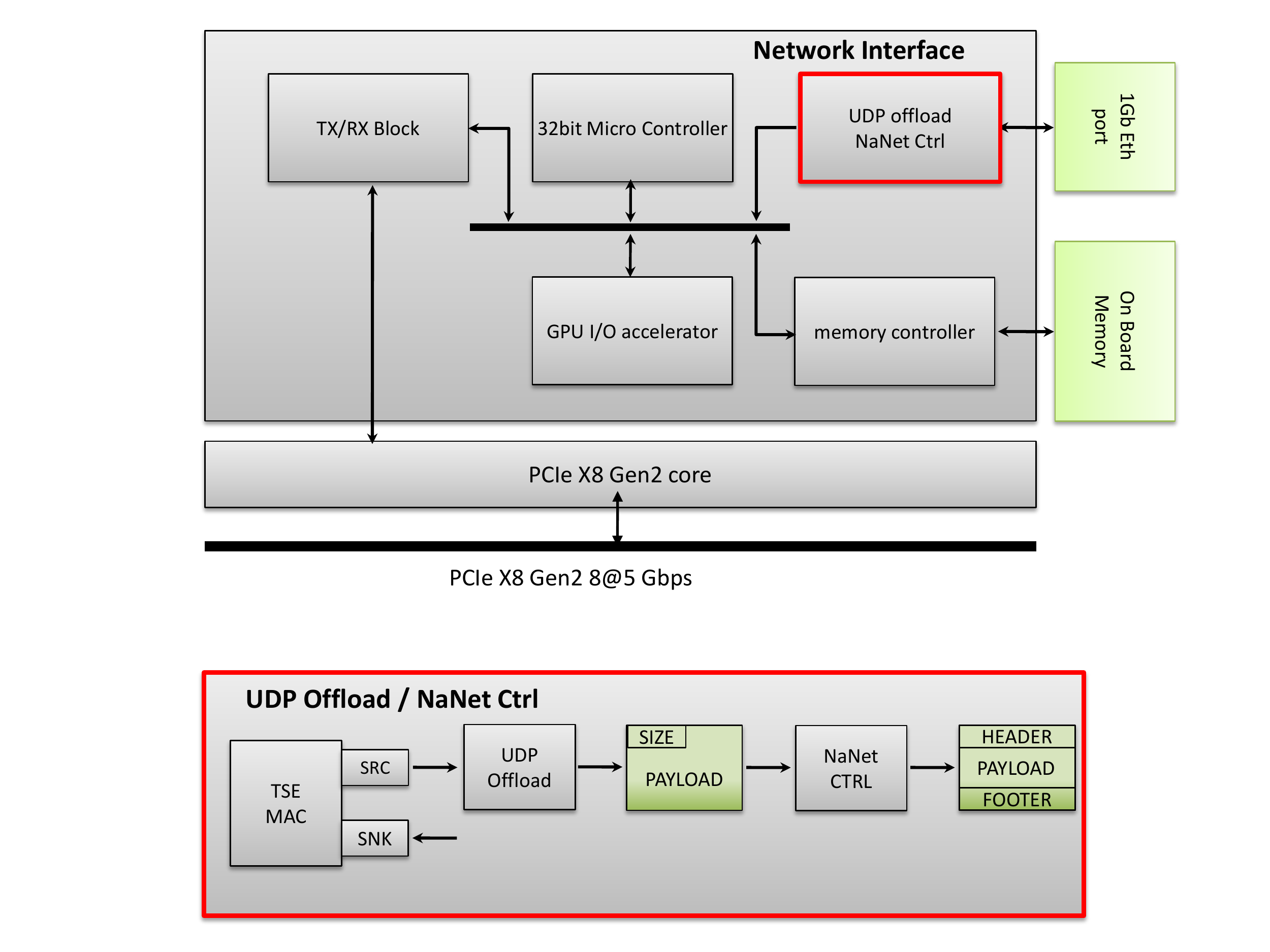}
  \caption{The NaNet Architecture}
  \label{fig:NaNet}
\end{figure}

%OSU~\cite{Traff:2012:OMB-GPU}
The \hw{UDP OFFLOAD} is an open core module \cite{Udp:2009:Online}.
It implements a method for UDP packet traffic offloading from a \nios
system such that can be processed in hardware rather than in
software. The open core design is built for a Stratix II 2SGX90
development board.
\hw{UDP OFFLOAD} collects data coming from the Avalon Streaming
Interface (\hw{Avalon-ST}) of the Altera Triple-Speed Ethernet
Megacore (\hw{TSE MAC}) and redirects UDP packets into a hardware
processing data path.
\nios subsystem executes the InterNiche TCP/IP stack to setup and tear
down high speed UDP packet streams which are processed in hardware at
the maximum data rate achievable over the GbE network.

Within the NaNet project we adapted the open core for our purposes.
Above all we ported the hardware code to the Stratix IV family. This
operation made possible to exploit the very improved performance of a
two technology steps forward FPGA.
In the Stratix II, the hardware components of the \nios subsystem are
synthesized at the operating frequency of 35~MHz. The limited
operating frequency forces to implement multiple channels to sustain
the data flow over a GbE Network.
The synthesis performed on a Stratix IV achieves the target frequency
of 200~MHz (in current implementation of \apenetp the \nios subsystem
operates at the same frequency).
The optimization allows to reduce the number of channels necessary to
sustain the data-flow. Current implementation of NaNet board provides
a single 32-bit width channel (at the present operating frequency it
achieves the capabilities of 6.4~gbps, 6 times greater of the GbE
requirements).

Data coming from the single channel of the modified \hw{UDP OFFLOAD}
are collected by the \hw{NaNet CTRL} (see figure \ref{fig:NaNet}).
\hw{NaNet CTRL} is an hardware module in charge of managing the GbE
flow by encapsulating packets in the typical \apenetp protocol
(\header, \payload, \footer).

The \hw{NaNet CTRL} main functionalities are:

\begin{itemize}
\item It implements an Avalon-ST Sink Interface (the hardware modules
of the \hw{UDP OFFLOAD} communicates through Avalon-ST protocol)
collecting the GbE data flow.
\item It generates the \header for the incoming data, analyzing the
UDP packet \header (message size) and several configuration registers
(see table \ref{tab:nanetconf} and table \ref{tab:nanetcmd}).
\item It parallelizes 32-bit data words coming from the \nios
subsystem into 128-bit \apenetp data words.
\item It redirects data-packets towards the corresponding FIFO (one
for \header and \footer and another for the \payload).  \hw{NaNet
HEADER FIFO} and \hw{NaNet DATA FIFO} are connected to the \hw{RX
BLOCK} that exploits the commands stored in the \header to perform the
GPU write transactions.
\end{itemize}

\begin{figure}[h!]
  \centering
  \includegraphics[trim=0mm 0mm 0mm 0mm,clip,width=0.8\textwidth]{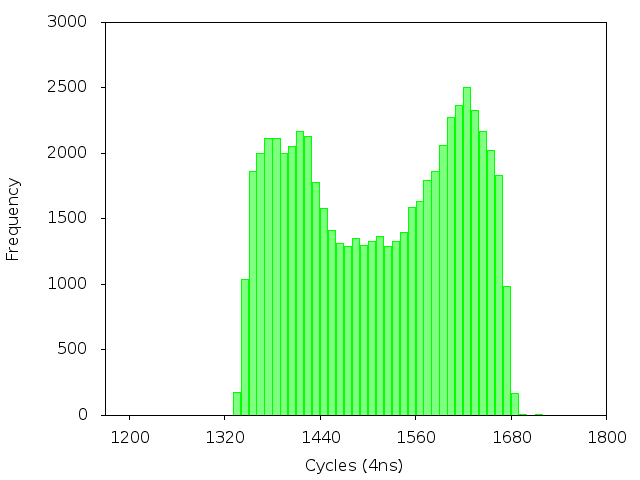}
  \caption{NaNet Latency; distribution plot over 60000 samples of a 
  NaNet packet traversal time}
  \label{fig:nanetlat}
\end{figure}

\begin{table}[!htb]
\centering
\begin{tabular}{|l|l|}
\hline
\hline
\multicolumn{2}{|c|}{\textbf{NaNet Configuration Register layout}} \\
\hline
Register Name             & Description\\
\hline 
NaNet RDMA virt addr LSB  & LSB of virtual address used in RDMA protocol\\
\hline
NaNet RDMA virt addr MSB  & MSB of virtual address used in RDMA protocol\\
\hline
NaNet RDMA command        & Commands to initialize the RDMA process\\
\hline
\hline
\end{tabular}
\caption{NaNet Configuration Register Overview.}
\label{tab:nanetconf}
\end{table}
\begin{table}[!htb]
\centering
\begin{tabular}{|l|l|l|}
\hline
\hline
\multicolumn{3}{|c|}{\textbf{NaNet RDMA command}}        \\
\hline
Bit Range & Name            & Description                \\
\hline
27 - 26   & port id         & process ID                 \\
\hline
28        & is gpu          & Identify a GPU packet      \\   
\hline
29        & gpuid           & Identify the target GPU    \\
\hline
30        & NaNet Last Frag & Identify the last part of a buffer \\
\hline
31        & NaNet Phys Addr & Enable the \nios bypass    \\ 
\hline
\hline
\end{tabular}
\caption{NaNet RDMA Command Register Layout.}
\label{tab:nanetcmd}
\end{table}
Preliminary benchmarks for latency and bandwidth were carried out.
Latency was measured adding timing info in the \apenetp \footer
with a resolution of 4~ns, and it shows that a packet traversal time 
ranges between 5.2~us and 6.8~us from input of \hw{NaNet CTRL} to the 
completion signal of the DMA transaction on the \PCIe bus 
(see figure \ref{fig:nanetlat}).

Currently we are investigating the spread of the measure.
Measured bandwidth was of about 120~MB/s, saturating the GbE channel 
capabilities.

We foresee several improvements on the \hw{NaNet} design:

\begin{itemize}
\item Adding a buffering stage to the \hw{NaNet CTRL}, enabling the
  coalescing of consecutive UDP payload data into a single \apenetp
  packet payload, improving bandwidth figure also with small-sized UDP
  packets.
\item Increasing the number of GbE channel in input, in order to
  sustain higher bandwidth demand coming from the experimental
  requirements.
\item Moving towards a 10-GbE data link interface.
\end{itemize}
%
%\vspace{5cm}

%% file: chapter4.tex
\section{Design of a Software-Programmable DNP architecture using ASIP technology}
\label{sec:target2}
\euretile's many-core platforms critically depend on efficient
packet-based communication between tiles in the 3D network.
Efficiency refers to both high-bandwidth and low-latency. To reduce
the latency, a significant acceleration of networking functions is
needed. Target and INFN are jointly investigating the use of ASIP
technology for this purpose, which will result in a low-latency
software-programmable 'DNP ASIP'.  This section of the report
describes the design of a first version of this DNP ASIP, which was
delivered in 2012. The design was made using IP Designer, a
retargetable tool-suite for the design and programming of ASIPs~\cite{Goossens:2011:ASIP}.

\subsection{Introduction}% (Target, Alessandro)}
The purpose of the new DNP ASIP is to accelerate time-critical network
tasks. By providing the right instruction set specialization, a
significant reduction can be obtained of the latency of key network
management tasks.  We focus on the APEnet+ network architecture. In
\apenetp, each network node has two compute processors: an Intel x86
processor and an NVidia GPU. Network related tasks are offloaded onto
a Distributed Network Processor (DNP). The DNP acts as an off-loading
network engine for the computing node, performing inter-node data
transfers. Currently, a DNP contains a \nios processor. The firmware
on the \nios implements the following tasks~\cite{euretile:D6_1}:

\begin{itemize}
\item Command processing. Commands are exchanged between the \nios and
the DNP core via command FIFOs.
\item Packet RX, reading incoming packet headers in \hw{FIFO REQ} and
computing the final destination memory address, be it on either the
host or the GPU, of the packet data.
\item GPU TX, reading the virtual address and length of the source GPU
memory buffer from \hw{FIFO P2P TX} and stimulating the GPU to send
the data to the \hw{GPU PKT CTRL}.
\end{itemize}

\subsection{The DLX template for architectural exploration}% (Target, Alessandro)}
A first step in the development of an ASIP architecture is the
selection of a suitable basic processor template. The characteristics
of the template should be aligned with the requirements of the ASIP
architecture. For the acceleration of the DNP tasks, the following
characteristics are important:

\begin{itemize}
\item The architecture should have a fairly deep pipeline. This
requirement is motivated by the fact that the architecture will be
mapped onto an FPGA. Compared to a standard cell based mapping, on an
FPGA the delay on the interconnections between architecture elements
is long. It is then required to insert pipeline registers on these
connections.
\item The applications use 32 bit and 64 bit data. The ideal template
architecture should therefore have hardware support for 32 bit and 64
bit operations.
\item The applications contain quite a lot of control code. The
template architecture should therefore be able to efficiently support
the full complement of C language constructs.
\item The applications make use of C library functions such a \malloc.
The template architecture should therefore come with a C run time
library that supports these functions.
\end{itemize}

Based on these requirements, we have selected the DLX processor
template. DLX is the RISC processor architecture that is used as an
example processor in the seminal book on computer architecture of John
Hennessy and David Patterson~\cite{Patterson:2011:Arch}. The DLX processor has been modeled
in nML, the processor description language supported by the IP
Designer tool-suite.

DLX is a reduced instruction set architecture (RISC) with the
following features:

\begin{itemize}
\item 32 bit wide data path, with an ALU, shifter and multiplier.
\item 32 bit wide instruction word with an orthogonal, easy to decode,
instruction encoding.
\item 32 field central register file.
\item a load/store architecture, which supports 8, 16 and 32 bit
memory transfers and an indexed addressing mode.
\item a 5 stage pipeline with separate stages for operand reads and
writes.
\end{itemize}

More details of the DLX architecture are provided below.

\subsubsection{Instruction Formats}
Being a RISC architecture, DLX has only three main instruction
formats. These are:
\begin{itemize}
\item The \textbf{R format}. This format is used for three register
instructions. It has two fields, s1 and s2, to specify operand
registers; one field, d, for the destination register, and a function
field to select the instruction. For all current R type instructions,
the opcode field is set to zero.
\item The \textbf{I format}. This format is used for two register
instructions with a 16 bit immediate operand. It has two fields, s1
and d, to specify operand or destination registers; and an opcode
field to select the instruction.
\item The \textbf{J format}. This format is used for control flow
instructions with a 26 bit immediate operand.
\end{itemize}

\begin{figure}[!hbt]
  \centering
  \includegraphics[trim=0mm 60mm 0mm 60mm,clip,width=\textwidth]{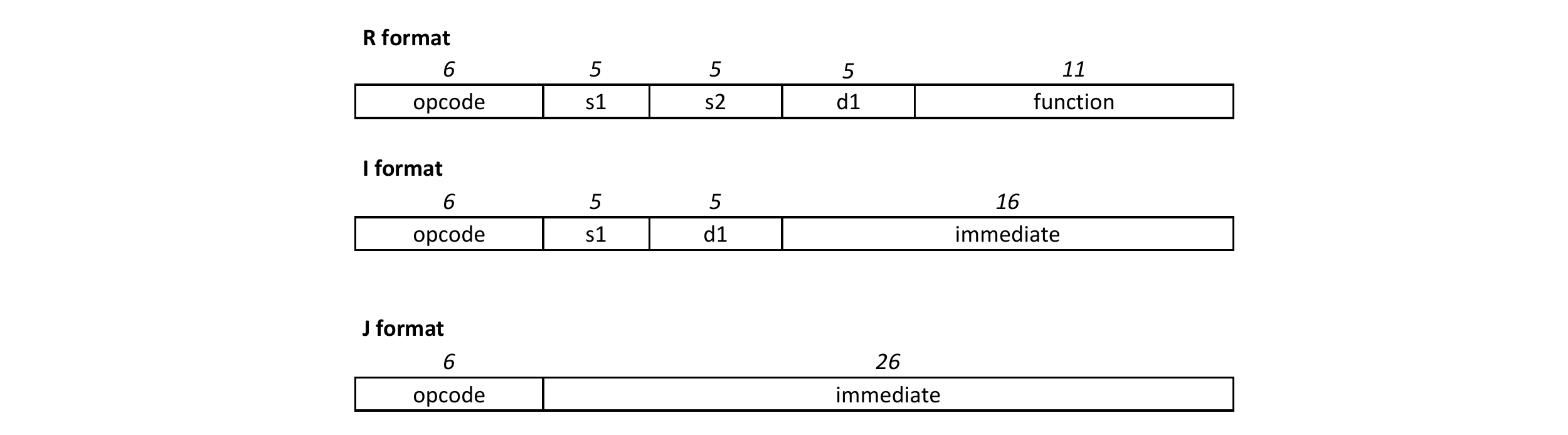}
  \caption{Instruction formats and encoding of DLX processor}
  \label{fig:DLXinst}
\end{figure}

The encoding of the opcode and function fields is specified in figure
~\ref{fig:DLXinst}.

\subsubsection{Pipeline}
DLX has a 5 stage pipeline, with the following stages:

\begin{enumerate}
\item \textbf{Fetch Stage (IF)}
  \begin{enumerate}
  \item A new instruction is fetched from program memory and is
  issued.
  \end{enumerate}
\item \textbf{Decode Stage (ID)}
  \begin{enumerate}
  \item The instruction is decoded and the operands are read from the
  register file.
  \item The unconditional jump instruction also executes in this
  stage: the target address is sent to program memory.
  \end{enumerate}
\item \textbf{Execute Stage (EX)}
  \begin{enumerate}
  \item This is the stage in which the ALU and shifter units execute
  their operation.
  \item The pipelined multiply unit starts operating in this stage
  (the multiplications are finished in the next stage).
  \item The multi-cycle iterative division is started in this stage.
  \item For memory load operations, the effective address is computed
  and is sent to the memory. For store operations both address and
  data are sent to the memory. The load or store operation is started.
  \item The conditional and indirect jump instructions execute in this
  stage.
  \end{enumerate}
\item \textbf{Memory Access Stage (ME)}
  \begin{enumerate}
  \item The multiply operations are completed.
  \item The result of memory load operations is available on the data
  bus.
  \end{enumerate}  
\item \textbf{Write Back Stage (WB)}
  \begin{enumerate}
  \item The result of ALU, shift, multiply or load operations is
  written to the destination field on the register file.
  \end{enumerate}  
\end{enumerate}

\subsubsection{Data Path}
\begin{figure}[!hbt]
  \centering
  \includegraphics[trim=50mm 45mm 50mm 45mm,clip,width=\textwidth]{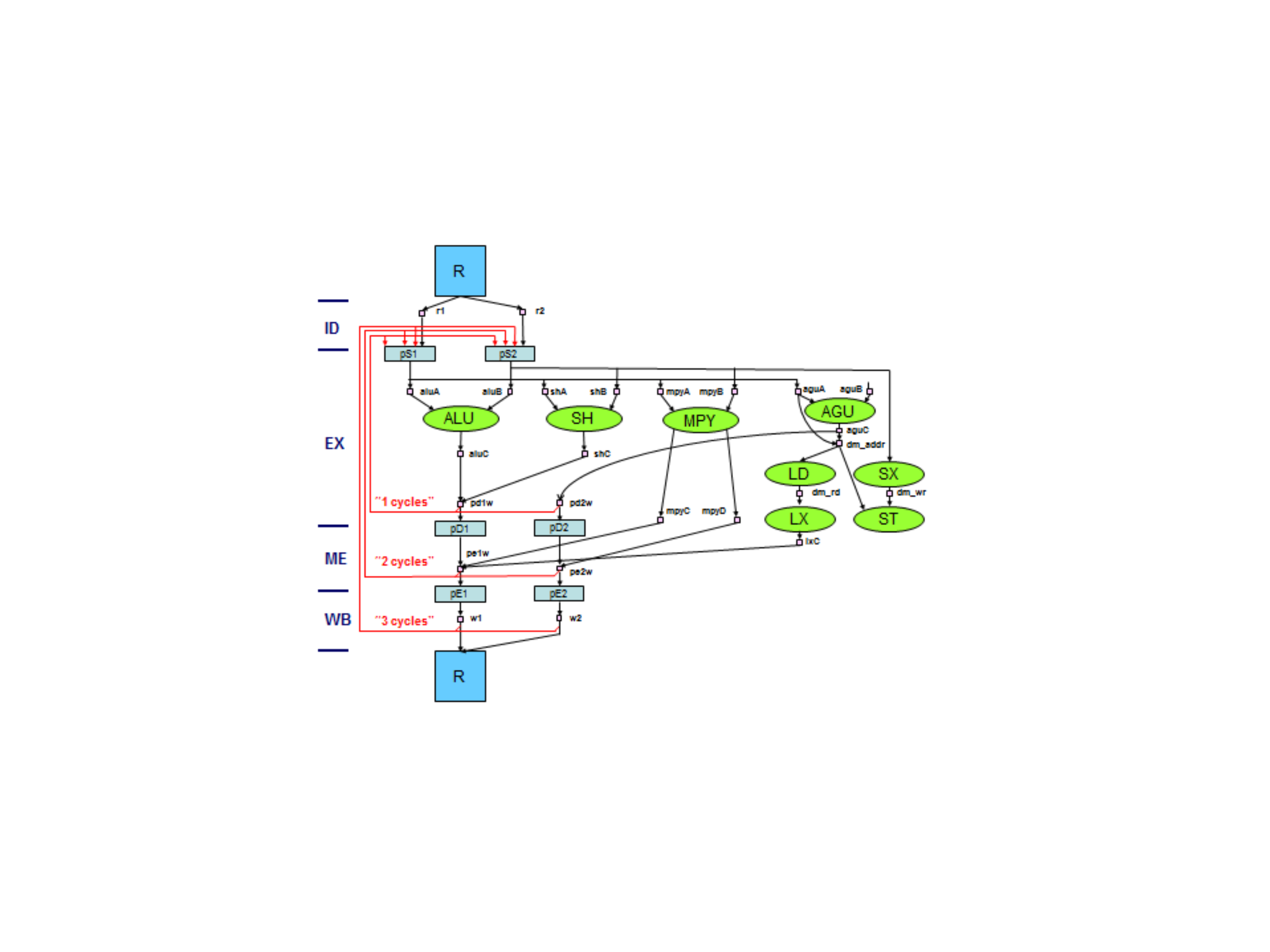}
  \caption{Data path of the DLX processor.}
  \label{fig:DLXdatapath}
\end{figure}

Figure \ref{fig:DLXdatapath} shows the data path of the DLX processor, with the names of
the main functional units and pipeline registers. In the ID stage the
operands are read from \hw{R}, using ports \hw{r1} and \hw{r2}, and
are stored in pipeline registers \hw{pS1} and \hw{pS2}.

In EX, the outputs of the pipeline registers are connected to the
inputs of the functional units, \hw{aluA} and \hw{aluB}, \hw{shA} and
\hw{shB}, \hw{mpyA} and \hw{mpyB} or \hw{aguA} and \hw{dm\_wr};
depending on the instruction that is being executed.

The \hw{ALU} and shifter (\hw{SH}) produce results at the end of EX.
Their results are then stored in \hw{pD1} and \hw{pE1} before being
written to the register file.

The 32x32 bit pipelined multiplier produces a full 64 bit result, with
the low part in \hw{mpyC} and the high part in \hw{mpyD}. These two
parts become available at the end of ME and are stored in \hw{pE1} and
\hw{pE2} before being written to the register file.

The result of a load is also available in ME. It goes through a
sign/zero extension unit \hw{SX}, and is then stored in \hw{pE1}
before it is written to the register file.

In case of a store operation the relevant part of the data is
extracted in the \hw{SX} unit.

For both load and store operations, the effective address is computed
on the \hw{AGU}. For the indexed addressing mode, an immediate offset
is assigned to \hw{aguB}, and is added to the address of \hw{aguA}.
The output \hw{aguC} is copied to the address bus \hw{dm\_addr}. For
indirect addressing with post-increment, the \hw{aguA} input is copied
to the address bus, and is also post-incremented by the value on
\hw{aguB}.

Operands are fetched from the register file in an early stage (ID) and
written back three stages later (in WB). Consider a pair of data
dependent instructions \textbf{A} and \textbf{B}: \textbf{A} produces
a result in \hw{r3}, and \textbf{B} uses this result. The \textbf{B}
instruction can only read its operand if it is scheduled as the fourth
instruction after \textbf{A}. Therefore three independent instructions
(possibly NOP instructions) must be scheduled between \textbf{A} and
\textbf{B}.

\subsubsubsection{A. Bypasses}
By adding register bypasses, it becomes possible to schedule
\textbf{A} and \textbf{B} closer to each other. Register bypasses are
connections that are added to the data path so that a result can be
fed back to the operand pipeline registers as soon as it is available.
For the \hw{ALU} and \hw{shifter} results, three bypasses can be
added:

\begin{itemize}
\item When the \textbf{A} instruction is in the EX stage, the result
is produced on the pd1w wire. It can be used by the \textbf{B}
instruction in the next cycle.  This requires that \hw{pd1w} is
written into \hw{pS1} or \hw{pS2}.
\item When the \textbf{A} and \textbf{B} instructions are scheduled
two cycles apart, the result is present on the \hw{pe1w} wire. It can
be used by the \textbf{B} instruction when \hw{pe1w} is written into
\hw{pS1} or \hw{pS2}.
\item When the \textbf{A} and \textbf{B} instructions are scheduled
three cycles apart, the result is present on the \hw{w1} transitory.
It can be used by the \textbf{B} instruction when \hw{w1} is written
into \hw{pS1} or \hw{pS2}.
\end{itemize}

\subsubsubsection{B. Stalls}
Not all data hazards can be avoided by means of bypasses. The result
of the pipelined multiplier, and the data that is loaded from memory
are available only in the ME stage. A 1 cycle bypass is therefore not
possible. The 2 and 3 cycles bypasses do apply here. To deal with the
data hazard for the case of a 1 cycle delay, hardware stalls are used.
A hardware stall means that the issuing of the dependent instruction
is postponed until the hazard is gone.

\subsubsection{Instructions}

The DLX instruction set is partitioned into three groups:
\begin{itemize}
\item arithmetic instructions;
\item load/store instructions;
\item control flow instructions.
\end{itemize}
These are explained in this section.

\subsubsubsection{Arithmetic instructions}

The following arithmetic instructions are supported:

\textit{Additive Instructions}:
\begin{itemize}
\item addition (\sw{add});
\item subtraction (\sw{sub});
\item addition with carry (\sw{addx});
\item subtraction with carry (\sw{subx});
\item addition and subtraction with a signed immediate (\sw{addi} and
\sw{subi});
\item addition and subtraction with a signed immediate and carry
(\sw{addix} and \sw{subix});
\item addition and subtraction with an unsigned immediate (\sw{addui}
and \sw{subui});
\item addition and subtraction with an unsigned immediate and carry
(\sw{adduix}, and \sw{subuix});
\end{itemize}

\textit{Bitwise Instructions}:
\begin{itemize}
\item bitwise \sw{AND};
\item bitwise \sw{OR};
\item bitwise \sw{exclusive OR};
\end{itemize}

\textit{Shift instructions}:
\begin{itemize}
\item logical shift left (\sw{sll}); 
\item arithmetic shift right (\sw{sra});
\item logical shift right (\sw{srl});
\end{itemize}

\textit{Compare instructions}:
\begin{itemize}
\item set when equal (\sw{seq} and \sw{seqi});
\item set when greater or equal (\sw{signed compare}) (\sw{sge} and
\sw{sgei});
\item set when greater or equal (\sw{unsigned compare}) (\sw{sgeu} and
\sw{sgeui});
\item set when greater (\sw{signed compare}) (\sw{sgt} and \sw{sgti});
\item set when greater (\sw{unsigned compare}) (\sw{sgtu} and
\sw{sgtui});
\item set when less or equal (\sw{signed compare}) (\sw{sle} and
\sw{slei});
\item set when less or equal (\sw{unsigned compare}) (\sw{sleu} and
\sw{sleui});
\item set when less (\sw{signed compare}) (\sw{slt} and \sw{slti});
\item set when less (\sw{unsigned compare}) (\sw{sltu} and
\sw{sltui});
\item set when not equal (\sw{sne} and \sw{snei});
\end{itemize}

\subsubsubsection{Load and Store Instructions} 
The data memory
supports 8, 16 and 32 bit wide load and store operations. Note that 16
bit values must be stored at addresses that are even, and 32 bit
values at addresses that are a multiple of 4.

Five different load operations are possible:
\begin{itemize}
\item word load (\sw{lw});
\item signed half word load (\sw{lh});
\item unsigned half word load (\sw{lhu});
\item signed byte load (\sw{lb});
\item unsigned byte load (\sw{lbu});
\end{itemize}

For the signed loads, the 8 or 16 bit value is sign extended and then
stored in the destination register. For the unsigned loads, the 8 or
16 bit value is zero extended.

Three different store operations are possible: word store (\sw{sw}),
half word store (\sw{sh}), and byte store (\sw{sb}). For the 8 and 16
bit stores, the least significant bits of the source register are
stored in memory.

The load and store operations can be combined with indexed addressing
mode, \sw{lw d1,i(s1)} or \sw{sw s2,i(s1)}, the 16 bit signed
immediate \sw{i} is added to the content of register \hw{s1}, to
obtain the effective address. Indexed load store instructions use
\textbf{format I}.

\subsubsubsection{Control flow instructions} 
The DLX core supports
conditional branch instructions with a PC-relative target address,
unconditional jumps with either PC-relative or register indirect
target address, and unconditional jumps with subroutine linkage.
\begin{itemize} 
\item The conditional branch instructions \sw{beqz s1,\#imm} and
\sw{bnez s1,\#imm} implement a conditional jump to the target
instruction T, at address PC + \sw{imm}, where \sw{imm} is a 16 bit signed
offset. The branch is taken when the operand register \hw{s1},
contains zero for the \sw{beqz} instruction, or a non-zero value for
the \sw{bnez} instruction. Conditional branches require three cycles
before the target instruction can be issued. In the first cycle, the
branch is issued; the following two cycles are filled with two delay
slot instructions. The \sw{imm} offset is relative to the address of
the second delay slot instruction.
\item The \sw{j \#imm} instruction implements an unconditional jump to
the target address PC + \sw{imm}, where \sw{imm} is a 26 bit signed
offset. It takes two cycles to complete and has one delay slot.  The
\sw{imm} offset is relative to the address of the delay slot
instruction.
\item For the \sw{jr} instruction, the absolute target address is
obtained from the register specified by \hw{s1}. The instruction takes
three cycles to complete and has two delay slots.
\item For the \sw{jal} instruction, the target address is obtained by
adding a 26 bit signed immediate to the PC. The instruction takes two
cycles to complete and has one delay slot. The \sw{imm} offset is
relative to the address of the delay slot instruction. The return
address is the address of the instruction after the delay slot
instruction and is stored in the \hw{LR} register.
\item For the \sw{jalr} instruction, the target address is obtained by
adding the register specified by \hw{s2} to the PC. The instruction
takes three cycles to complete and has two delay slots.  The return
address is the address of the instruction after the two delay slot
instructions and is stored in the \hw{LR} register.
\end{itemize}

\subsection{The Buffer Management Application}
Buffer management is a part of Remote Direct Memory Access (RDMA).
RDMA allows to directly read/write information in the memory of
another processor with minimal demands on memory bus bandwidth and CPU
processing overhead. The \mbox{micro-controller} in the DNP simplifies
the architecture by implementing in firmware some RDMA tasks. Its role
is to manage the RDMA look-up table allocated in the on-board memory,
with the following basic operations:

\begin{itemize}
\item To add/delete entries in case of register/unregister buffer
operations;
\item To retrieve the appropriate entry to satisfy buffer info
requests for the incoming \sw{DNP\_put} or \sw{DNP\_get} operands.
\end{itemize}

A buffer is characterized by the following record

\begin{verbatim}
typedef struct buf_desc {
        addr64_t      VirtAddr;
        u32           Len;
        u32           Flags;
        addr64_t      MagicWord;
} buf_desc_t
\end{verbatim}

Elements of this type are stored in some form of storage. In a pure
software based implementation they are stored in a double linked list
of which the nodes are dynamically allocated. In the ASIP
implementation, we will add a dedicated hardware structure to store
the \sw{buf\_desc} information.

There is a software API for manipulating the pool of buffer
descriptors. The most important functions of that API are the
following:

\begin{itemize}
\item Add an element to the pool of buffers;
\begin{verbatim}
int append_persistent_buf(buf_lists_t *bls, 
      addr64_t VirtAddr, 
      u32 Len, 
      u32 Flags, 
      addr64_t MagicWord)
\end{verbatim}
\item Check if an address range given by a start and end address lies
within one of the buffers of the bool, and return the properties of
that buffer.
\begin{verbatim}
int search_persistent_buf_by_range(buf_lists_t *bls, 
                                   addr64_t start_addr,
                                   addr64_t end_addr,
                                   buf_desc_t *buf)   
\end{verbatim}
\item Search for a buffer based on the parameters start address and
length, and remove the buffer from the pool when found.
\begin{verbatim}
int remove_persistent_buf(buf_lists_t *bls, 
                          addr64_t start_addr, 
                          size_t len)
\end{verbatim}
\end{itemize}

For the development and the bench marking of an ASIP architecture, it
is useful to have a benchmark program for these API calls. The
benchmark that we use in our design contains the following API calls:

\begin{itemize}
\item First 32 buffers are added (\sw{append\_persistent\_buf});
\item Next three buffers are searched:
  \begin{itemize}
  \item the buffer that was added first;
  \item the buffer that was added as 16th buffer;
  \item the buffer that was added last;
  \end{itemize}
\item The first, 16th and final buffers are removed;
\item The 16th buffer, which is no longer present in the pool is
searched again;
\end{itemize}

This benchmark will give a good impression of the average, minimum and
maximum execution times for the three main operations.
Table~\ref{tab:cyclecnt} shows the cycle counts for the execution of
the benchmark program on different processor architectures, for a
typical set of data. These results will be discussed further in the
next sections.

\begin{table}[!hbt]
\centering
\setlength\extrarowheight{2pt}
\begin{tabular}{|c|cccccc|}
\hline
\hline
\textbf{Operations}       & \textbf{\nios} & \textbf{DLX} & \textbf{D64} & \textbf{D64AC} & \textbf{D64SB} & \textbf{D64OPT}  \\
\hline
\textbf{Append 0...31}    &  6839          &  4867        &  4836        & 4836           &   418          & 419  \\
\textbf{Search 0, 16, 31} &  1802          &  2454        &  1938        & 1644           &   794          & 304  \\
\textbf{Remove 0, 16, 31} &  1192          &  3523        &  2732        & 2438           &  1504          & 560  \\
\textbf{Search 16}        &   596          &   787        &   517        &  455           &   483          & 166  \\
\textbf{Total}            & 10433          & 11631        & 10023        & 9373           &  3199          & 1449 \\
\hline
\hline
\end{tabular}
\caption{Measured Cycle counters for benchmark program on different
processor architecture.}
\label{tab:cyclecnt}
\end{table} 

\subsection{Architectural exploration using IP Designer}
The IP Designer tool-suite was used to design and optimize the DNP
ASIP architecture, starting from the initial DLX implementation. The
architectural exploration process is described in the next sections.

\subsubsection{Implementation on \nios and DLX32}
For the pure software implementation, the pool of buffers is
implemented as a double linked list, of which the elements are
dynamically allocated on the heap. The reference code makes use of the
type agnostic list implementation of the Linux kernel code
(\sw{include/linux/list.h}). In order to implement the double linked
list, a \sw{list\_head} member is added to the buffer descriptor

\begin{verbatim}
struct list_head {
        struct list_head *next, *prev;
};
typedef struct buf_desc {
        struct list_head node;
        addr64_t      VirtAddr;
        u32           Len;
        u32           Flags;
        addr64_t      MagicWord;
} buf_desc_t;
\end{verbatim}

The results obtained for the benchmark code on the \nios architecture
are captured in first column of Table~\ref{tab:cyclecnt}.

The results obtained for the benchmark code on the DLX architecture
are captured in second column of Table~\ref{tab:cyclecnt}. These
results will serve as a reference to measure the speedup of
architecture specializations.

From the profiling information that is produced by the ChessDE
debugger (Table~\ref{tab:chessde}), we can learn that a lot of time,
almost 35\%, of the total number of cycles, is spent in the memory
allocation when adding new buffers to the pool.

\begin{table}[!hbt]
\centering
\setlength\extrarowheight{2pt}
\begin{tabular}{|l|cc|}
\hline
\hline
\textbf{Function}                       & \textbf{Cycles} & \textbf{\% of total}  \\
\hline
\sw{malloc}                             & 2215            & 34.65\%               \\
\sw{search\_buf\_by\_range\_nocopy}     & 1528            & 23.90\%               \\
\sw{append\_buf}                        &  940            & 14.70\%               \\
\sw{memcpy}                             &  702            & 10.98\%               \\
\sw{main \_main}                        &  490            &  7.66\%               \\
\sw{remove\_buf}                        &  214            &  3.35\%               \\ 
\hline
\hline
\end{tabular}
\caption{ChessDE profiling information.}
\label{tab:chessde}
\end{table} 

Another important observation is that the application contains many 64
bit operations. These are emulated as dual precision operations,
making use of 32 bit instructions. For example, the important address
range check

\begin{verbatim}
if(p->VirtAddr<=start_addr && end_addr<=(p->VirtAddr+p->Len-1))
\end{verbatim}

requires 23 DLX instructions to implement it.

\subsubsection{Implementation on D64}
A first important step that is required to achieve a speedup of the
buffer management code is the extension of the data width of the
processor. A variant of the DLX architecture, which we call
\textbf{D64}, has been created to achieve this purpose. The
\textbf{D64} has the following features:

\begin{itemize}
\item The width of the central register file is increased from 32 to
64 bit. This allows to store 64 bit data in a single register field.
The 64 bit values can be used to store the virtual addresses that are
present in the packet headers.
\item The width of the ALU has been increased from 32 to 64 bit.
While all 32 bit op codes are still supported, new opcodes have been
added to the following operations:
  \begin{itemize}
  \item long addition (\sw{ladd});
  \item long subtraction (\sw{lsub});
  \item long addition with 16 bit signed and unsigned immediate
  operand (\sw{laddi}, \sw{laddui});
  \item long subtraction with 16 bit signed and unsigned immediate
  operand (\sw{lsubi}, \sw{lsubui});
  \item long logical left shift (\sw{lsll});
  \item long logical left shift with immediate shift factor
  (\sw{lsll});
  \item long signed and unsigned compare instructions (\sw{lsge}, 
  \sw{lsgeu}, \sw{lsgt}, \sw{lsgtu}, \sw{lsle}, \sw{lsleu}, \sw{lslt},
  \sw{lsltu}).
  \item long signed and unsigned compare instructions with immediate
  operand (\sw{lsgei}, \sw{lsgeui}, \sw{lsgti}, \sw{lsgtui},
  \sw{lslei}, \sw{lsleui}, \sw{lslti}, \sw{lsltui}).
  \end{itemize}
\item The bitwise logical operations \sw{AND}, \sw{OR} and \sw{XOR},
are only present in a 64 bit form. These 64 operations can also be
used to implement 32 bit bitwise operations.
\item The arithmetic and logical right shift operations are only
present in a 64 bit form. These 64 operations can also be used to
implement 32 bit shift operations.
\item The size of the multiplier is kept at 32x32 bit. Indeed,
multiplication is not a critical operation in the buffer manipulation
code. Allocating a 64x64 bit multiplier would significantly increase
the size of the design. It would also have a negative impact on the
achievable clock frequency, or would require a deeper pipeline.
\item The width of the data memory interface has been extended to 64
bit, and 64 bit load and store instructions have been added.
\end{itemize}

Concerning the use of the 64 bit architecture by the C compiler,
following decisions were made:

\begin{itemize}
\item The 64 bit data type is used to implement the C built in types
\sw{long long} and \sw{unsigned long long}. The \sw{unsigned long
long} type in its turn, is used in the buffer manipulation code to
represent 64 bit addresses.
\item The 32 bit type is kept for representing the \sw{int},
\sw{unsigned}, \sw{long} and \sw{unsigned long} C types, as well as
pointers (the D64 still has a 32 bit address bus).
\end{itemize}

The results obtained for the benchmark code on the D64 architecture
are captured in third column of table Table~\ref{tab:cyclecnt}.
\begin{itemize}
\item The append function does not benefit from the 64 bit
instruction. This is because most of the time is spent in \sw{malloc}.
The \sw{malloc} function still exclusively uses 32 bit data and
instructions. This is so because addresses on D64 are still 32 bit,
which is sufficient because the data structures that need to be
manipulated are limited in size.
\item The search and remove functions do benefit from the 64 bit
instructions. The address range check now requires 10 instructions.
\end{itemize}

\subsubsection{Address Compare Intrinsic}
To further specialize the D64 architecture we will add an address
compare instruction. A new functional unit, called \hw{ACU},
specialized for doing 64 bit address range compares is added. The unit
implements the following operation:

\begin{verbatim}
w64 check_addr_in_range(w64 start, w64 end, w64 VirtAddr, w64 Len)
{
   return VirtAddr <= start && end <= (VirtAddr + Len - 1);
}
\end{verbatim}

The register file of the DLX architecture has only two read ports. The
address compare unit on the other hand, needs four operands. Because
it would be too costly to add two additional read ports for accessing
the complete register file, we have decided to read the \sw{VirtAddr}
and \sw{Len} parameters from dedicated fields R12 and R13. This is
modeled as follows in the processor model:

\begin{verbatim}
opn rdS1(r: c5u) { 
   action: stage ID: pS1=r1=R[r]; syntax: "r"r; image: r; 
}
opn rdS2(r: c5u) { 
   action: stage ID: pS2=r2=R[r]; syntax: "r"r; image: r;
}
opn rdS12() { 
   action: stage ID: pS12=r12=R12; syntax: "r12"; 
}
opn rdS13() { 
   action: stage ID: pS13=r13=R13; syntax: "r13"; 
}

opn alu_check_addr(d1: wrD1, s1: rdS1, s2: rdS2, s12: rdS12, s13: rdS13)
{
    action {
    stage ID:   // read operands
        s1;
        s2;
        s12;
        s13;
    stage EX:  // execute range check
        acuA=pS1;        
        acuB=pS2;
        acuC=pS12;
        acuD=pS13;
        acuX = check_addr_in_range(acuA,acuB,acuC,acuD);
        pd1w = acuX;
    stage EX..WB:  // write back result
        d1;
    }
    syntax : "ac " d1 "," s1 "," s2 "," s12 "," s13;
    image  : opcode.function::s1::s2::d1::function_code.ac;
}
\end{verbatim}

As can be seen in column D64AC of table Table~\ref{tab:cyclecnt},
adding the \hw{ACU} has little impact in the cycle count. The unit
will however be an important component in the following architecture.

\subsubsection{Dedicated Storages for the Buffer Table}
The main inefficiencies that still remain can be attributed to the
fact that the pool of buffers is still allocated as a linked list in
memory.
\begin{itemize}
\item A first consequence is that the append operation calls
\sw{malloc}, which is time consuming.
\item Secondly, the \sw{VirtAddr} and \sw{Len} inputs of the address
range check need to be loaded from memory. Since DLX has only one
memory read port, these reads are executed sequentially.
\item Going from one list element to the next during a search 
operation requires an additional load of the \sw{next} pointer.
\end{itemize}

\begin{figure}[!hbt]
  \centering
  \includegraphics[trim=40mm 65mm 40mm 65mm,clip,width=\textwidth]{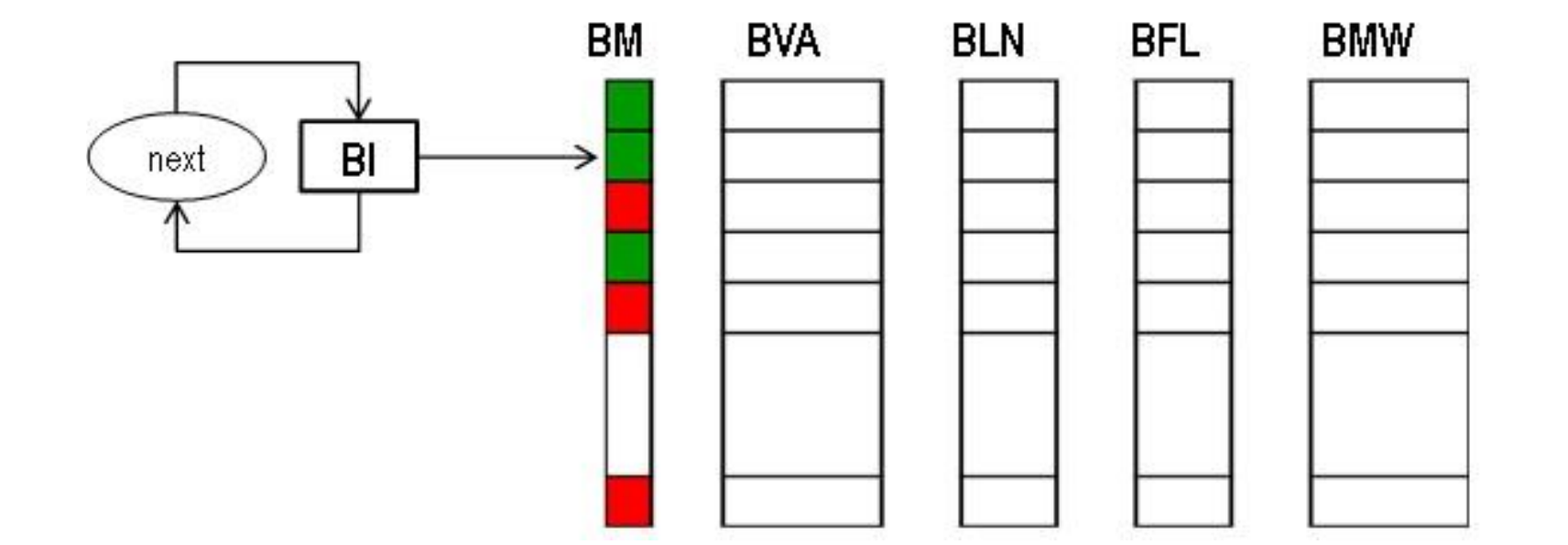}
  \caption{Register structure for buffer storage.}
  \label{fig:regfile}
\end{figure}

A fundamental solution to the problem is to allocate the buffer
descriptors in a dedicated storage, in such a way that there is
sufficient bandwidth to access all operands of the range check in
parallel. For the application domain of scientific computing, it has
been established that typical applications use only a limited number
of large buffers. Typically, a network node will need to keep track of
10 to 20 buffers. Based on this property, we decided to implement the
\sw{buf\_desc} record in dedicated registers files of 32 fields
each. These register files, \sw{BVA}, \sw{BLN}, \sw{BFL} and \sw{BMW}
are shown in figure~\ref{fig:regfile}.

The \sw{BVA} register file is 64 bit wide and holds the virtual
address of a buffer (the \sw{VirtAddr} member of \sw{buf\_desc}).
 
The \sw{BLN} register file is 32 bit wide and holds the length of a
buffer (the \sw{Len} member of \sw{buf\_desc}).

The \sw{BFL} register file is 32 bit wide and holds the flags
associated to a buffer (the \sw{Flags} member of \sw{buf\_desc}).
 
The \sw{BMW} register file is 64 bit wide and holds the magic word of
a buffer (the \sw{MagicWord} member of \sw{buf\_desc}).
 
In addition to these data registers, there is a single 32 bit register
\sw{BM} (\sw{buffer mask}). Each \sw{BM} bit indicates whether the
corresponding fields in the \sw{BVA}, \sw{BLN}, \sw{BFL} and \sw{BMW}
register are in use.
 
The \sw{BI} register is a 5 bit pointer register that is used to
select a common field in the \sw{BVA}, \sw{BLN}, \sw{BFL} and \sw{BMW}
registers, as well as select a bit of the BM mask.

Finally, there is a condition register \sw{BC}. \sw{BC} is set or
cleared by the instructions that manipulate the buffer storages to
indicate whether the instruction was successful or not.

\subsubsection{Instructions}
In this section, we describe the instructions that were added to 
manipulate the buffer storages.

\textit{Instructions needed by the append operation}.

\begin{itemize}
\item \textbf{buffree} This instruction inspects the \sw{BM} registers
and looks for the first unused entry. When an unused entry is found,
its index is written to \sw{BI} and the corresponding \sw{BM} bit is
set to indicate that a new buffer descriptor will be written to this
location. The instruction indicates that a free location was found by
setting the \sw{BC} condition register. When no unused entry was
found, a false condition is written to \sw{BC}. The C language
intrinsic for this instruction is \sw{buf\_free()}.
\item \textbf{bufadd} This instruction implements a 4-way parallel
write, from the central register file, to the buffer descriptor fields
that are pointed to by \sw{BI}. The \sw{VirtAddr} and \sw{Len}
parameters can be read from any field of the central register file.
The \sw{Flags} and \sw{MagicWord} parameters are read from dedicated
registers \hw{R12} and \hw{R13} respectively.
\item \textbf{bbc/bnbc} This is a conditional jump, which jumps if the
\sw{BC} register is set (bbc) or not set (bnbc).
\end{itemize}

The C implementation of the add operation is given below:

\begin{verbatim}
buf_add(unsigned long long virt_addr,
        unsigned len,
	   unsigned flags,
	   unsigned long long magic_word)
{
    bool ok = buf_free();
    if (ok) {
        *pVA = virt_addr; // pVA is a pointer to BVA
        *pLN = len;
        *pFL = flags;
        *pMW = magic_word;
    }
    return ok;
}
\end{verbatim}

This C code maps into the following instruction sequence:

\begin{verbatim}
buffree
bnbc #next
bufadd r4,r5,r12,r13
next: ...
\end{verbatim}

\textit{Instructions needed by the append operation}.

\begin{itemize}
\item \textbf{mv bi,<imm5>} This instruction loads a 5 bit immediate
value into the \sw{BI} register. It is mainly used at the start of a
search to set \sw{BI} to zero.
\item \textbf{bufnxt} This instruction inspects the BM registers and
looks for the first used entry. When a used entry is found, its index
is written to \sw{BI}. The instruction then also sets the \sw{BC}
register. When no used entry was found, a false condition is written
to \sw{BC}. The C intrinsic for this instruction is called
\sw{buf\_next()}.
\item \textbf{bufmv Rn,[bva,bln]} This instruction moves the fields of
the \sw{BVA} and \sw{BLN} registers filed that are pointer to by the
current \sw{BI} index to a filed in the central register file.
\item \textbf{bufrng} This instruction performs the address range 
check on operands that are stored in the central register file. The
matching condition is written to the central register file. The C 
intrinsic for this instruction is called \sw{buf\_check\_addr\_in\_range()}.
\end{itemize}

The C implementation of the search operation is given below:

\begin{verbatim}
inline bool buf_search(unsigned long long start_addr, 
                       unsigned long long end_addr,
		             unsigned long long& virt_addr,
		             unsigned& len,
		             unsigned& flags,
		             unsigned long long& magic_word)
{
    buf_first();
    int ret = 0;
    for (int i = 0; i<32; i++) {
        if (buf_next()) {
            if (buf_check_addr_in_range(start_addr, 
                                        end_addr, 
                                        *pVA, 
                                        *pLN)) {
                 virt_addr  = *pVA;
                 len        = *pLN;
                 flags      = *pFL;
                 magic_word = *pMW;
                 ret = 1;
                 break;
	          }
	      }
    }
    return ret;
}
\end{verbatim}

This C code maps into the following instruction sequence:

\begin{verbatim}
bufnxt                   ; locate next used buffer entry
bnbc, #13                ; skip range check if not found  
nop
nop
bufmv r12,bva            ; move buffer parameters to R
bufmv r13,bln
bufrng r11,r2,r8,r12,r13 ; range check
beqz r11,#7              ; skip if no match
nop
nop
bufmv r11,bln            ; copy parameters when match
bufmv r14,bfl
bufmv r8,bmw
ori r16,r0,#1
j #7
bufmv r2,bva
addi r9,r9,#1
slti r11,r9,#32
bnez r11,#-20            ; loop back
\end{verbatim} 

\textit{Instructions needed by the remove operation}.
\begin{itemize}
\item \textbf{bufeq} This instruction evaluates whether two sets of
address parameters (\sw{virtual address} and \sw{length}) are equal.
It is used to check if a certain buffer is present in the \hw{BVA/BLN}
registers. The C intrinsic for this instruction is called
\sw{buf\_check\_buf\_equal()}.
\item \textbf{bufrem} This instruction clears the bit in the \sw{BM}
register that is pointed to by the \sw{BI} index. The C intrinsic for
this instruction is called \sw{buf\_rem()}.
\end{itemize}

The C implementation of the remove operation is given below:

\begin{verbatim}
inline bool buf_remove(unsigned long long virt_addr, 
                       unsigned len)
{
    buf_first();
    for (int i = 0; i<32; i++) {
        if (buf_next()) {            
           if(buf_check_buf_equal(virt_addr,len,*pVA,*pLN)) {
              buf_rem();
              return 1;
           }
        }
    }
    return 0;
}
\end{verbatim}

This C code maps into the following instruction sequence:

\begin{verbatim}
bufnxt                   ; locate next used buffer entry
bnbc, #9                 ; skip equal check if not found  
nop
nop
bufmv r12,bva            ; move buffer parameters to R
bufmv r13,bln
bufeq r11,r0,r14,r12,r13 ; equal test
beqz r11,#3
nop
nop
j #6
bufrem                   ; remove by clearing mask register
addi r2,r2,#1
slti r11,r2,#32
bnez r11,#-16
\end{verbatim}

\subsubsubsection{Results} 
The results obtained for the benchmark code
on the architecture with the dedicated storages are captured in fifth
column "D64SB" of Table~\ref{tab:cyclecnt}. This architecture
update has a significant impact on all operations that are in the
benchmark code.
\begin{itemize}
\item The append function no longer calls \sw{malloc()}. The
\textbf{buffree} and \textbf{bufadd} instructions are single cycle
instructions. This has led to a speedup of a factor 11.6.
\item For the search and remove functions, the dedicated storages and
the hardware intrinsics for range and equal tests have led to a 
speedup of 3.1 and 2.4 respectively.
\end{itemize}

The overall improvement for the complete benchmark is a factor 3.6.
In the following section, we will present further architecture
improvements that will improve the speedup even further.

\subsubsection{Further Architecture Improvements}
The D64SB architecture still has three areas of inefficiency:

\begin{itemize}
\item The address compare unit reads operands from the central
register file.  The two operands that are related to the buffer pool
therefore have to be moved from the \sw{BVA} and \sw{BLN} registers to
the central register file. This is done in separate instructions.
\item The results of a search operation (\sw{virtual address},
\sw{buffer length}, \sw{flags} and \sw{magic word}) need to be copied
from the dedicated registers to the data structures of the application
program, which typically reside in memory. Currently this is done in
two steps:
  \begin{itemize}  
  \item from the dedicated registers to the central register file;
  \item from the central register file to memory;
  \end{itemize}
\item The conditional branch instruction requires three cycles to
fetch the target instruction. One cycle to execute the branch and two
cycles which are implemented as delay slots. In the buffer search
code, there are no useful instructions that can be scheduled in the
delay slots. We will therefore look for architecture improvements with
which we can reduce the number of conditional branches.
\end{itemize}

Following instructions have been added in this architecture variant:

\begin{itemize}
\item \textbf{bufrngbd} and \textbf{bufeqbd}. These are variants of
bufrng and bufeq, but differ in two ways:
  \begin{itemize}
  \item The operands are read directly from the \sw{BVA} and \sw{BLN}
  registers. This will avoid register moves.
  \item The resulting conditions are not stored in the central
  register file, but are written to a second dedicated condition
  register RC.
  \end{itemize}
\item \sw{sva}, \sw{sln}, \sw{sfl} and \sw{smv}. These are store
instructions, which directly store a value from a \sw{BVA}, \sw{BLN},
\sw{BFL} or \sw{BWM} register to the memory. The indexed addressing
mode of DLX is used to provide the memory address.
\item \textbf{bend} and \textbf{bnend}. The bend instruction is a
conditional branch which tests the \sw{BC} condition register (next
buffer found) and the \sw{RC} condition register (range or equal
match), according to the following predicate:
\begin{verbatim}
{!BC || BC & RC}.
\end{verbatim} 
It is used to terminate the search loop when all new buffers have been
tested (\sw{!BC}) or when a match is found (\sw{BC \& RC}). The
\sw{bnend} instruction tests the complementary predicate.
\item \textbf{band} and \textbf{bnand}. The \sw{bend} instruction is a
conditional branch which tests the \sw{BC} and the \sw{RC} condition
registers according to the following predicate: \sw{BC \& RC}. It is
used after the search loop, to test if valid match was found. The
\sw{bnend} instruction tests the complementary predicate.
\end{itemize}

The C implementation of the search operation can now be programmed 
as follows:

\begin{verbatim}
inline bool buf_search(unsigned long long start_addr, 
                       unsigned long long end_addr,
                       unsigned* presult)
{
    buf_first();
    int ret = 0;
    dlx_primitive::uint1 found;
    dlx_primitive::uint1 match;
    do {
        found = buf_next_bool();
        match = check_addr_in_range_bool(start_addr,
                                         end_addr, 
                                         *pVA, 
                                         *pLN);
    } while (!buf_end(found,match));
    if (buf_and(found,match)) {
        *((unsigned long long*)presult) = *pVA;
        presult += 2;
        *((unsigned*)presult) = *pLN;
        presult += 1;
        *((unsigned*)presult) = *pFL;
        presult += 1;
        *((unsigned long long*)presult) = *pMW;
        ret = 1;
    }
    return ret;
}       
\end{verbatim}

Note that the \sw{buf\_search} function now has an argument which is a
pointer result, to the user data structure that holds the buffer
parameters; and that the results are copied directly from the buffer
storages to the user space.

Given these new instructions, the search loop reduces to the 
following instructions:

\begin{verbatim}
bufnxt                   ; locate next used buffer entry
bufrngbd r2,r5           ; range check
bnend, #-4               ; loop back when not found    
nop
nop
\end{verbatim}

The results obtained for the benchmark code on the architecture with
the dedicated storages are captured in the sixth column of
Table~\ref{tab:cyclecnt}. Compared to the D64SB versions, there is an
improvement of a factor 2.6 for the search and remove operations.

\subsection{Hardware Characteristics}

In order to gain insight in the hardware cost of the architecture
specialization, in a first experiment an RTL implementation of the
different DLX variants was generated with the IP Designer tool-suite
and subsequently synthesized into a standard cell based
implementation.  The target library uses a 65nm general purpose
technology. As a synthesis tool, Synopsys DC Expert was used. For a
target frequency of 200MHz, the gate count results shown in
Table~\ref{tab:gatecnt} were obtained.

\begin{table}[!hbt]
\centering
\setlength\extrarowheight{2pt}
\begin{tabular}{|ccccc|}
\hline
\hline
\textbf{DLX}  & \textbf{D64} & \textbf{D64AC} & \textbf{D64SB} & \textbf{D64OPT}  \\
34722         &  42700       &  44152         &  101987        & 107712           \\
\hline
\hline
\end{tabular}
\caption{Gate counts for different ASIP architectures.}
\label{tab:gatecnt}
\end{table} 

The gate count increase from the 32 bit to the 64 bit data path and
register file is 23\%, which is reasonable. The cost of the \hw{ACU}
is 3\% of the total cost of D64AC, which can also be considered a
small increase. A big increase in gate count, of a factor 2.3, can be
found in the designs to which the dedicated storages were added.
However, this hardware cost pays off since these architectures also
give the best acceleration of the application code.

Note that, while these results were obtained for a clock frequency of
200 MHz, the maximal achievable frequency for this design in the 65nm
technology is 330 MHz.

The D64OPT design was also mapped onto FPGAs. For a first mapping on
Altera Cyclone V (device type 5CEFA7F23I7), the logic utilization was
at 13\%, while the maximum frequency was 93 MHz (table
\ref{tab:CycloneV_synt_summary}).

\begin{table}[!htb]
\centering
\begin{tabular}{|l|l|}
\hline\hline
Logic utilization (in ALMs)& 7,601 / 56,480 (13\%) \\
Combinational ALUTs&  9,455 / 112,960 (8\%) \\
Memory ALUTs & 7,664 / 56,480 (14\%) \\
Dedicated logic registers & 9,455 / 112,960 (8\%) \\
Total pins&  7 / 240 (3\%) \\
Total block memory bits&  393,216 / 7,024,640 (6\%) \\
DSP block 18-bit elements & 0 / 156 (0\%) \\
Total PLLs & 1 / 7 (14\%) \\
Fmax  & 93,56 Mhz  \\
\hline
\end{tabular}
\caption{Logic usage summary of Target VHDL Code on CycloneV 5CEFA7F23I7}
\label{tab:CycloneV_synt_summary}
\end{table}

For a second mapping on Altera Stratix V, 
the maximum clock frequency is 180 MHz, which is
approximately the same frequency at which the \nios core is clocked.

\begin{table}[!htb]
\centering
\begin{tabular}{|l|l|}
\hline\hline
Logic utilization (in ALMs)& 7,323 / 234,720 (3\%) \\
Combinational ALUTs&  7 / 232,960 (30\%) \\
Memory ALUTs &9,513 / 469,440 (<1\%) \\
Dedicated logic registers & 9,513 / 469,440 (<1\%) \\
Total pins&  7 / 864 (<1\%) \\
Total block memory bits&  393,216 / 52,428,800 (<1\%) \\
DSP block 18-bit elements & 0 / 156 (0\%) \\
Total PLLs & 0 / 80 (0\%) \\
Fmax  & 179,99 Mhz  \\
\hline
\end{tabular}
\caption{Logic usage summary of Target VHDL Code on StratixV}
\label{tab:StratixV_synt_summary}
\end{table}

\subsection{Conclusion and Next Steps}
In this chapter of the deliverable report, we described the design of
an ASIP to accelerate the DNP functionality, with the aim of reducing
the latency of RDMA transfers. The new DNP ASIP architecture includes
the basic functionality of a 32-bit microprocessor. The architecture
has been further optimized in a stepwise refinement process, analyzing
\eg the cycle count and storage profiles produced by the IP Designer
tool-suite.

Key architectural features of the DNP ASIP include support for 64-bit
operations, a dedicated register-file structure for efficient data
buffer storage, and a dedicated address generation unit with a 64-bit
range instruction.

The DNP ASIP can be programmed in C using IP Designer's retargetable C
compiler. For buffer management, intrinsic function calls are
available in C code, \eg to locate the next active entry in the buffer
table, or to perform address range comparisons.
 
Benchmark buffer-search programs were compiled on the new DNP ASIP
architecture. Results show a cycle-count reduction by more than 85\%
compared to the existing implementation on the \nios microprocessor.
The clock frequency of the ASIP is comparable to that of \nios.
Therefore the buffer search already executes significantly faster on
the ASIP compared to NIOS. Further cycle-time improvements are within
reach.

Further extensions of the DNP ASIP are planned in the next project
period. First we will investigate to what extent dedicated functional
units, storages and instructions can speed up the execution of
virtual-to-physical memory address translation. Secondly we will
investigate if more data computation (as opposed to data
communication) support can be integrated in the DNP ASIP.  Possible
extensions could include the addition of a floating-point unit or a
bit-manipulation unit in the ASIP.

%\subsection{Specification (Target, Alessandro)}
%\subsubsection{We run their benchmark on NIOS (Laura)}
%\subsection{Target Activity (System Design, Benchmarking) (Target)}
%
%\subsection{Target sent VHDL for synthesis}
% \subsubsection{Synthesis of Target VHDL Code on Stratix IV/Cyclone V}
%
%The VHDL code produced by Target is synthesized on CycloneV and
%StratixV with Quartus II software.   and \ref{tab:StratixV_synt_summary}
%show resource utilization information for both devices.

%% file: chapter5.tex
\section{Development on 28nm FPGA}
\label{sec:28nm}
\subsection{DNP on Altera Development kit Stratix V} % - Non optimized (preliminary) - Francesca

Stratix V FPGA is an Altera 28-nm device, which has an enhanced core
architecture and offers up to 66 integrated 14.1 Gbps transceivers.
Moreover, its design advancements such as the \PCIe HARD IP
implementation, partial reconfiguration and programmable power
technology, ensure that designs implemented in the Stratix V GX FPGAs
operate faster, with lower power consumption than in previous FPGA
families.

We ran a preliminary synthesis of \apenetp on the Stratix V;
some changes to the \apenetp firmware have been performed to adapt
the hardware code to a different FGPA family. 
%
%to do that, some changes have been made to the \apenetp firmware.
%
%Using a configurable hard IP block, rather than programmable
%logic, saves significant FPGA resources

The main responsible of the management of \PCIe interface are Altera
\PCIe HARD IP and PLDA core. For the past generation of FPGA, Altera
provided the ALTGX megafunction as a general purpose transceiver 
solution.
On the contrary, the Stratix V family offers protocol-specific IPs,
which guarantee high performance and simplify the parameterization
process.
The \PCIe HARD IP supports Gen3, Gen2, and Gen1 end point and root port
up to x8 lane configurations.
%
%http://www.alterawiki.com/uploads/b/b5/Stratix_V_hard_IP_For_PCIe.pdf
%
Current implementation provides a \PCIe HARD IP in Gen2 mode 
(5.0 Gbps\slash lane), but we plan the development of a new version
that takes advantage of Gen3 data rate (8.0 Gbps\slash lane).
%
%in the next future we will use \PCIe configuration for Gen3 data rate.
%
% a PCI Express link Gen2 communication protocol
%
% rx_st_empty[1:0] For 128-bit data, only bit 0 applies-
%
%Altera recommends resetting the Stratix V Hard IP for PCI Express when an uncorrectable (double-bit) error correction coding ECC error is
%detected (When an uncorrectable ECC error is detected, rx_st_err is asserted for at least 1 cycle while rx_st_valid is asserted).
%coreclkout_hip: This is a fixed frequency clock used by the Data Link and Transaction Layers. To meet PCI
%Express link bandwidth constraints, this clock has minimum frequency requirements as listed
%in Table 8–3 on page 8–6 (×8 Gen2 avalon@128bit =>250 MHz)
%reset_status: When asserted, this signal indicates that the Hard IP clock is in reset.
%You should use reset_status to drive the reset of your application.
%hotrst_exit: This signal should cause the Application Layer to be reset
%You can use the Altera Transceiver Reconfiguration Controller to dynamically reconfigure analog settings in Stratix V devices.
%pin_perst!!!
%
%The PCI Express edge connector also has a presence detect feature for the
%motherboard to determine if a card is installed. A jumper is provided to optionally
%connect PRSNT1n to any of the three PRSNT2n pins found within the x8 connector
%definition. This is to address issues on some PC systems that would base the
%link-width capability on the presence detect pins versus a query operation.
%

Stratix V HARD IP for \PCIe includes an embedded reset controller
and provides an Avalon interface to access to the Application Layer
(for more information about the Avalon protocol refer to 
http:\slash\slash www.altera.com\slash literature\slash manual\slash mnl\_avalon\_spec.pdf).

The multi-port DMA engine for \PCIe for Altera Devices is a PLDA IP;
the core PLDA wraps around Altera \PCIe HARD IP and leverages on the 
renowned PLDA EZDMA interface, providing multi-channel DMA capability. 

\begin{figure}[!hbt]
  \centering
  \includegraphics[width=\textwidth]{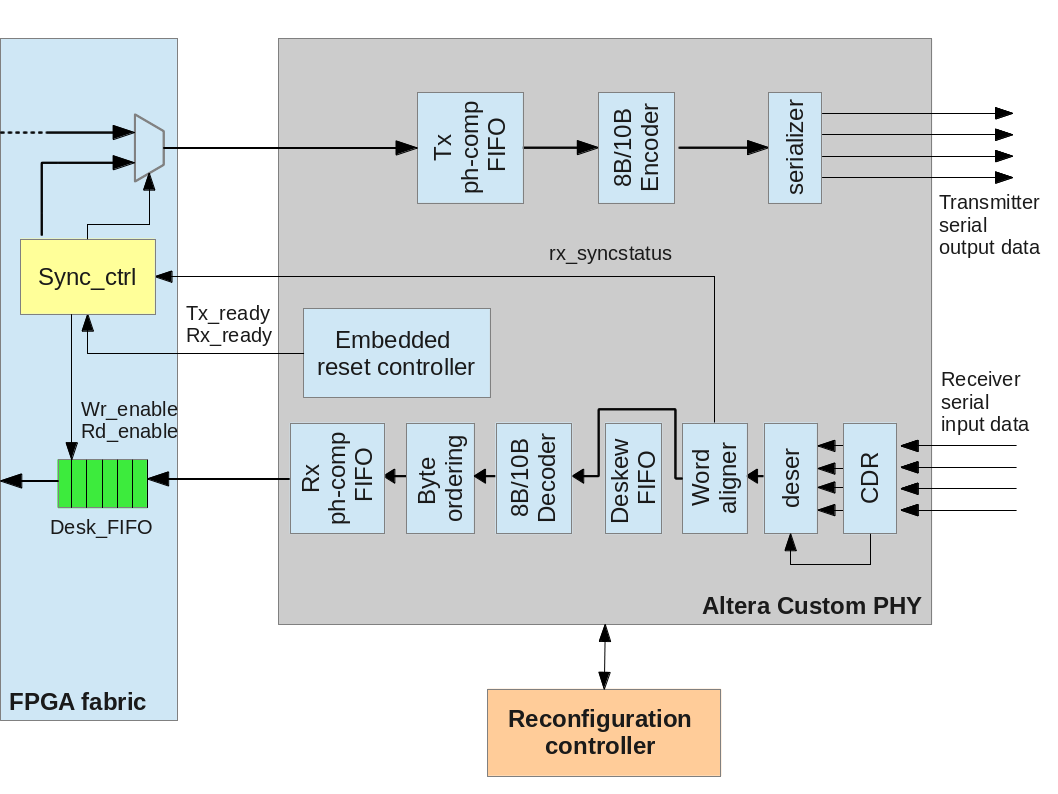}
  \caption{Architecture of \apenetp's channel.}
  \label{fig:channel}
\end{figure}

The \hw{TORUS LINK} block is composed by an Altera Custom IP
Core with the correspondent reconfiguration block, and a proprietary
channel control logic (\hw{SYNC\_CTRL}) (as shown in figure \ref{fig:channel}).
%
%StratixV GX channels have two receiver equalizer gain bandwidth modes: half-bandwidth and full-bandwidth.
%The half-bandwidth data rate is up to 6.25 Gbps; the full-bandwidth data rate is from 6.25 Gbps to 14.1 Gbps.
%You can select the mode in the Assignment Editor of the Quartus II software (Receiver Equalizer Gain
%Bandwidth Select).
%Each Stratix V transceiver is divided into two blocks: physical medium attachment (PMA) and physical coding
%sublayer (PCS). The PMA block connects the FPGA to the channel, generates the required clocks, and
%converts the data from parallel to serial or serial to parallel. The PCS block performs digital processing logic
%between the PMA and the FPGA core.
%
%Custom PHY Correspondences between Stratix IV GX Device and Stratix V Device Signals
%Avalon-ST TX Interface
%tx_datain tx_parallel_data [<d><n>-1:0]
%tx_ctrlenable tx_datak [<d><n>-1:0]
%rx_ctrldetect rx_datak
%
The Altera transceiver supports proprietary protocols and implements
4 bi-directional lanes bonded into a single channel.
%
%bonded= a single clock drives multiple lanes, reducing clock skew.
%
On the receiving side each transceiver includes \hw{Word Aligner}, 
\hw{Deskew FIFO}, 8B\slash10B decoder, \hw{Byte Ordering Block} and 
\hw{RX phase compensation FIFO}. 
%
%CDR LTD mode (handbook)
%LTD mode to recover the clock from the incoming serial data.
%
%DESER (handbook)
The \hw{Deserializer} clocks serial input data using the high-speed
serial clock recovered from the receiver clock data recovery (\hw{CDR}) 
block, and deserializes the data using the low-speed parallel recovered
clock.
%
%compensation FIFO
%The high-speed serial clock and low-speed parallel clock skew between channels and unequal latency in the
%transmitter phase compensation FIFO contribute to transmitter channel-to-channel skew. In bonded channel
%configurations the parallel clock is generated by a central clock divider for all channels, rather than using a
%local clock divider for each transmitter channel. Also, the transmitter phase compensation FIFO in all bonded
%channels shares common pointers and control logic generated in the central clock divider, resulting in equal
%latency in the transmitter phase compensation FIFO of all bonded channels. The lower transceiver clock
%skew and equal latency in the transmitter phase compensation FIFOs in all channels provide lower
%channel-to-channel skew in bonded channel configurations.
%
%CDR
%You can configure the CDR in either automatic lock mode or manual lock mode. By default, the Quartus II
%software configures the CDR in automatic lock mode.
%You can switch the CDR to manual mode by writing the
%pma_rx_setlocktodata or pma_rx_set_locktoref registers to 1.
%
The \hw{Word Aligner} receives parallel data from the \hw{Deserializer} 
and restores the word boundary based on a pre-defined alignment pattern
that must be received during link synchronization.
In Stratix IV design the word alignment operation was manually 
controlled by the input signal \hw{rx\_enapatternalign}: a rising edge
on the \hw{rx\_enapatternalign} signal triggers the \hw{Word Aligner} to 
look for the word alignment pattern in the received data stream.
%
%To access control and status registers in the Stratix V Custom PHY, an
%embedded controller with an Avalon-MM master interface is required.
%Also rx\_enapatternalign is asserted using the Avalon-MM interface (writing to register address 0x085).
%embedded controller with an Avalon-MM master interface is required.
%

In Stratix V Custom IP \hw{rx\_enapatternalign} is a status register
that can be asserted using the Avalon interface. In the current
implementation of the DNP on Stratix V, we used the automatic 
synchronization State Machine mode \hw{Word Aligner}. 
%
%alignment is controlled by a programmable state machine
%
%The Word alignment pattern used is K.28.5 †	188	BC	001111 1010 (RD-)	110000 0101 (RD+)
%
The status ports \hw{rx\_syncstatus} high indicates that 
synchronization is acquired.
%
%presence or absence of synchronization on the RX interface.
%
The \hw{Deskew FIFO} circuitry is able to align the data across
multiple lanes. Unfortunately this circuitry is available only in XAUI
protocol, thus we implement our DESKEW FIFOs connected to the lanes of
the channels. 
The rate match FIFO %(clock rate compensation) 
compensates for small
clock frequency differences between the upstream transmitter and the
local receiver clocks.
The 8B/10B decoder receives 10-bit data from the \hw{Word Aligner} and
decodes it into 2-bit control identifier and 8-bit data; this data may
or may not match the original byte ordering of the transmitted data.
The \hw{byte ordering} looks for the user-programmed byte ordering 
pattern in the parallel data:
%
%You must select a byte ordering pattern that you know is at the LSBytes position of the parallel transmitter
%data. 
if it finds the byte ordering pattern in the MSB position of the
data, it inserts pad bytes to push the byte ordering pattern to the 
LSByte(s) position, thereby restoring proper byte ordering.
Finally, the \hw{RX phase compensation FIFO} compensates for the phase
difference between the parallel receiver clock and the FPGA fabric 
clock.
Similarly on the transmitter side each transceiver includes 
\hw{TX phase compensation FIFO}, \hw{8B\slash10B encoder} and 
\hw{Serializer}. 
%
%Transmitter PLL
%I (transceiver user guide)
%For Stratix V devices, you can select either the CMU or ATX PLL.
%The CMU PLL has a larger frequency range than the ATX PLL. The
%ATX PLL is designed to improve jitter performance and achieves
%lower channel-to-channel skew; however, it supports a narrower
%range of data rates and reference clock frequencies. Another
%advantage of the ATX PLL is that it does not use a transceiver
%channel, while the CMU PLL does.
%
%II (handbook)
%When you use the channel PLL as aCMU PLL, that particular channel cannot be used as a receiver; however,
%that channel can be used as a transmitter in conjunction with a different transmitter PLL. If all transmitters
%and receivers within the transceiver block are required, you must use an ATX PLL or a clock from another
%transceiver block.
%
%SER (handbook)
%The serializer converts the incoming low-speed parallel data from the transceiver PCS or FPGA fabric to
%high-speed serial data and sends the data to the transmitter buffer.
%
%What is sent over the link is 8B10B-encoded to maintain the DC balance in the serial data transmitted.

To ensure the correct initialization of the Custom IP Core, a right
reset sequence has to be followed. %reliable
In Stratix IV design we implemented a custom reset controller block to 
reset the ALTGX Transceiver; Stratix V's Custom IP allows to 
implement an external reset controller logic with a sequence very 
similar  to the Stratix IV ALTGX's one, but it also supplies an 
embedded reset control which automatically performs the entire 
transceiver's reset sequence.
%
% whenever the phy\_mgmt\_clk_reset signal is triggered. 
% Assert the mgmt_rst_reset signal of the transceiver reconfiguration controller before or at the
% same time as phy_mgmt_clk_reset to start calibration
% During device power-up, mgmt_rst_reset and phy_mgmt_clk_reset must be asserted to
% initialize the reset sequence. phy_mgmt_clk_reset holds the transceiver blocks in reset and
% mgmt_rst_reset is required to start the calibration IPs. Both these signals should be held asserted
% for a minimum of two phy_mgmt_clk clock cycles. Deassert phy_mgmt_clk_reset at the same
% time as mgmt_rst_reset.
%
%Another option to reset the transceivers might be to use Altera's Transceiver PHY Reset Controller.
%
After reset sequence is completed the \hw{tx\_ready} and \hw{rx\_ready}
status signals are asserted, and the \hw{Sync\_ctrl} can start the 
synchronization's procedure: transceivers of different nodes connected
together have to be align to allow a correct data transmission.

%
%after the transmitter and the receiver calibration. 
%the transmitter is ready to transmit data, the receiver is ready to receive data.

\hw{Sync\_ctrl} sends the alignment pattern \hw{Align Word} for a 
constant number of clock cycles (\checkalign phase), at the 
same time looking for the same word in the received data stream. 
If \hw{Align Word} is received, this block signals that synchronization
is acquired by sending the \hw{Align\_ok Word}, otherwise the 
\checkalign phase restarts after the \restartalign state.
When both nodes are aligned and receive the \hw{Align\_ok Word}, 
channel alignment is assured and the deskewing operation starts: 
\hw{Desk\_FIFO} write enable is asserted, while the read enable signal
is asserted when all FIFOs are no longer empty.
%
%Beyond Alignment and Deskewing,  the  link alignment procedure has an Acknowledge 
%phase. Once they get deskewed, transceivers of different nodes connected together have 
%to communicate each other their  accomplished synchronization; this is done by another 
%specific acknowledge word. Once got, channel is ready to be used.
%
%Modifying programmable values within transceiver buffers can be performed by a single reconfiguration
%controller for the entire FPGA,
%additional circuitry to improve signal integrity
%Pre-emphasis can maximize the data eye opening at the far-end receiver.

\begin{figure}[!htb]
  \centering
  \includegraphics[width=0.8\textwidth]{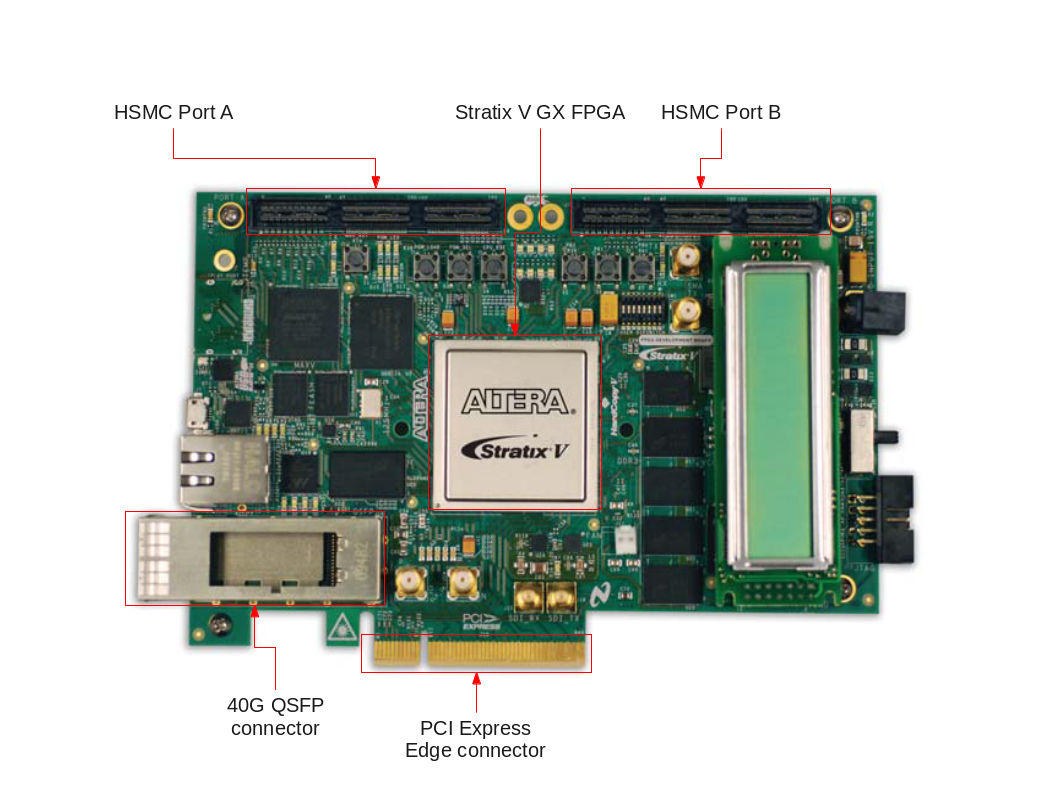}
  \caption{Architecture of \apenetp's channel.}
  \label{fig:board}
\end{figure}

We developed the \apenetp design on the Altera DK-DEV-5SGXEA7N 
development kit, a complete design environment that features a Stratix
V GX FPGA (5SGXEA7K2F40C2N). This kit also includes a \PCIe slot, a 
40G QSFP connector and two HSMC ports (figure \ref{fig:board}). 
%
%On-Board USB-Blaster II is the default method for configuring the FPGA at any
%time using the Quartus II Programmer in JTAG mode with the supplied
%micro-USB cable.
%By default, the on-board USB-Blaster II clocks TCK at 24 MHz. For the on-board
%USB-Blaster II to function correctly, you must set the Quartus II clock constraint on the
%internal_tck input signal to 24 MHz.

The \PCIe edge connector supports up to x8 signaling in either Gen1,
Gen2, or Gen3 mode.
%
%PCIe
% PCIE_REFCLK = Motherboard reference clock (Off-Board Clock Inputs) LVDS input from the PCI Express edge connector
% 100-MHz differential input that is driven from the PC motherboard to the board through the PCI Express edge connector.
% This signal connects directly to a Stratix V GX FPGA REFCLK input pin pair using DC coupling.
% This clock is terminated on the motherboard and therefore, no on-board termination is required.
% The I/O standard is High-Speed Current Steering Logic (HCSL).
% PCIE_PERSTN = Reset
%
%
%%36 transceivers (12.5 Gbps)
% QSFP (V35) 282.500 MHz clkin_425_x
% qsfp_modsel,qsfp_reset,qsfp_modprs,qsfp_int,qsfp_lpmode COMMENTATI (ma controllati=> OK!!)

The QSFP connector uses four transceiver lanes from the FPGA device;
we used this component to achieve the X+ channel. 
%
%takes in serial data from the Stratix V GX FPGA device and transform them to optical signals.
%
The signal generated by a 282.500 MHz on-board oscillator is used as a
reference clock; in the next versions we will use the 644.53125 MHz 
on-board oscillator.
%
%REFCLK2_QL1_P    AB34
%REFCLK2_QL1_N    AB35

The development board contains also two High-Speed Mezzanine Card 
(HSMC) connectors, called port A and port B, to provide 8 channels in
port A and 4 channels in port B of 10.0 Gbps transceivers.
Both ports use a 625.000 MHz on-board oscillator as reference clock,
and can be removed or included in the active JTAG chain by means of a 
DIP switch.  
%
%(SW3)
% To connect a device or interface in the chain, their corresponding switch must be in the OFF position.
%  2  ON : Bypass HSMC port A.    OFF : HSMC port A in-chain.
%  3  ON : Bypass HSMC port B.    OFF : HSMC port B in-chain.
%

In Table \ref{tab:StratixV_apenet} is shown the Altera Quartus II 
synthesis report for our system.

\begin{table}[htbp]
\centering
\begin{tabular}{|l|l|}
\hline
\hline
Logic utilization (in ALMs)   & 37,345 / 234,720 ( 16\%)        \\
%Combinational ALUTs&  \\
%Memory ALUTs & \\
Dedicated logic registers     & 56,834 (12\%)                   \\
Total pins                    & 215 / 864 ( 25 \% )             \\
Total block memory bits       & 7,450,168 / 52,428,800 ( 14 \% )\\
DSP block                     & 2 / 256 ( < 1 \% )              \\
Total PLLs                    & 28 / 80 ( 35 \% )               \\
%Fmax  & Mhz  \\
\hline
\end{tabular}
\caption{Stratix V logic usage overview.}
\label{tab:StratixV_apenet}
\end{table}

%\subsection{Measures (Link Speed, Clock freq muP, Area) - Ottorino}
%\subsection{2xGEN2 - Ottorino}
%
%Stratix V devices have up to 4 instances of the Hard IP for PCI Express.

%% file: glossary.tex
\section{Glossary}
The following table contains acronyms and terms used in this document.
\begin{longtable}{|l|p{0.85\textwidth}|}
\hline
\hline
AED & Abstract Execution Device.\\
\hline
ALUT & Adaptive LookUp Table.\\
\hline
API & Application Programming Interface.\\
\hline
\apenetp & An FPGA-based card for low latency, high bandwidth direct network interconnection based on the DNP.\\
\hline
ASIP & Application Specific Instruction Set Processor.\\
\hline
BER & Bit Error Rate.\\
\hline
BFS & Breadth-First Search.\\
\hline
BML & Byte Management Layer, framework of the OpenMPI library.\\
\hline
BTL & Byte Transfer Layer, framework of the OpenMPI library.\\
\hline
CDR & Clock Data Recovery.\\
\hline
CPLD & Complex Programmable Logic Device.\\
\hline
CQ & Completion Queue.\\
\hline
CRC & Cyclic Redundancy Check.\\
%\hline
%CPU & Central Processing Unit.\\
\hline
DAL & Distributed Application Layer.\\
\hline
DFM & DNP Fault Manager.\\
\hline
DNAOS & DNA is Not just Another Operating System.\\
\hline
DNP & Distributed Network Processor, it is the core of the \apenetp board.\\
\hline
DPSNN & Distributed Polychronous Spiking Neural Networks (a PSNN code natively redesigned and rewritten to exploit parallel/distributed computing systems).\\
\hline
DWR & DNP Watchdog Register.\\
\hline
ECC & Error Correcting Code.\\
\hline
ELF & Extensible Linking Format.\\
\hline
\euretile & EUropean REference TILed architecture Experiment.\\
\hline
%FIFO & First In First Out.\\
%\hline
FIT & Failures In Time.\\
\hline
FLOPS & FLoating-point Operations Per Second.\\
\hline
FM & Fault Manager.\\
\hline
FPGA & Field-Programmable Gate Array.\\
\hline
HAL & Hardware Abstraction Layer.\\
\hline
HCA & Host Channel Adapters.\\
\hline
HdS & Hardware dependent Software.\\
\hline
HFM & Host Fault Manager.\\
\hline
HPC & High Performance Computing.\\
\hline
HW & Hardware.\\
\hline
HWR & Host Watchdog Register.\\
\hline
IC & Integrated Circuit.\\
\hline
IDE & Integrated Development Environment (tools to develop and debug embedded software, integrated in a GUI).\\
\hline
INFN & Istituto Nazionale di Fisica Nucleare (National Institute for Nuclear Physics).\\
\hline
I/O & Input/Output.\\
\hline
IOCTL & Input/Output Control, is a system call for device-specific input/output operations and other operations which cannot be expressed by regular system calls.\\
\hline
IP & Intellectual Property.\\
\hline
IP Designer & TARGET's tool-suite for the design and programming of ASIPs.\\
\hline
IPMI & Intelligent Platform Management Interface.\\
\hline
ISS & Instruction Set Simulator.\\
\hline
IT & Information Technology.\\
\hline
LDM & LiFaMa Diagnostic Message.\\
\hline
LiFaMa & Link Fault Manager.\\
\hline
\lofamo & Local Fault Monitor, an HW/SW approach to obtain Systemic Fault Awareness.\\
\hline
LQCD & Lattice Quantum-ChromoDynamics.\\
\hline
LSB & Least Significant Bit.\\
\hline
LTP & Long Term synaptic Potentiation.\\
\hline
LUT & Look-Up Table.\\
\hline
MM & Memory Management.\\
\hline
MPI & Message Passing Interface.\\
\hline
MP-SoC &  MultiProcessor System-on-Chip.\\
\hline
MSB & Most Significant Byte.\\
\hline
MTL & Matching Transport Layer, framework of the OpenMPI library.\\
\hline
NIC & Network Interface Controller.\\
\hline
NIOS II & 32-bit microprocessor available as soft-core in Altera FPGAs (often shorthanded as "NIOS").\\
\hline
nML & Not a Modelling Language (a processor architectural description language).\\
\hline
OMPI & Open MPI, an implementation of the MPI standard.\\
\hline
OPAL & Point-to-point Message Layer.\\
\hline
ORTE & Open Run-time Environment, part of the OpenMPI library.\\
\hline
OS & Operating System.\\
\hline
OSI & Open Systems Interconnection.\\
\hline
OSU & Ohio State University.\\
\hline
OMB & OSU Micro-Benchmarks.\\
%\hline
%PCI & Peripheral Component Interconnect.\\
\hline
PCS & Physical Coding Sublayer.\\
\hline
PCU & Processor Control Unit.\\
\hline
PDC & Pin-Down Cache.\\
\hline
PDG & Primitives Definition and Generation.\\
\hline
PMA & Physical Medium Attachment.\\
\hline
PML & Point-to-point Message Layer, framework of the OpenMPI library.\\
\hline
Presto & MPI-like library for APEnet+/DNP.\\
\hline
PRBS & Pseudorandom Binary Sequence.\\
\hline
PSNN & Polychronous Spiking Neural Network.\\
\hline
P2P & peer-to-peer.\\
\hline
QUonG & LQCD on GPU platform.\\
\hline
RB & Ring Buffer.\\
\hline
RDMA & Remote Direct Memory Access.\\
\hline
RDMA GET & RDMA READ operation that implies an handshake between the sender and the receiver.\\
\hline
RDMA PUT & RDMA WRITE operation that implies an handshake between the sender and the receiver.\\
\hline
RTL & Register Transfer Level (also used to refer to register-transfer languages, such as VHDL or Verilog)\\
\hline
RX & Receive.\\
\hline
SIMD & Single Instruction Multiple Data (also known as vector processing, a processor architectural concept to implement data-level parallelism).\\
\hline
SIP & Software Interface Protocol.\\
\hline
SNET & Service Network.\\
\hline
STDP & synaptic Spiking Time Dependent Plasticity.\\
\hline
SW & Software.\\
\hline
TCL & Tool Command Language.\\
\hline
T\&D & Test \& Debug.\\
\hline
TX & Transmission.\\
\hline
UDP & User Datagram Protocol.\\
\hline
V2P & Virtual To Physical.\\
\hline
VEP & Virtual EURETILE Platform.\\
\hline
VHDL & VHSIC Hardware Description Language.\\
\hline
WD & WatchDog.\\
\hline
\hline
\end{longtable}
\pagebreak